\documentclass[12pt,twoside,a4paper]{article}
\pdfoutput=1

\usepackage{amsmath, amsthm, amsfonts, amssymb}
\usepackage{graphicx}
\usepackage{multirow}
\usepackage{float}
\usepackage{afterpage}
\usepackage{color}
\usepackage[width=\textwidth,font=small,labelfont=bf,
labelsep=endash]{caption}

\normalsize
\long\def\@makecaption#1#2{%
  \vskip\abovecaptionskip
  \sbox\@tempboxa{\small #1. #2}%
  \ifdim \wd\@tempboxa >\hsize
    \small #1. #2\par
  \else
    \global \@minipagefalse
    \hb@xt@\hsize{\hfil\box\@tempboxa\hfil}%
  \fi
  \vskip\belowcaptionskip}

\newcommand\Lame {Lam\'e\ }

\newcommand\Schrodinger {Schr\"{o}dinger }

\newtheorem{theorem}{Theorem}[section]
\newtheorem{problem}{Problem}[section]

\renewcommand{\thesection}{}
\renewcommand{\thesubsection}{\arabic{section}.\arabic{subsection}}
\renewcommand{\theequation}{\arabic{section}.\arabic{equation}}

\makeatletter
\def\@seccntformat#1{\csname #1ignore\expandafter\endcsname\csname the#1\endcsname\quad}
\let\sectionignore\@gobbletwo
\let\latex@numberline\numberline
\def\numberline#1{\if\relax#1\relax\else\latex@numberline{#1}\fi}
\makeatother

\begin{document}

\title{Four Lectures on Weierstrass Elliptic Function and Applications in Classical and Quantum Mechanics}
\author{Georgios Pastras$^1$}
\date{$^1$NCSR ``Demokritos'', Institute of Nuclear and Particle Physics\\15310 Aghia Paraskevi, Attiki, Greece\\\texttt{pastras@inp.demokritos.gr}}

\vskip .5cm

\maketitle

\abstract{In these four lectures, aiming at senior undergraduate and junior graduate Physics and Mathematics students, basic elements of the theory of elliptic functions are presented. Simple applications in classical mechanics are discussed, including a point particle in a cubic, sinusoidal or hyperbolic potential, as well as simple applications in quantum mechanics, namely the $n = 1$ \Lame potential. These lectures were given at the School of Applied Mathematics and Physical Sciences of the National Technical University of Athens in May 2017.}

\newpage

\tableofcontents

\newpage

\setcounter{equation}{0}
\section{Lecture 1: Weierstrass Elliptic Function}
\label{sec:elliptic}
\subsection{Prologue}
\label{subsec:introduction}
In the present lectures, basic elements of the theory of elliptic functions are presented and simple applications in classical and quantum mechanics are discussed. The lectures target an audience of senior undergraduate or junior graduate Physics and Mathematics students. For lectures 1 and 2, the audience is required to know basic complex calculus including Cauchy's residue theorem. For lecture 3, basic knowledge on classical mechanics is required. For lecture 4, the audience is required to be familiar with basic quantum mechanics, including \Schrodinger 's equation and Bloch's theorem.

The original constructions of elliptic functions are due to Weierstrass \cite{Weierstrass} and Jacobi \cite{Jacobi}. In these lectures, we focus on the former. Excellent pedagogical texts on the subject of elliptic functions are the classic text by Watson and Whittaker\cite{Whittaker} and the more specialized text by Akhiezer \cite{Akhiezer}. Useful reference handbooks with many details on transcendental functions including those used in these lectures are provided by Bateman and Erd\'{e}lyi, \cite{Bateman}, which is freely available online, as well as the classical reference by Abramowitz and Stegun \cite{Abramowitz}.

On the applications in quantum mechanics we will meet the \Lame equation. Historically, this equation was studied by \Lame towards completely different applications \cite{Lame}. An excellent treatment of this class of ordinary differential equations in given by Ince in \cite{Ince}

The presenter took advantage of the experience acquired during his recent research on classical string solutions and minimal surfaces to prepare these lectures. The applications of elliptic functions in physics extend to many, much more interesting directions.

\subsection{Elliptic Functions}
\label{subsec:elliptic}
\subsubsection*{Basic Definitions}
Assume a complex function of one complex variable $f\left( z \right)$ obeying the property
\begin{equation}
f\left( {z + 2{\omega _1}} \right) = f\left( z \right),\quad f\left( {z + 2{\omega _2}} \right) = f\left( z \right) ,
\label{eq:doubly_periodic}
\end{equation}
for two complex numbers $\omega_1$ and $\omega_2$, whose ratio is not purely real (thus, they correspond to different directions on the complex plane). Then, the function $f$ is called \emph{doubly-periodic} with periods $2\omega_1$ and $2\omega_2$. A meromorphic, doubly-periodic function is called an \emph{elliptic} function.

The complex numbers $0$, $2 \omega_1$, $2 \omega_2$ and $2 \omega_1 + 2 \omega_2$ define a parallelogram on the complex plane. Knowing the values of the elliptic function within this parallelogram completely determines the elliptic function, as a consequence of \eqref{eq:doubly_periodic}. However, instead of $2 \omega_1$ and $2 \omega_2$, one could use any pair of linear combinations with integer coefficients of the latter, provided that their ratio is not real. In the general case the aforementioned parallelogram can be divided to several identical cells. If $2 \omega_1$ and $2 \omega_2$ have been selected to be ``minimal'', or in other words if there is no $2 \omega$ within the parallelogram (boundaries included, vertices excepted), such that $f \left( z + 2 \omega \right) = f \left( z \right)$, then the parallelogram is called a \emph{fundamental period parallelogram}. Two points $z_1$ and $z_2$ on the complex plane whose difference is an integer multiple of the periods
\begin{equation}
z_2 - z_1 = 2m \omega_1 + 2n  \omega_2 , \quad m,n\in \mathbb{Z},
\end{equation}
are called \emph{congruent} to each other. For such points we will use the notation
\begin{equation}
{z_1} \sim {z_2} .
\end{equation}
Obviously, by definition, the elliptic function at congruent points takes the same value,
\begin{equation}
{z_1} \sim {z_2} \Rightarrow f \left( z_1 \right) = f \left( z_2 \right).
\end{equation}

A parallelogram defined by the points $z_0$, $z_0 + 2 \omega_1$, $z_0 + 2 \omega_2$ and $z_0 + 2 \omega_1 + 2 \omega_2$, for any $z_0$, is called a ``\emph{cell}''. It is often useful to use the boundary of an arbitrary cell instead of the fundamental period parallelogram to perform contour integrals, when poles appear at the boundary of the latter.
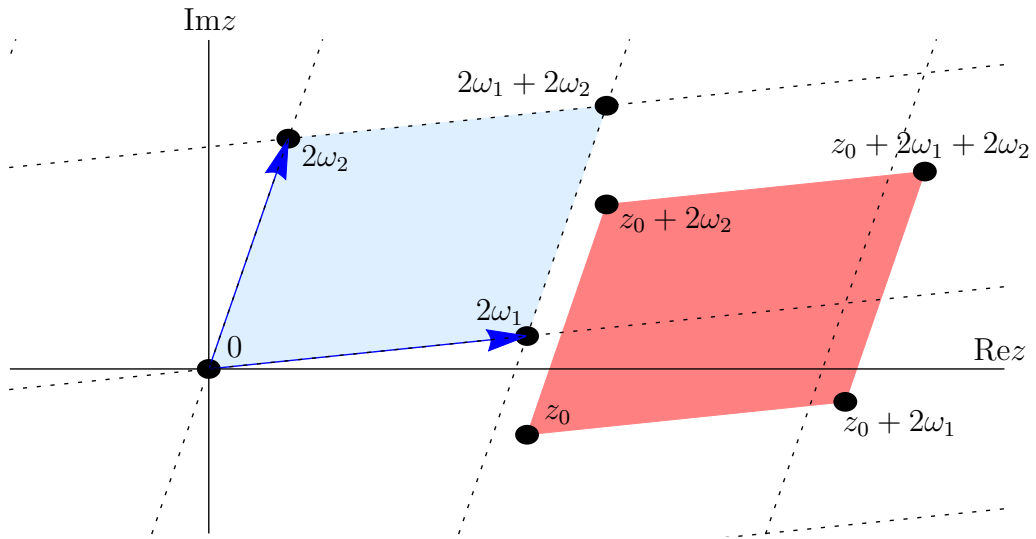
\begin{figure}[h]
\vspace{10pt}
\begin{center}
\begin{picture}(100,43)
\put(10,2){\includegraphics[width = 0.8\textwidth]{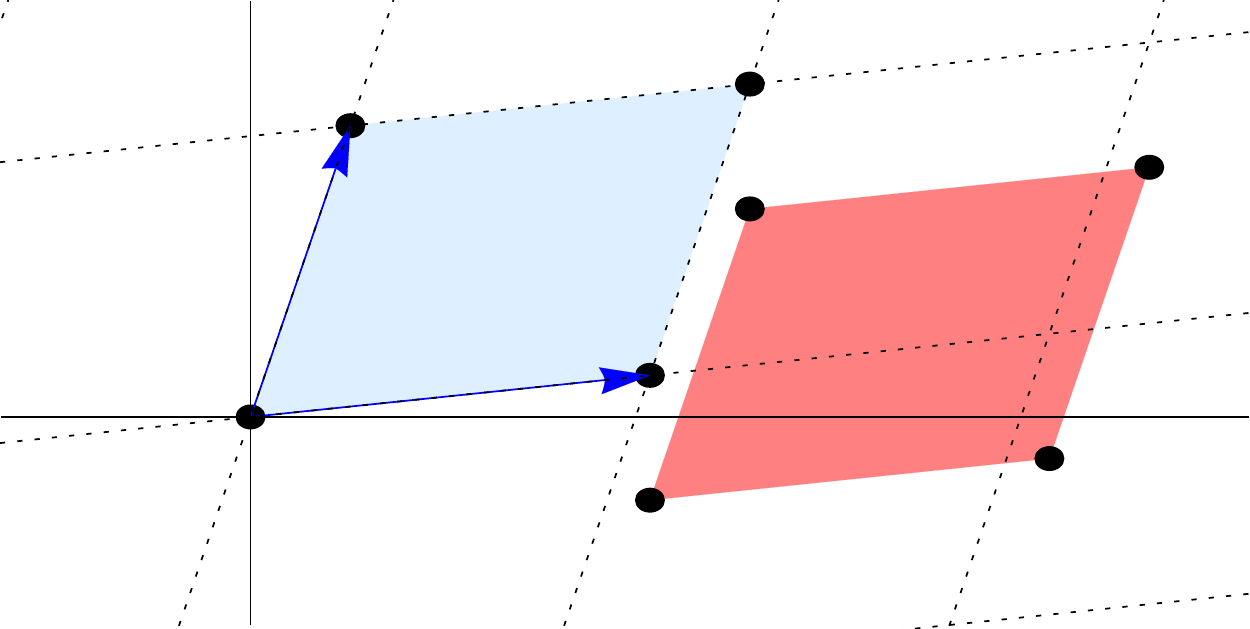}}
\put(87.5,16.25){Re$z$}
\put(24,43){Im$z$}
\put(27.5,16.5){0}
\put(47.5,19.5){$2\omega_1$}
\put(33.5,32){$2\omega_2$}
\put(46,38){$2\omega_1+2\omega_2$}
\put(53,11.5){$z_0$}
\put(77,10.5){$z_0+2\omega_1$}
\put(59,27){$z_0+2\omega_2$}
\put(76,33){$z_0+2\omega_1+2\omega_2$}
\end{picture}
\end{center}
\vspace{-10pt}
\caption{The fundamental period parallelogram is shaded with blue and an arbitrary cell is shaded with red. The dashed lines define the period parallelograms.}
\vspace{5pt}
\label{fig:lattice}
\end{figure}

Knowing the roots and poles of an elliptic function within a cell suffices to describe all roots and poles of the elliptic function, as all other roots and poles are congruent to the former. As such, a set of roots and poles congruent to those within a cell is called an \emph{irreducible set of roots or poles}, respectively. 

\subsubsection*{Modular Transformations}
Given two fundamental periods $2 \omega_1$ and $2 \omega_2$, one can define two different fundamental periods as,
\begin{align}
{\omega _1}' &= a{\omega _1} + b{\omega _2} ,\\
{\omega _2}' &= c{\omega _1} + d{\omega _2} ,
\end{align}
where $a,b,c,d \in \mathbb{Z}$. Any period in the lattice defined by ${\omega _1}'$ and ${\omega _2}'$, $2\omega = 2m' {\omega_1}' + 2n' {\omega_2}'$ is obviously a period of the old lattice, but is the opposite also true? In order for the opposite statement to hold, the area of the fundamental period parallelogram defined by the new periods ${2 \omega _1}'$ and ${2 \omega _2}'$ has to be equal to the area of the fundamental period parallelogram defined by the original ones ${2 \omega _1}$ and ${2 \omega _2}$. The area of the parallelogram defined by two complex numbers $z_1$ and $z_2$ is given by
\begin{equation}
A = \left| {{\mathop{\rm Im}\nolimits} \left( {{z_1}{{\bar z}_2}} \right)} \right| .
\end{equation}
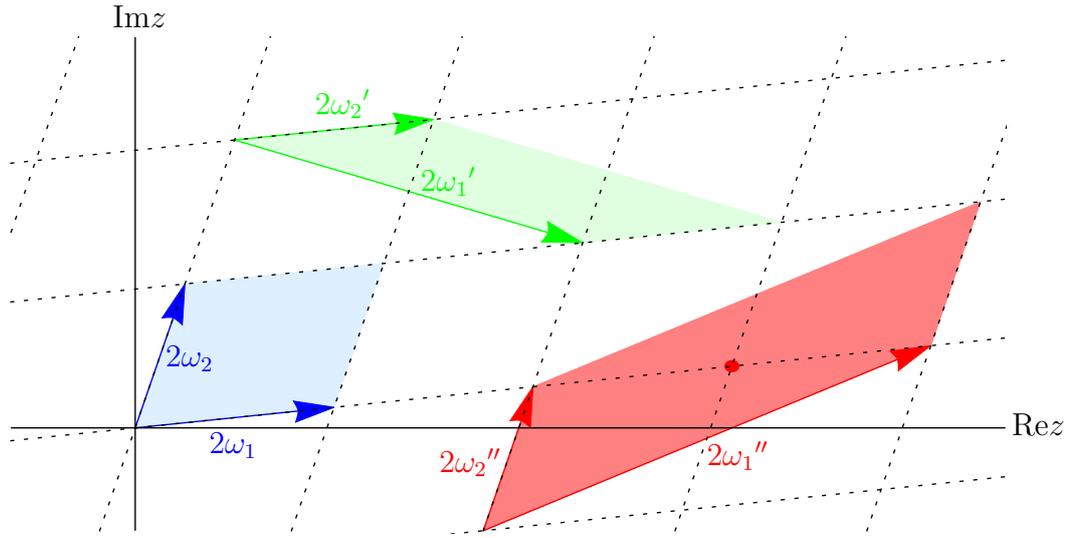
\begin{figure}[ht]
\vspace{10pt}
\begin{center}
\begin{picture}(100,45)
\put(10,2){\includegraphics[width = 0.8\textwidth]{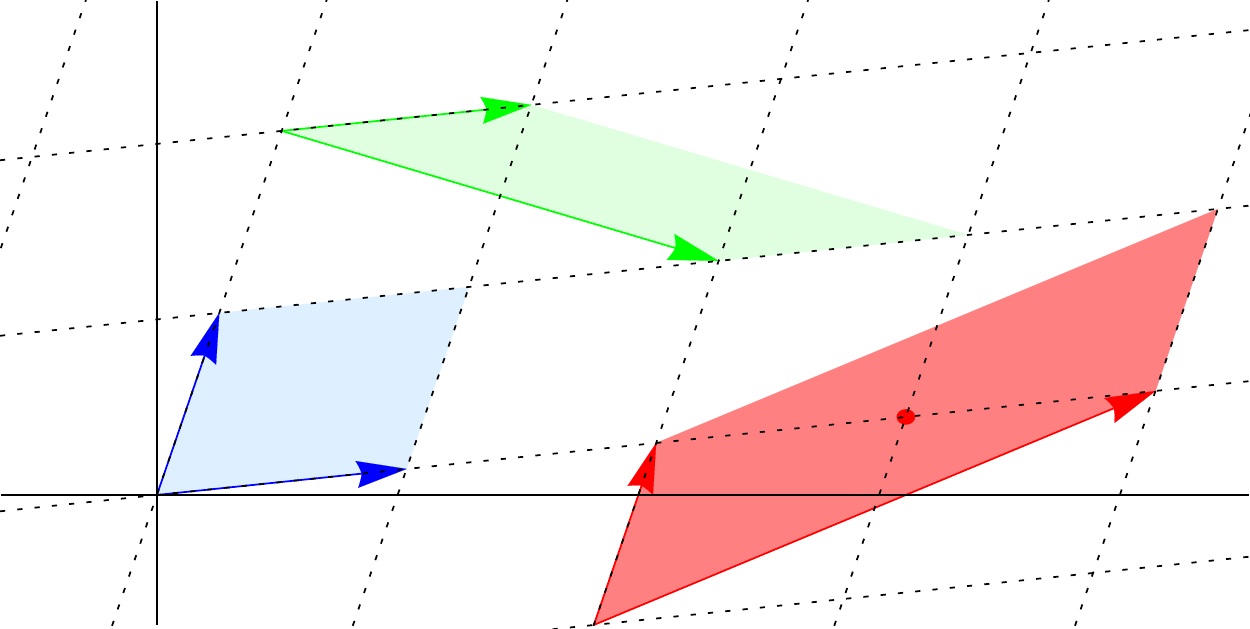}}
\put(90.5,10){Re$z$}
\put(18.25,43){Im$z$}
\put(26,8.5){\color{blue}$2\omega_1$}
\put(22.5,15,5){\color{blue}$2\omega_2$}
\put(43,29.75){\color{green}${2\omega_1}'$}
\put(34.5,36){\color{green}${2\omega_2}'$}
\put(66,7.5){\color{red}${2\omega_1}''$}
\put(44.5,7){\color{red}${2\omega_2}''$}
\end{picture}
\end{center}
\vspace{-10pt}
\caption{The green periods $2 {\omega_1}' = 4 \omega_1 - 2 \omega_2$ and $2 {\omega_2}' = 2 \omega_1$ can generate the whole original lattice. The red periods $2 {\omega_1}'' = 4 \omega_1 + 2 \omega_2$ and $2 {\omega_2}'' = 2 \omega_2$ generate only half of the original lattice. For example, there is no way to reach the red-dotted vertex by a linear combination of $2 {\omega_1}''$  and $2 {\omega_2}''$ with integer coefficients. The area of the parallelogram defined by $2 {\omega_1}'$  and $2 {\omega_2}'$ is equal to the area of the parallelogram defined by the original periods, whereas the one defined by $2 {\omega_1}''$ and $2 {\omega_2}''$ has double this area.}
\vspace{5pt}
\label{fig:modular}
\end{figure}

It is a matter of simple algebra to show that
\begin{equation*}
{\mathop{\rm Im}\nolimits} \left( {{\omega _1}'{{\bar \omega }_2}'} \right) = \left( {ad - bc} \right){\mathop{\rm Im}\nolimits} \left( {{\omega _1}{{\bar \omega }_2}} \right) .
\end{equation*}
Thus, the new periods can generate the original lattice if
\begin{equation}
\left| {\begin{array}{*{20}{c}}
a&b\\
c&d
\end{array}} \right| =  \pm 1 ,
\end{equation}
or in other words if
\begin{equation}
\left( {\begin{array}{*{20}{c}}
a&b\\
c&d
\end{array}} \right) \in SL\left( {2,\mathbb{Z}} \right) .
\end{equation}

It is a direct consequence that an elliptic function necessarily obeys
\begin{equation}
f\left( {z;{\omega _1},{\omega _2}} \right) = f\left( {z;{\omega _1}',{\omega _2}'} \right) ,
\end{equation}
when
\begin{equation}
\left( {\begin{array}{*{20}{c}}
{{\omega _1}'}\\
{{\omega _2}'}
\end{array}} \right) = S\left( {\begin{array}{*{20}{c}}
{{\omega _1}}\\
{{\omega _2}}
\end{array}} \right) ,
\end{equation}
where $S \in SL\left( {2,\mathbb{Z}} \right)$.

\subsubsection*{Basic Properties of Elliptic Functions}

\begin{theorem}
The sum of residues over an irreducible set of poles of an elliptic function vanishes.
\label{th:residues_vanish}
\end{theorem}
To demonstrate this, we use Cauchy's residue theorem over the boundary of a cell.

\begin{multline*}
\sum\limits_{{z_i} \in {\rm{irreducible} \; \rm{set} \; \rm{of} \; \rm{poles}}} {{\mathop{\rm Res}\nolimits} \left( {f,{z_i}} \right)} = \frac{1}{{2\pi i}}\oint_{\partial \rm{cell}} {f\left( z \right)dz} \\
= \frac{1}{{2\pi i}}\int_{{z_0}}^{{z_0} + 2{\omega _1}} {f\left( z \right)dz}  + \frac{1}{{2\pi i}}\int_{{z_0} + 2{\omega _1}}^{{z_0} + 2{\omega _1} + 2{\omega _2}} {f\left( z \right)dz} \\
 + \frac{1}{{2\pi i}}\int_{{z_0} + 2{\omega _1} + 2{\omega _2}}^{{z_0} + 2{\omega _2}} {f\left( z \right)dz}  + \frac{1}{{2\pi i}}\int_{{z_0} + 2{\omega _2}}^{{z_0}} {f\left( z \right)dz} .
\end{multline*}
Shifting $z$ by $2 \omega_1$ in the second integral and by $2 \omega_2$ in the third, we yield
\begin{multline*}
\sum\limits_{{z_i} \in {\rm{irreducible} \; \rm{set} \; \rm{of} \; \rm{poles}}} {{\mathop{\rm Res}\nolimits} \left( {f,{z_i}} \right)} = \frac{1}{{2\pi i}}\int_{{z_0}}^{{z_0} + 2{\omega _1}} {\left[ {f\left( z \right) - f\left( {z + 2{\omega _2}} \right)} \right]dz} \\
 - \frac{1}{{2\pi i}}\int_{{z_0}}^{{z_0} + 2{\omega _2}} {\left[ {f\left( z \right) - f\left( {z + 2{\omega _1}} \right)} \right]dz},
 \end{multline*}
which vanishes as a consequence of $f$ being a doubly periodic function. Therefore,
\begin{equation}
\sum\limits_{{z_i} \in {\rm{irreducible} \; \rm{set} \; \rm{of} \; \rm{poles}}} {{\mathop{\rm Res}\nolimits} \left( {f,{z_i}} \right)} = 0.
\label{eq:residues_sum_zero}
\end{equation}

\begin{theorem}
An elliptic function with an empty irreducible set of poles is a constant function.
\label{th:no_poles_constant}
\end{theorem}
An elliptic function with no poles in a cell, necessarily has no poles at all, as a pole outside a cell necessarily would have a congruent pole within the cell. Consequently, such a function is not just meromorphic, but rather it is analytic. Furthermore, an analytic function in a cell is necessarily bounded within the cell. A direct consequence of property \eqref{eq:doubly_periodic} is that an analytic elliptic function is bounded everywhere. But a bounded analytic function is necessarily a constant function.

\emph{The number of roots of the equation}
\begin{equation}
f \left( z  \right) = z_0
\end{equation}
\emph{is the same for all $z_0 \in \mathbb{C}$.} Before demonstrating this, we will review some properties of Cauchy's integral.

Consider a meromorphic function $g$, with a number of poles $u_i$ and roots $w_j$ with multiplicities $r_i$ and $s_j$ respectively, within a region bounded by a closed contour $C$. The Laurent series of the function $g$ at the regime of a pole or a root $z_0$ is
\begin{equation*}
g\left( z \right) \simeq {c_m}{\left( {z - z_0} \right)^m} + {c_{m + 1}}{\left( {z - z_0} \right)^{m + 1}} +  \ldots 
\end{equation*}
with $c_m \neq 0$. In the case $z_0$ is a root $w_j$, then $m = s_j > 0$, while in the case $z_0$ is a pole $u_i$, then $m = -r_i < 0$. The derivative of $g$ in the regime of a pole or root is
\begin{equation*}
g'\left( z \right) \simeq m{c_m}{\left( {z - z_0} \right)^{m - 1}} + \left( {m + 1} \right){c_{m + 1}}{\left( {z - z_0} \right)^m} +  \ldots 
\end{equation*}
Furthermore, assume an analytic function $h$. Its Laurent series at the regime of a pole or root of $g$ is trivially
\begin{equation*}
h\left( z \right) \simeq h\left( z_0 \right) + h'\left( z_0 \right)\left( {z - z_0} \right) +  \ldots 
\end{equation*}
It is a matter of simple algebra to show that the Laurent series of the function $h g' / g$ at the region of $z_0$ is
\begin{equation*}
h\left( z \right)\frac{{g'\left( z \right)}}{{g\left( z \right)}} \simeq \frac{{mh\left( z_0 \right)}}{{z - z_0}} + mh'\left( z_0 \right) + {c_{m + 1}}\left( {m + 1 - \frac{1}{{c_m^2}}} \right)h\left( z_0 \right) +  \mathcal{O} \left( z - z_0 \right)^2 . 
\end{equation*}
Thus, at any root or pole of the function $g$, the function $h g' / g$ has a first order pole with residue $m h \left( z_0 \right)$. It is a direct consequence of Cauchy's residue theorem that
\begin{equation}
\frac{1}{{2\pi i}}\oint_C {h\left( z \right)\frac{{g'\left( z \right)}}{{g\left( z \right)}}}  = \sum\limits_j {{s_j}h\left( {{w_j}} \right)}  - \sum\limits_i {{r_i}h\left( {{u_i}} \right)} .
\label{eq:Cauchy_integral}
\end{equation}

Let's now return to the case of an elliptic function $f$. We would like to calculate the contour integral of formula \eqref{eq:Cauchy_integral} with $h \left( z \right) = 1$, $g \left( z \right) = f \left( z \right) - z_0$ and $C$ being the boundary of a cell, namely \begin{equation*}
I = \frac{1}{{2\pi i}}\oint_{\partial \rm{cell}} {\frac{{f'\left( z \right)}}{{f\left( z \right) - z_0}}} .
\end{equation*}
The function $f - z_0$ is trivially elliptic, while differentiating equations \eqref{eq:doubly_periodic}, one yields
\begin{equation}
f'\left( {z + 2{\omega _1}} \right) = f'\left( z \right),\quad f'\left( {z + 2{\omega _2}} \right) = f'\left( z \right) ,
\label{eq:derivative_periodic}
\end{equation}
implying that $f'$ is also an elliptic function with the same periods as $f$. In an obvious manner, the function ${f'\left( z \right)} / \left( {f\left( z \right) - z_0} \right)$ is an elliptic function with the same periods as $f$. A direct application of the fact that the sum of the residues of an elliptic function over a cell vanishes \eqref{eq:residues_sum_zero} is
\begin{equation*}
I = \frac{1}{{2\pi i}}\oint_{\partial \rm{cell}} {\frac{{f'\left( z \right)}}{{f\left( z \right) - z_0}}}  = 0 .
\end{equation*}
Finally, using equation \eqref{eq:Cauchy_integral}, we get
\begin{equation}
\sum\limits_j {{s_j}}  - \sum\limits_i {{r_i} = 0} .
\end{equation}
Therefore,
\begin{theorem}
the number of roots of the equation $f \left( z  \right) = z_0$ in a cell is equal to the number of poles of $f$ in a cell (weighted by their multiplicity), independently of the value of $z_0$.
\label{th:number_poles_equals_number_roots}
\end{theorem}
This number is called the \emph{order} of the elliptic function $f$.

\begin{theorem}
The order of a non-constant elliptic function cannot be equal to 1.
\label{th:minimum_order_is_two}
\end{theorem}
A non-constant elliptic function has necessarily at least one pole in a cell as a consequence of theorem \ref{th:no_poles_constant}, thus its order is at least 1. However, an elliptic function of order 1 necessarily has only a single first order pole in a cell. In such a case though, the sum of the residues of the elliptic function in a cell equals to the residue at this single pole, and, thus, it cannot vanish. This contradicts \eqref{eq:residues_sum_zero} and therefore an elliptic function cannot be of order one. The lowest order elliptic functions are of order 2. Such a function can have either a single second order pole in a cell or two first order poles with opposite residues.

\begin{theorem}
The sum of the locations of an irreducible set of poles (weighted by their multiplicity) is congruent to the sum of the locations of an irreducible set of roots (also weighted by their multiplicity).
\label{th:average_root_congruent_to_average_pole}
\end{theorem}
To demonstrate this, we will calculate Cauchy's integral with $h \left( z \right) = z$, $g \left( z \right) = f \left( z \right)$ and as contour of integration $C$ the boundary of a cell. The left hand side of \eqref{eq:Cauchy_integral} equals
\begin{multline*}
I = \frac{1}{{2\pi i}}\oint_{\partial \rm{cell}} {\frac{{zf'\left( z \right)}}{{f\left( z \right)}}}  = \frac{1}{{2\pi i}}\int_{{z_0}}^{{z_0} + 2{\omega _1}} {\frac{{zf'\left( z \right)}}{{f\left( z \right)}}dz}  + \frac{1}{{2\pi i}}\int_{{z_0} + 2{\omega _1}}^{{z_0} + 2{\omega _1} + 2{\omega _2}} {\frac{{zf'\left( z \right)}}{{f\left( z \right)}}dz}  \\
+ \frac{1}{{2\pi i}}\int_{{z_0} + 2{\omega _1} + 2{\omega _2}}^{{z_0} + 2{\omega _2}} {\frac{{zf'\left( z \right)}}{{f\left( z \right)}}dz}  + \frac{1}{{2\pi i}}\int_{{z_0} + 2{\omega _2}}^{{z_0}} {\frac{{zf'\left( z \right)}}{{f\left( z \right)}}dz} .
\end{multline*}
We shift $z$ by $2 \omega_1$ in the second integral and by $2 \omega_2$ in the third one to yield
\begin{multline*}
I = \frac{1}{{2\pi i}}\int_{{z_0}}^{{z_0} + 2{\omega _1}} {\left( {\frac{{zf'\left( z \right)}}{{f\left( z \right)}} - \frac{{\left( {z + 2{\omega _2}} \right)f'\left( {z + 2{\omega _2}} \right)}}{{f'\left( {z + 2{\omega _2}} \right)}}} \right)dz}  \\
- \frac{1}{{2\pi i}}\int_{{z_0}}^{{z_0} + 2{\omega _2}} {\left( {\frac{{zf'\left( z \right)}}{{f\left( z \right)}} - \frac{{\left( {z + 2{\omega _1}} \right)f'\left( {z + 2{\omega _1}} \right)}}{{f'\left( {z + 2{\omega _1}} \right)}}} \right)dz} .
\end{multline*}
Using the periodicity properties of $f$, \eqref{eq:doubly_periodic} and $f'$ \eqref{eq:derivative_periodic}, we yield
\begin{multline*}
I = \frac{1}{{2\pi i}}\int_{{z_0}}^{{z_0} + 2{\omega _1}} {\left( {\frac{{zf'\left( z \right)}}{{f\left( z \right)}} - \frac{{\left( {z + 2{\omega _2}} \right)f'\left( z \right)}}{{f'\left( z \right)}}} \right)dz}  \\
- \frac{1}{{2\pi i}}\int_{{z_0}}^{{z_0} + 2{\omega _2}} {\left( {\frac{{zf'\left( z \right)}}{{f\left( z \right)}} - \frac{{\left( {z + 2{\omega _1}} \right)f'\left( z \right)}}{{f'\left( z \right)}}} \right)dz} \\
=  - \frac{{{\omega _2}}}{{\pi i}}\int_{{z_0}}^{{z_0} + 2{\omega _1}} {\frac{{f'\left( z \right)}}{{f\left( z \right)}}dz}  + \frac{{{\omega _1}}}{{\pi i}}\int_{{z_0}}^{{z_0} + 2{\omega _2}} {\frac{{f'\left( z \right)}}{{f\left( z \right)}}dz} \\
=  - \frac{{{\omega _2}}}{{\pi i}}\left( {\ln f\left( {{z_0} + 2{\omega _1}} \right) - \ln f\left( {{z_0}} \right)} \right) + \frac{{{\omega _1}}}{{\pi i}}\left( {\ln f\left( {{z_0} + 2{\omega _2}} \right) - \ln f\left( {{z_0}} \right)} \right) .
\end{multline*}
Although $z_0 + 2 \omega_1 \sim z_0 \sim z_0 + 2 \omega_2$, due to the branch cut of the logarithmic function, in general we have that ${\ln f\left( {{z_0} + 2{\omega _1}} \right) - \ln f\left( {{z_0}} \right)} = 2 m \pi$ and ${\ln f\left( {{z_0} + 2{\omega _2}} \right) - \ln f\left( {{z_0}} \right)} = 2 n \pi$, with $m,n \in \mathbb{Z}$. Thus,
\begin{equation*}
I = 2m{\omega _1} + 2n{\omega _2} \sim 0.
\end{equation*}
Applying property \eqref{eq:Cauchy_integral} we get
\begin{equation*}
I = \sum\limits_i {{r_i}{u_i}}  - \sum\limits_j {{s_j}{w_j}} ,
\end{equation*}
implying
\begin{equation}
\sum\limits_j {{r_j}{u_j}} \sim \sum\limits_i {{s_i}{w_i}} ,
\end{equation}
which is the proof of theorem 1.5.

\subsection{The Weierstrass Elliptic Function}
\label{subsec:WeierstrassP}
\subsubsection*{Definition}
As we showed in previous section, the lowest possible order of an elliptic function is 2. One possibility for such an elliptic function is a function having a single second order pole in each cell. It is actually quite easy to construct such a function. It suffices to sum an infinite set of copies of the function $1 / z^2$ each one shifted by $2 m \omega_1 + 2 n \omega_2$ for all $m,n \in \mathbb{Z}$. The usual convention includes the addition of a constant cancelling the contributions of all these functions at $z = 0$ (except for the term with $m=n=0$), so that the Laurent series of the constructed function at the region of $z=0$ has a vanishing zeroth order term. Following these directions, we define,
\begin{equation}
\wp \left( z \right) := \frac{1}{{{z^2}}} + \sum\limits_{\left\{ {m,n} \right\} \ne \left\{ {0,0} \right\}} {\left( {\frac{1}{{{{\left( {z + 2m{\omega _1} + 2n{\omega _2}} \right)}^2}}} - \frac{1}{{{{\left( {2m{\omega _1} + 2n{\omega _2}} \right)}^2}}}} \right)} .
\label{eq:Weierstrass_definition}
\end{equation}
By construction, this function is doubly periodic with fundamental periods equal to $2\omega_1$ and $2\omega_2$.
\begin{equation}
\wp \left( {z + {2\omega _1}} \right) = \wp \left( z \right) ,\quad \wp \left( {z + {2\omega _2}} \right) = \wp \left( z \right) .
\end{equation}
This function is called \emph{Weierstrass elliptic function}.

\subsubsection*{Basic Properties}
A direct consequence of the definition \eqref{eq:Weierstrass_definition} is the fact that the Weierstrass elliptic function is an even function
\begin{equation}
\wp \left( { - z} \right) = \wp \left( z \right) .
\end{equation}

Let's acquire the Laurent series of the Weierstrass elliptic function at the regime of $z=0$. It is easy to show that
\begin{equation*}
\frac{1}{{{{\left( {z + w} \right)}^2}}} - \frac{1}{{{w^2}}} = \frac{1}{{{w^2}}}\sum\limits_{k = 1}^\infty  {\left( {k + 1} \right){{\left( { - \frac{z}{w}} \right)}^k}} .
\end{equation*}
Consequently,
\begin{equation*}
\wp \left( z \right) = \frac{1}{{{z^2}}} + \sum\limits_{k = 1}^\infty  {\sum\limits_{\left\{ {m,n} \right\} \ne \left\{ {0,0} \right\}} {\frac{{\left( {k + 1} \right){{\left( { - 1} \right)}^k}}}{{{{\left( {m{\omega _1} + n{\omega _2}} \right)}^{k + 2}}}}{z^k}} }  = \frac{1}{{{z^2}}} + \sum\limits_{k = 1}^\infty  {{a_k}{z^k}} ,
\end{equation*}
where
\begin{equation*}
{a_k} = \left( {k + 1} \right){\left( { - 1} \right)^k}\sum\limits_{\left\{ {m,n} \right\} \ne \left\{ {0,0} \right\}} {\frac{1}{{{{\left( {2m{\omega _1} + 2n{\omega _2}} \right)}^{k + 2}}}}} .
\end{equation*}
The fact that $\wp$ is even implies that only the even indexed coefficients do not vanish,
\begin{align}
{a_{2l + 1}} &= 0 ,\\
{a_{2l}} &= \left( {2l + 1} \right)\sum\limits_{\left\{ {m,n} \right\} \ne \left\{ {0,0} \right\}} {\frac{1}{{{{\left( {2m{\omega _1} + 2n{\omega _2}} \right)}^{2\left( {l + 1} \right)}}}}} .
\end{align}
For reasons that will become apparent later, we define $g_2$ and $g_3$ so that
\begin{equation}
\wp \left( z \right) = \frac{1}{{{z^2}}} + \frac{{{g_2}}}{{20}}{z^2} + \frac{{{g_3}}}{{28}}{z^4} + \mathcal{O}\left( {{z^6}} \right) ,
\end{equation}
implying that
\begin{align}
{g_2} &:= 60\sum\limits_{\left\{ {m,n} \right\} \ne \left\{ {0,0} \right\}} {\frac{1}{{{{\left( {2m{\omega _1} + 2n{\omega _2}} \right)}^4}}}} , \label{eq:g2_definition}\\
{g_3} &:= 140\sum\limits_{\left\{ {m,n} \right\} \ne \left\{ {0,0} \right\}} {\frac{1}{{{{\left( {2m{\omega _1} + 2n{\omega _2}} \right)}^6}}}} . \label{eq:g3_definition}
\end{align}

\subsubsection*{Weierstrass Differential Equation}
Directly differentiating equation \eqref{eq:Weierstrass_definition}, the derivative of Weierstrass function can be expressed as
\begin{equation}
\wp '\left( z \right) =  - 2\sum\limits_{m,n} {\frac{1}{{{{\left( {z + m{\omega _1} + n{\omega _2}} \right)}^3}}}} .
\label{eq:Weierstrass_derivative_series}
\end{equation}
It follows that the derivative of Weierstrass elliptic function is an odd function
\begin{equation}
\wp '\left( { - z} \right) =  - \wp '\left( z \right) .
\end{equation}

The Laurent series of $\wp '\left( z \right)$, $\wp ^3 \left( z \right)$ and $\wp {'^2}\left( z \right)$ at the regime of $z=0$ are
\begin{align*}
\wp \left( z \right) &= \frac{1}{{{z^2}}} + \frac{{{g_2}}}{{20}}{z^2} + \frac{{{g_3}}}{{28}}{z^4} + \mathcal{O}\left( {{z^6}} \right) ,\\
\wp '\left( z \right) &=  - \frac{2}{{{z^3}}} + \frac{{{g_2}}}{{10}}z + \frac{{{g_3}}}{7}{z^3} + \mathcal{O}\left( {{z^5}} \right) ,\\
{\wp ^3}\left( z \right) &= \frac{1}{{{z^6}}} + \frac{{3{g_2}}}{{20}}\frac{1}{{{z^2}}} + \frac{{3{g_3}}}{{28}} + \mathcal{O}\left( {{z^2}} \right) ,\\
\wp {'^2}\left( z \right) &= \frac{4}{{{z^6}}} - \frac{{2{g_2}}}{5}\frac{1}{{{z^2}}} - \frac{{4{g_3}}}{7} + \mathcal{O}\left( {{z^2}} \right).
\end{align*}
It is not difficult to show that there is a linear combination of the above, which is not singular at $z=0$ and furthermore it vanishes there. One can eliminate the sixth order pole by taking an appropriate combination of $\wp{'^2}$ and $\wp^3$. This leaves a function with a second order pole. Taking an appropriate combination of the latter combination and $\wp$ allows to write down a function with no poles at $z=0$. Trivially, adding an appropriate constant results in a non-singular function vanishing at $z=0$. The appropriate combination turns out to be
\begin{equation*}
\wp {'^2}\left( z \right) - 4{\wp ^3}\left( z \right) + {g_2}\wp \left( z \right) + {g_3} = \mathcal{O}\left( {{z^2}} \right) .
\end{equation*}
But, the derivative, as well as powers of an elliptic function are elliptic functions with the same periods. Therefore, the function $\wp {'^2}\left( z \right) = 4{\wp ^3}\left( z \right) - {g_2}\wp \left( z \right) - {g_3}$ is an elliptic function with the same periods as $\wp \left( z \right)$. Since the latter has no pole at $z=0$, it does not have any pole at all, and, thus, it is an elliptic function with no poles. According to theorem \ref{th:no_poles_constant}, elliptic functions with no poles are necessarily constants and since $\wp {'^2}\left( z \right) = 4{\wp ^3}\left( z \right) - {g_2}\wp \left( z \right) - {g_3}$ vanishes at the origin, it vanishes everywhere. This implies that Weierstrass elliptic function obeys the differential equation,
\begin{equation}
\wp {'^2}\left( z \right) = 4{\wp ^3}\left( z \right) - {g_2}\wp \left( z \right) - {g_3} = 0 .
\end{equation}
This differential equation is of great importance in the applications of Weierstrass elliptic function in physics. For a physicist it is sometimes useful to even conceive this differential equation as the definition of Weierstrass elliptic function.

It turns out that the Weierstrass elliptic function is the general solution of the differential equation
\begin{equation}
{\left( {\frac{{dy}}{{dz}}} \right)^2} = 4{y^3} - {g_2}y - {g_3} .
\label{eq:Weierstrass_equation}
\end{equation}
Performing the substitution $ y = \wp \left( w \right) $, the equation \eqref{eq:Weierstrass_equation} assumes the form
\begin{equation*}
{\left( {\frac{{dw}}{{dz}}} \right)^2} = 1,
\end{equation*}
which obviously has the solutions, $ w =  \pm z + z_0$. This implies that $y = \wp \left( { \pm z + z_0} \right)$ and since the Weierstrass elliptic function is even, \emph{the general solution of Weierstrass equation \eqref{eq:Weierstrass_equation} can be written in the form}
\begin{equation}
y = \wp \left( {z + z_0} \right) .
\end{equation}

In the following, we will deduce an integral formula for the inverse function of $\wp$. In order to do so, we define the function $z \left( y \right)$ as
\begin{equation}
z \left( y \right) := \int_y^\infty  {\frac{1}{{\sqrt {4{t^3} - {g_2}t - {g_3}} }}dt} .
\label{eq:integral_predefinition}
\end{equation}
Differentiating with respect to $z$ one gets
\begin{equation*}
1 =  - \frac{{dy}}{{dz}} \frac{{1}}{{\sqrt {4{y^3} - {g_2}y - {g_3}} }} \Rightarrow {\left( {\frac{{dy}}{{dz}}} \right)^2} = 4{y^3} - {g_2}y - {g_3} .
\end{equation*}
We just showed that the general solution of this equation is
\begin{equation*}
y = \wp \left( {z + z_0} \right) .
\end{equation*}
Since the integral in \eqref{eq:integral_predefinition} converges, it should vanish at the limit $y \to \infty$, or equivalently, $\mathop {\lim }\limits_{y \to \infty } z\left( y \right) = 0$. This implies that $z = z_0$ is the position of a pole, or in other words it is congruent to $z = 0$. This means that 
\begin{equation*}
y = \wp \left( {z + 2m{\omega _1} + 2n{\omega _2}} \right) = \wp \left( z \right)
\end{equation*}
Substituting the above into the equation \eqref{eq:integral_predefinition} yields the integral formula for Weierstrass elliptic function, 
\begin{equation}
z = \int_{\wp \left( z \right)}^\infty  {\frac{1}{{\sqrt {4{t^3} - {g_2}t - {g_3}} }}dt} .
\label{eq:integral_formula}
\end{equation}

Once should wonder, how the above formula is consistent with the fact that $\wp$ is an elliptic function, and, thus, all numbers congruent to each other should be mapped to the same value of $\wp$. The answer to this question is that the integrable quantity in \eqref{eq:integral_formula} has branch cuts. Depending on the selection of the path from $\wp \left( z \right)$ to infinity and more specifically depending on how many times the path encircles each branch cut, one may result in any number congruent to $z$ or $-z$. A more precise expression of the integral formula is
\begin{equation}
\int_{\wp \left( z \right)}^\infty  {\frac{1}{{\sqrt {4{t^3} - {g_2}t - {g_3}} }}dt} \sim \pm z.
\label{eq:integral_formula_sim}
\end{equation}

\subsubsection*{The Roots of the Cubic Polynomial}
\emph{We define the values of the Weierstrass elliptic function at the half-periods $\omega_1$, $\omega_2$ and $\omega_3 := \omega_1 + \omega_2$ as}
\begin{equation}
{e_1} := \wp \left( {{\omega _1}} \right) , \quad {e_2} := \wp \left( {{\omega _3}} \right) , \quad {e_3} := \wp \left( {{\omega _2}} \right) .
\end{equation}
The permutation between the indices of $\omega$'s and $e$'s is introduced for notational reasons that will become apparent later. The periodicity properties of $\wp$ combined with the fact that the latter is an even function, imply that $\wp$ is stationary at the half periods. For example,
\begin{equation*}
\wp '\left( {{\omega _1}} \right) =  - \wp '\left( { - {\omega _1}} \right) =  - \wp '\left( {2{\omega _1} - {\omega _1}} \right) =  - \wp '\left( {{\omega _1}} \right) ,
\end{equation*}
implying that $\wp '\left( {{\omega _1}} \right) = 0$. Similarly one can show that
\begin{equation}
\wp '\left( {{\omega _1}} \right) = \wp '\left( {{\omega _2}} \right) = \wp '\left( {{\omega _3}} \right) = 0 .
\end{equation}

Substituting a half-period into Weierstrass equation \eqref{eq:Weierstrass_equation}, we yield
\begin{equation}
4e_i^3 - {g_2}{e_i} - {g_3} = 0 .
\label{eq:roots_are_roots}
\end{equation}
The derivative of $\wp$, as shown in equation \eqref{eq:Weierstrass_derivative_series} have a single third order pole in each cell, congruent to $z = 0$. Thus, $\wp '$ is an elliptic function of order 3 and therefore it has exactly three roots in each cell. Since $\omega_1$, $\omega_2$ and $\omega_3$ all lie within the fundamental period parallelogram, they cannot be congruent to each other, and, thus, there is no other root within the latter. This also implies that $\omega_1$, $\omega_2$ and $\omega_3$ are necessarily first order roots of $\wp '$. All other roots of $\wp'$ are congruent to those. Finally, when equation \eqref{eq:roots_are_roots} has a double root, the solution of the differential equation \eqref{eq:Weierstrass_equation} cannot be an elliptic function.

An implication of the above is the fact that the locations $z=\omega_1$, $z=\omega_2$ and $z=\omega_3$ are the only locations within the fundamental period parallelogram, where the Laurent series of the function $\wp \left( z \right) - \wp \left( z_0 \right)$ have a vanishing first order term at the region of $z_0$. Consequently the equation $\wp \left( z \right) = f_0$ has a double root only when $f_0$ equals any of the three roots $e_1$, $e_2$ or $e_3$. Since $\wp$ is an order two elliptic function, \emph{the complex numbers $e_1$, $e_2$ and $e_3$ are the only ones appearing only once in a cell, whereas all other complex numbers appear twice}.

Finally, equation \eqref{eq:roots_are_roots} implies that $e_i$ are the three roots of the polynomial appearing in the right hand side of Weierstrass equation, namely
\begin{equation}
Q \left( t \right) := 4{t^3} - {g_2}t - {g_3} = 4\left( {t - {e_1}} \right)\left( {t - {e_2}} \right)\left( {t - {e_3}} \right) .
\end{equation}
This directly implies that $e_i$ obey
\begin{align}
{e_1} + {e_2} + {e_3} &= 0,\\
{e_1}{e_2} + {e_2}{e_3} + {e_3}{e_1} &=  - \frac{{{g_2}}}{4} , \label{eq:g2_roots}\\
{e_1}{e_2}{e_3} &= \frac{{{g_3}}}{4} . \label{eq:g3_roots}
\end{align}

\subsubsection*{Other Properties}
The Weierstrass elliptic function obeys the homogeneity relation
\begin{equation}
\wp \left( {z;{g_2},{g_3}} \right) = {\mu ^2}\wp \left( {\mu z;\frac{{{g_2}}}{{{\mu ^4}}},\frac{{{g_3}}}{{{\mu ^6}}}} \right) .
\label{eq:Weierstras_homogeneity_wp}
\end{equation}
For the specific case $\mu = i$, the above relation assumes the form
\begin{equation}
\wp \left( {z;{g_2},{g_3}} \right) = - \wp \left( {i z ; g_2 , - g_3} \right) .
\end{equation}

Finally, when two of the roots $e_1$, $e_2$ and $e_3$ coincide, the Weierstrass elliptic function degenerates to a simply periodic function. Assuming that the moduli $g_2$ and $g_3$ are real, then the existence of a double root implies that all roots are real. When the double root is larger than the simple root, the Weierstrass elliptic function takes the form
\begin{equation}
\wp \left( z ; 12e_0^2 , - 8 e_0^3 \right) = {e_0} + \frac{{3{e_0}}}{{{{\sinh }^2}\left( {\sqrt {3{e_0}} z} \right)}} ,
\label{eq:two_large_roots_coincide}
\end{equation}
whereas when the double root is smaller than the simple root, it takes the form
\begin{equation}
\wp \left( z ; 12e_0^2 , 8 e_0^3 \right) =  - {e_0} + \frac{{3{e_0}}}{{{{\sin }^2}\left( {\sqrt {3{e_0}} z} \right)}} .
\label{eq:two_small_roots_coincide}
\end{equation}
If there is only one triple root, then it must be vanishing, since the three roots sum to zero. In this case, the Weierstrass elliptic function degenerates to a function that is not periodic at all, namely
\begin{equation}
\wp \left( z ; 0 , 0 \right) = \frac{1}{z^2} .
\label{eq:three_roots_coincide}
\end{equation}

The proofs of the homogeneity relation, as well as the double root limits of the Weierstrass elliptic function are left as an exercise for the audience.

\newpage
\subsection*{Problems}

\begin{problem}
Show, using the integral formula for the Weierstrass elliptic function \eqref{eq:integral_formula} that when $g_2$ and $g_3$ are real and all roots $e_1$, $e_2$ and $e_3$ are also real, the half-period corresponding to the largest root $e_1$ is congruent to a real number, whereas the half-period corresponding to the smallest root $e_3$ is congruent to a purely imaginary number.

Then, show that when there is one real root and two complex ones, the half-period corresponding to the real root $e_2$ is congruent to both a real and a purely imaginary number.
\label{pr:reality_omega_1}
\end{problem}

\begin{problem}
Show that at the limit two of the roots $e_1$, $e_2$ and $e_3$ coincide, the Weierstrass elliptic function degenerates to a simply periodic function and can be expressed in terms of trigonometric or hyperbolic functions as described by formulae \eqref{eq:two_large_roots_coincide} and \eqref{eq:two_small_roots_coincide}. Find the value of the unique period in terms of the double root. Also show that at the limit all three roots $e_1$, $e_2$ and $e_3$ coincide, the Weierstrass elliptic function degenerates to the non-periodic function of equation \eqref{eq:three_roots_coincide}.
\label{pr:degenerate cases}
\end{problem}

\begin{problem}
Use the definition \eqref{eq:Weierstrass_definition} to deduce the homogeneity property of the Weierstrass elliptic function \eqref{eq:Weierstras_homogeneity_wp}.
\label{pr:homogeneity_wp}
\end{problem}

\newpage

\setcounter{equation}{0}
\section{Lecture 2: Weierstrass Quasi-periodic Functions}
\label{sec:Weierstrass}
In the previous lecture, we used several times the fact that the derivative of an elliptic function is also an elliptic function with the same periods. However, the opposite statement is not correct; the indefinite integral of an elliptic function is not necessarily an elliptic function. Such non-elliptic functions typically expose other interesting quasi-periodicity properties. In the following we will study two such quasi-periodic functions that are derived from Weierstrass elliptic function and have numerous important applications.

\subsection{Quasi-periodic Weierstrass Functions}
\label{subsec:Weierstrass_quasi}
\subsubsection*{The Weierstrass $\zeta$ Function}
\emph{The Weierstrass $\zeta$ function is defined as}
\begin{equation}
\frac{{d\zeta \left( z \right)}}{{dz}} :=  - \wp \left( z \right) ,
\label{eq:zeta_definition}
\end{equation}
\emph{furthermore satisfying the condition}
\begin{equation}
\mathop {\lim }\limits_{z \to 0} \left( {\zeta \left( z \right) - \frac{1}{z}} \right) := 0 ,
\label{eq:zeta_definition_condition}
\end{equation}
which fixes the integration constant. 

Using the definition of Weierstrass $\wp$ function, we acquire
\begin{equation}
\zeta \left( z \right) = \frac{1}{z} + \sum\limits_{\left\{ {m,n} \right\} \ne \left\{ {0,0} \right\}} {\left( {\frac{1}{{z + 2m{\omega _1} + 2n{\omega _2}}} - \frac{1}{{2m{\omega _1} + 2n{\omega _2}}} + \frac{z}{{{{\left( {2m{\omega _1} + 2n{\omega _2}} \right)}^2}}}} \right)} .
\label{eq:zeta_series}
\end{equation}

Weierstrass $\zeta$ function is an odd function.
\begin{equation}
\zeta \left( { - z} \right) = - \zeta \left( {z} \right) .
\end{equation}
Notice that the definition \eqref{eq:zeta_definition} coupled with the fact that $\wp$ is an even function implies that $\zeta$ is an odd function up to a constant. The condition \eqref{eq:zeta_definition_condition} fixes this constant to zero, so that $\zeta$ is an odd function.

\subsubsection*{Quasi-periodicity of the Function $\zeta$}
Equation \eqref{eq:zeta_series} implies that in each cell defined by the periods of the corresponding $\wp$ function, the function $\zeta$ has only a first order pole with residue equal to one. As such, it cannot be an elliptic function with the same periods as $\wp$. Actually, it cannot be an elliptic function with any periods, since it is not possible to define a cell where the sum of the residues would vanish. \emph{The Weierstrass $\zeta$ function is quasi-periodic.} Its quasi-periodicity properties can be deduced from the periodicity of Weierstrass $\wp$ function. More specifically, integrating the relation $\wp \left( {z + 2{\omega _i}} \right) = \wp \left( z \right)$, we find
\begin{equation*}
\zeta \left( {z + 2{\omega _i}} \right) = \zeta \left( z \right) + c .
\end{equation*}
The above relation for $z = - \omega_i$ yields $ \zeta \left( {{\omega _i}} \right) = \zeta \left( { - {\omega _i}} \right) + c $. Since $\zeta$ is an odd function, the above implies that $ c = 2\zeta \left( {{\omega _i}} \right) $, which in turn yields
\begin{equation}
\zeta \left( {z + 2{\omega _i}} \right) = \zeta \left( z \right) + 2\zeta \left( {{\omega _i}} \right) .
\label{eq:zeta_quasi_periodicity_step}
\end{equation}
One can easily show inductively that
\begin{equation}
\zeta \left( {z + 2m{\omega _1} + 2n{\omega _2}} \right) = \zeta \left( z \right) + 2m\zeta \left( {{\omega _1}} \right) + 2n\zeta \left( {{\omega _2}} \right) .
\label{eq:zeta_quasi_periodicity}
\end{equation}

The quantities $\zeta \left( {{\omega _1}} \right)$ and $\zeta \left( {{\omega _2}} \right)$ are related with an interesting property. Consider the contour integral
\begin{equation*}
I = \frac{1}{{2\pi i}}\oint_{\partial \rm{cell}} {\zeta \left( z \right)dz}
\end{equation*}
Since $\zeta$ has only a first order pole with residue equal to one within a cell, Cauchy residue theorem implies that
\begin{equation*}
I = 1 .
\end{equation*}
Performing the contour integral along the boundary of a cell, we get
\begin{equation*}
2 \pi i I = \int_{{z_0}}^{{z_0} + 2{\omega _1}} {\zeta \left( z \right)dz}  + \int_{{z_0} + 2{\omega _1}}^{{z_0} + 2{\omega _1} + 2{\omega _2}} {\zeta \left( z \right)dz}
+ \int_{{z_0} + 2{\omega _1} + 2{\omega _2}}^{{z_0} + 2{\omega _2}} {\zeta \left( z \right)dz}  + \int_{{z_0} + 2{\omega _2}}^{{z_0}} {\zeta \left( z \right)dz} .
\end{equation*}
Shifting $z$ by $2 \omega_1$ in the second integral and by $2 \omega_2$ in the third, we yield
\begin{equation*}
\begin{split}
2 \pi i I &= \int_{{z_0}}^{{z_0} + 2{\omega _1}} {\left( {\zeta \left( z \right) - \zeta \left( {z + 2{\omega _2}} \right)} \right)dz}  - \int_{{z_0}}^{{z_0} + 2{\omega _2}} {\left( {\zeta \left( z \right) - \zeta \left( {z + 2{\omega _1}} \right)} \right)dz} \\
 &= \int_{{z_0}}^{{z_0} + 2{\omega _1}} {\left( {\zeta \left( z \right) - \zeta \left( z \right) - 2\zeta \left( {{\omega _2}} \right)} \right)dz}  - \int_{{z_0}}^{{z_0} + 2{\omega _2}} {\left( {\zeta \left( z \right) - \zeta \left( z \right) - 2\zeta \left( {{\omega _1}} \right)} \right)dz} \\
 &=  - 4{\omega _1}\zeta \left( {{\omega _2}} \right) + 4{\omega _2}\zeta \left( {{\omega _1}} \right) .
\end{split}
\end{equation*}
As a result, $\zeta \left( {{\omega _i}} \right)$ and $\zeta \left( {{\omega _i}} \right)$ are related as
\begin{equation}
{\omega _2}\zeta \left( {{\omega _1}} \right) - {\omega _1}\zeta \left( {{\omega _2}} \right) = \frac{{\pi i}}{2} .
\label{eq:z1_z2_relation}
\end{equation}

\subsubsection*{The Weierstrass $\sigma$ Function}
Since Weierstrass elliptic function has a single second order pole in each cell, integrating it once resulted in a function (the Weierstrass $\zeta$ function) with a single first order pole in each cell. Integrating once more would lead to a logarithmic singularity in each cell. To avoid this, we define the next quasi-periodic function as the exponential of the integral of the Weierstrass $\zeta$ function.

\emph{The Weierstrass $\sigma$ function is defined as}
\begin{equation}
\frac{{\sigma '\left( z \right)}}{{\sigma \left( z \right)}} := \zeta \left( z \right) ,
\label{eq:sigma_definition}
\end{equation}
\emph{together with the condition}
\begin{equation}
\mathop {\lim }\limits_{z \to 0} \frac{{\sigma \left( z \right)}}{z} := 1,
\label{eq:sigma_definition_condition}
\end{equation}
which fixes the integration constant.


Integrating equation \eqref{eq:zeta_series} term by term results in the following expression for $\sigma$ function.
\begin{equation}
\sigma \left( z \right) = z {\prod\limits_{\left\{ {m,n} \right\} \ne \left\{ {0,0} \right\}} \left( {\frac{{z + 2m{\omega _1} + 2n{\omega _2}}}{{2m{\omega _1} + 2n{\omega _2}}}}  e^{{-\frac{z}{{2m{\omega _1} + 2n{\omega _2}}}} + {\frac{z^2}{{{{2\left( {2m{\omega _1} + 2n{\omega _2}} \right)}^2}}}}} \right)}.
\end{equation}
This implies that $\sigma$ is analytic. At the locations of the poles of $\wp$, it has first order roots.

Similarly to the definition of $\zeta$ function, definition \eqref{eq:sigma_definition} implies that $\sigma$ is odd up to a constant, which is set to zero by the defining condition \eqref{eq:sigma_definition_condition}, so that $\sigma$ is an odd function,
\begin{equation}
\sigma \left( { - z} \right) = - \sigma \left( z \right) .
\end{equation}

\subsubsection*{Quasi-periodicity of the Function $\sigma$}
\emph{The Weierstrass $\sigma$ function is a quasi-periodic function.} Its quasi-periodicity properties follow from the corresponding properties of the Weierstrass $\zeta$ function. Integrating the equation 
\eqref{eq:zeta_quasi_periodicity_step}, one yields
\begin{equation*}
\ln \sigma \left( {z + 2{\omega _i}} \right) = \ln \sigma \left( z \right) + 2\zeta \left( {{\omega _i}} \right)z + c ,
\end{equation*}
or
\begin{equation*}
\sigma \left( {z + 2{\omega _i}} \right) = C{e^{2\zeta \left( {{\omega _i}} \right)z}}\sigma \left( z \right) ,
\end{equation*}
where $C=e^c$. Substituting $z = - \omega_i$, we get
\begin{equation*}
\sigma \left( {{\omega _i}} \right) = C{e^{ - 2\zeta \left( {{\omega _i}} \right){\omega _i}}}\sigma \left( { - {\omega _i}} \right) =  - C{e^{ - 2\zeta \left( {{\omega _i}} \right){\omega _i}}}\sigma \left( {{\omega _i}} \right) ,
\end{equation*}
implying, $ C =  - {e^{ 2\zeta \left( {{\omega _i}} \right){\omega _i}}} $. This means that the quasi-periodicity of the function $\sigma$ is
\begin{equation}
\sigma \left( {z + 2{\omega _i}} \right) =  - {e^{2\zeta \left( {{\omega _i}} \right)\left( {z + {\omega _i}} \right)}}\sigma \left( z \right) .
\label{eq:sigma_quasi_periodicity_step}
\end{equation}
Equation \eqref{eq:sigma_quasi_periodicity_step} can be used to prove inductively that
\begin{equation}
\sigma \left( {z + 2m{\omega _1} + 2n{\omega _2}} \right) = {\left( { - 1} \right)^{m + n + mn}}{e^{\left( {2m\zeta \left( {{\omega _1}} \right) + 2n\zeta \left( {{\omega _2}} \right)} \right)\left( {z + m{\omega _1} + n{\omega _2}} \right)}}\sigma \left( z \right) .
\label{eq:sigma_quasi_periodicity}
\end{equation}

\subsection{Expression of any Elliptic Function in Terms of Weierstrass Functions}
\label{subsec:elliptic_terms_Weierstrass}
The functions $\wp$, $\zeta$ or $\sigma$ can be used to express any elliptic function with the same periods. In the following, we will derive such expressions and deduce interesting properties of a general elliptic function.

\subsubsection*{Expression of any Elliptic Function in Terms of $\wp$ and $\wp'$}
Assume an elliptic function $f\left( z \right)$. Then, it can be written as
\begin{equation*}
f\left( z \right) = \frac{{f\left( z \right) + f\left( { - z} \right)}}{2} + \frac{{f\left( z \right) - f\left( { - z} \right)}}{{2\wp '\left( z \right)}}\wp '\left( z \right).
\end{equation*}
Both functions $\frac{{f\left( z \right) + f\left( { - z} \right)}}{2}$ and $\frac{{f\left( z \right) - f\left( { - z} \right)}}{{2\wp '\left( z \right)}}$ are even. Thus, in order to express an arbitrary elliptic function in terms of $\wp$ and $\wp'$, it suffices to find an expression of an arbitrary even elliptic function in terms of $\wp$ and $\wp'$.

Assume an even elliptic function $g\left( z \right)$. As it is even, any irreducible set of poles of $g\left( z \right)$ can be divided to a set of points $u_i$ with multiplicity $r_i$ and another set of points congruent to $-u_i$ with equal multiplicities. In a similar manner, an irreducible set of roots of $g\left( z \right)$ can be divided to a set of points $w_i$ with multiplicities $s_i$ and another set of points congruent to $-w_i$ with equal multiplicities. Now consider the function
\begin{equation*}
h\left( z \right) := g\left( z \right)\frac{{\prod\limits_i {{{\left( {\wp \left( z \right) - \wp \left( {{u_i}} \right)} \right)}^{{r_i}}}} }}{{\prod\limits_j {{{\left( {\wp \left( z \right) - \wp \left( {{w_j}} \right)} \right)}^{{s_j}}}} }} .
\end{equation*}
This function is obviously an elliptic function. Furthermore, it has no poles since, the poles of $g\left( z \right)$ are cancelled by the roots of the product in the numerator, the poles of the numerator are cancelled with the poles of the denominator, and the poles due to the zero's of the denominator are cancelled by the zeros of $g\left( z \right)$. As a result, the function $h$ is constant by Liouville's theorem and therefore,
\begin{equation}
g\left( z \right) = C\frac{{\prod\limits_j {{{\left( {\wp \left( z \right) - \wp \left( {{w_j}} \right)} \right)}^{{s_j}}}} }}{{\prod\limits_i {{{\left( {\wp \left( z \right) - \wp \left( {{u_i}} \right)} \right)}^{{r_i}}}} }} .
\end{equation}

Summarizing, \emph{any elliptic function can be written in terms of $\wp \left( z \right)$ and $\wp' \left( z \right)$ with the same periods. This expression is rational in $\wp \left( z \right)$ and linear in $\wp' \left( z \right)$.}

The above result has some interesting consequences. Consider two elliptic functions $f\left( z \right)$ and $g\left( z \right)$ with the same periods. Then, they can be both expressed as functions of $\wp$ and $\wp'$ with the same periods,
\begin{align*}
f\left( z \right) &= F \left[\wp\left( z \right) , \wp' \left( z \right) \right] , \\
g\left( z \right) &= G \left[\wp\left( z \right) , \wp' \left( z \right) \right] .
\end{align*}
Both functions $F$ and $G$ are rational in their first argument and linear in the second. Bearing in mind that $\wp \left( z \right)$ and $\wp' \left( z \right)$ are also connected through the Weierstrass equation, 
\begin{equation*}
\wp {'^2}\left( z \right) = 4{\wp ^3}\left( z \right) - {g_2}\wp \left( z \right) - {g_3} = 0,
\end{equation*}
one can eliminate $\wp$ and $\wp'$ and result in an algebraic relation between $f\left( z \right)$ and $g\left( z \right)$. This means that \emph{any pair of elliptic functions with the same periods are algebraically connected.} Two implications of the above sentence are
\begin{enumerate}
\item \emph{There is an algebraic relation between any elliptic function and its derivative.}
\item There is an algebraic relation between any elliptic function and the same elliptic function with shifted argument.
\end{enumerate}

An algebraic relation between an elliptic function $f$ and the same elliptic function with shifted argument reads
\begin{equation*}
\sum\limits_{k = 1}^n {{f^k}\left( {z + w} \right)\left( {\sum\limits_{l = 1}^n {{c_l}\left( w \right){f^l}\left( z \right)} } \right)}  = 0 ,
\end{equation*}
where ${c_l}\left( w \right)$ are unspecified functions of $w$. If one interchanges $z$ and $w$, they will get
\begin{equation*}
\sum\limits_{k = 1}^n {{f^k}\left( {z + w} \right)\left( {\sum\limits_{l = 1}^n {{c_l}\left( z \right){f^l}\left( w \right)} } \right)}  = 0 ,
\end{equation*}
implying that the unknown functions ${c_l}\left( w \right)$ are necessarily powers of ${{f}\left( w \right)}$, so that the sums ${\sum\limits_{l = 1}^n {{c_l}\left( w \right){f^l}\left( z \right)} }$ are symmetric polynomials in ${{f}\left( z \right)}$ and ${{f}\left( w \right)}$. This implies that the above equations take the form of an algebraic relation between ${{f}\left( z \right)}$, ${{f}\left( w \right)}$ and ${{f}\left( z + w \right)}$, i.e. \emph{an algebraic addition theorem}.

\subsubsection*{Expression of any Elliptic Function in Terms of $\zeta$ and its derivatives}

Consider an elliptic function $f\left( z \right)$. Assume that $u_i$ is an irreducible set of poles of $f\left( z \right)$, which have multiplicities $r_i$. Furthermore, assume that the principal part of the Laurent series at the region of a pole is given by
\begin{equation*}
f\left( z \right) \simeq \frac{{{c_{i,{r_i}}}}}{{{{\left( {z - {u_i}} \right)}^{{r_i}}}}} +  \ldots  + \frac{{{c_{i,2}}}}{{{{\left( {z - {u_i}} \right)}^2}}} + \frac{{{c_{i,1}}}}{{z - {u_i}}} + \mathcal{O}\left( {{{\left( {z - {u_i}} \right)}^0}} \right) .
\end{equation*}

Weierstrass $\zeta$ function has a single first order pole with residue equal to one in locations congruent to $z = 0$. It is obvious that the $n$-th derivative of $\zeta$ has a single $\left( n + 1 \right)$-th order pole and the principal part of Laurent series in the regime of $z = 0$ is
\begin{equation*}
\frac{{{d^n \zeta \left( {z} \right)}}}{{d{z^n}}} \simeq \frac{{{{\left( { - 1} \right)}^n}n!}}{{{ {z} ^{n + 1}}}} + \mathcal{O}\left( {{{\left( {z} \right)}^0}} \right) .
\end{equation*}

It follows that the function
\begin{multline*}
g\left( z \right) := f\left( z \right) - \sum\limits_i \left( {{c_{i,1}}\zeta \left( {z - {u_i}} \right) - {c_{i,2}}\frac{d \zeta \left( {z - {u_i}} \right)}{{dz}} +  \ldots  \phantom{\frac{{{{\left( { - 1} \right)}^{{r_i} - 1}}{c_{i,{r_i}}}}}{{\left( {{r_i} - 1} \right)!}}}} \right.\\
\left. {+ \frac{{{{\left( { - 1} \right)}^{{r_i} - 1}}{c_{i,{r_i}}}}}{{\left( {{r_i} - 1} \right)!}}\frac{{{d^{{r_i} - 1} \zeta \left( {z - {u_i}} \right)}}}{{d{z^{{r_i} - 1}}}}} \right)
\end{multline*}
has no poles. It is not obvious though whether $g\left( z \right)$ is an elliptic function, since the function $\zeta$ is not an elliptic function. However, all derivatives of $\zeta$ are elliptic functions, and, thus,
\begin{equation*}
g\left( {z + 2{\omega _i}} \right) - g\left( z \right) = \sum\limits_i {{c_{i,1}}\left( {\zeta \left( {z - {u_i}} \right) - \zeta \left( {z + 2{\omega _i} - {u_i}} \right)} \right)}  =  - 2\zeta \left( {{\omega _i}} \right)\sum\limits_i {{c_{i,1}}}  = 0 ,
\end{equation*}
since $\sum\limits_i {{c_{i,1}}}$ is the sum of the residues of the elliptic function $f\left( z \right)$ in a cell.

Thus, $g\left( z \right)$ is an elliptic function with no poles, and, thus, by Liouville's theorem, it is a constant function. This means that the original elliptic function $f\left( z \right)$ can be written as
\begin{equation}
f\left( z \right) = C + \sum\limits_i {\sum\limits_{j = i}^{{r_i}} {\frac{{{{\left( { - 1} \right)}^{j - 1}}{c_{i,j}}}}{{\left( {j - 1} \right)!}}\frac{{{d^{j - 1}}\zeta \left( {z - {u_i}} \right)}}{{d{z^{j - 1}}}}} } .
\label{eq:elliptic_terms_zeta}
\end{equation}

It follows that \emph{an elliptic function is completely determined by the principal parts of its Laurent series at an irreducible set of poles, up to an additive constant}.

\subsubsection*{Expression of any Elliptic Function in Terms of $\sigma$}

Consider an elliptic function $f\left( z \right)$ having an irreducible set of poles $u_i$ with multiplicities $r_i$ and an irreducible set of roots $w_j$ with multiplicities $s_j$. We remind the reader that $ \sum\limits_i {{s_i}{w_i}} \sim \sum\limits_j {{r_j}{u_j}} $. It is always possible to select the irreducible sets of poles and roots so that $ \sum\limits_i {{s_i}{w_i}} = \sum\limits_j {{r_j}{u_j}} $. In the following, we assume that we have made such a selection.

Assume the function
\begin{equation*}
g\left( z \right) := f\left( z \right)\frac{{\prod\limits_i {{\sigma ^{{r_\iota }}}\left( {z - {u_i }} \right)} }}{{\prod\limits_j {{\sigma ^{{s_j}}}\left( {z - {w_j}} \right)} }} .
\end{equation*}
Since the function $\sigma$ has a unique first order root in each cell congruent to the origin, it is obvious that $g\left( z \right)$ has neither poles nor roots in a cell.

It turns out that the quasi-periodicity property \eqref{eq:sigma_quasi_periodicity_step} of function $\sigma$ implies that $g\left( z \right)$ is an elliptic function. Indeed,
\begin{multline*}
g\left( {z + 2{\omega _k}} \right) = f\left( z \right)\frac{{\prod\limits_i {{\sigma ^{{r_\iota }}}\left( {z - {u_\iota }} \right)} }}{{\prod\limits_j {{\sigma ^{{s_j}}}\left( {z - {w_j}} \right)} }}\frac{{{{\prod\limits_i {\left( { - {e^{2\zeta \left( {{\omega _k}} \right)\left( {z - {u_i} - {\omega _k}} \right)}}} \right)} }^{{r_i}}}}}{{\prod\limits_j {{{\left( { - {e^{2\zeta \left( {{\omega _k}} \right)\left( {z - {w_j} - {\omega _k}} \right)}}} \right)}^{{s_j}}}} }}\\
 = g\left( z \right){\left( { - 1} \right)^{\sum\limits_i {{r_i}}  - \sum\limits_j {{s_j}} }}{e^{2\zeta \left( {{\omega _k}} \right)\left[ {\left( {z - {\omega _k}} \right)\left( {\sum\limits_i {{r_i}}  - \sum\limits_j {{s_j}} } \right) - \left( {\sum\limits_i {{r_i}{u_i}}  - \sum\limits_j {{s_j}{w_j}} } \right)} \right]}} = g\left( z \right) .
\end{multline*}
Since $g\left( z \right)$ is an elliptic function with no poles, it is a constant function and the original elliptic function $f\left( z \right)$ can be expressed as
\begin{equation}
f\left( z \right) = C\frac{{\prod\limits_j {{\sigma ^{{s_j}}}\left( {z - {w_j}} \right)} }}{{\prod\limits_i {{\sigma ^{{r_\iota }}}\left( {z - {u_\iota }} \right)} }} .
\label{eq:elliptic_terms_sigma}
\end{equation}

This implies that \emph{an irreducible set of poles and roots of an elliptic function completely determines it, up to a multiplicative constant}.

\subsection{Addition Theorems}
\label{subsec:addition_theorem}
\subsubsection*{Addition Theorem for $\wp$}
Above, we showed that there is an algebraic addition theorem for every elliptic function. Therefore such an addition theorem exists for Weierstrass elliptic function, too.

We define two functions of two complex variables ${c_1}\left( {z,w} \right)$ and ${c_2}\left( {z,w} \right)$ as
\begin{align*}
\wp '\left( z \right) &= {c_1}\left( {z,w} \right)\wp \left( z \right) + {c_2}\left( {z,w} \right) ,\\
\wp '\left( w \right) &= {c_1}\left( {z,w} \right)\wp \left( w \right) + {c_2}\left( {z,w} \right) 
\end{align*}
or in other words,
\begin{align*}
{c_1}\left( {z,w} \right) &:= \frac{{\wp '\left( z \right) - \wp '\left( w \right)}}{{\wp \left( z \right) - \wp \left( w \right)}} ,\\
{c_2}\left( {z,w} \right) &:= \frac{{\wp \left( z \right)\wp '\left( w \right) - \wp \left( w \right)\wp '\left( z \right)}}{{\wp \left( z \right) - \wp \left( w \right)}} .
\end{align*}
We also define the function $f\left( x \right)$ of one complex variable as
\begin{equation*}
f\left( x \right) := \wp '\left( x \right) - {c_1}\left( {z,w} \right)\wp \left( x \right) - {c_2}\left( {z,w} \right) .
\end{equation*}
Clearly, the function $f$ has only one third order pole in each cell congruent to $x = 0$. Moreover, by definition it has two roots at $x=z$ and $x=w$. Theorem \ref{th:number_poles_equals_number_roots} implies that there is another first order root, which is not congruent to $x=z$ or $x=w$. Theorem \ref{th:average_root_congruent_to_average_pole} implies that the position of this pole is congruent to $x = - z - w$,
\begin{equation*}
f\left( { - z - w} \right) = 0 .
\end{equation*}

The function $g\left( x \right)$ of one complex variable defined as,
\begin{equation*}
g\left( x \right) := \wp {'^2}\left( x \right) - {\left( {{c_1}\left( {z,w} \right)\wp \left( x \right) + {c_2}\left( {z,w} \right)} \right)^2} ,
\end{equation*}
clearly vanishes everywhere $f\left( x \right)$ vanishes, therefore,
\begin{equation*}
g\left( z \right) = g\left( w \right) = g\left( { - z - w} \right) = 0 .
\end{equation*}
Using Weierstrass differential equation \eqref{eq:Weierstrass_equation}, one can write the function $g\left( x \right)$ as,
\begin{equation*}
g\left( x \right) = 4{\wp ^3}\left( x \right) - c_1^2\left( {z,w} \right){\wp ^2}\left( x \right) - \left( {2{c_1}\left( {z,w} \right){c_2}\left( {z,w} \right) + {g_2}} \right)\wp \left( x \right) - \left( {c_2^2\left( {z,w} \right) + {g_3}} \right) .
\end{equation*}
The fact that $g\left( x \right)$ has the three roots $x=z$, $x=w$ and $x=-z-w$, implies that the third order polynomial
\begin{equation}
Q\left( P \right) := 4{P^3} - c_1^2\left( {z,w} \right){P^2} - \left( {2{c_1}\left( {z,w} \right){c_2}\left( {z,w} \right) + {g_2}} \right)P - \left( {c_2^2\left( {z,w} \right) + {g_3}} \right)
\label{eq:addition_polynomial_1}
\end{equation}
has the roots $P=\wp \left( z \right)$, $P=\wp \left( w \right)$ and $P=\wp \left( -z-w \right)$,
\begin{equation*}
Q\left( {\wp \left( z \right)} \right) = Q\left( {\wp \left( w \right)} \right) = Q\left( {\wp \left( { - z - w} \right)} \right) = 0 .
\end{equation*}
In other words, the polynomial $Q\left( P \right)$ can be written as
\begin{equation*}
Q\left( P \right) = 4\left( {P - \wp \left( z \right)} \right)\left( {P - \wp \left( w \right)} \right)\left( {P - \wp \left( { - z - w} \right)} \right) .
\label{eq:addition_polynomial_2}
\end{equation*}
Comparing the coefficients of the second order term of the polynomial $Q\left( P \right)$ in expressions \eqref{eq:addition_polynomial_1} and \eqref{eq:addition_polynomial_2}, we find
\begin{equation*}
c_1^2\left( {z,w} \right) = 4\left( {\wp \left( z \right) + \wp \left( w \right) + \wp \left( { - z - w} \right)} \right)
\end{equation*}
or
\begin{equation}
\wp \left( {z + w} \right) =  - \wp \left( z \right) - \wp \left( w \right) + \frac{1}{4}{\left( {\frac{{\wp '\left( z \right) - \wp '\left( w \right)}}{{\wp \left( z \right) - \wp \left( w \right)}}} \right)^2} ,
\label{eq:wp_addition}
\end{equation}
which is the desired addition theorem for the Weierstrass elliptic function. Although, expression \eqref{eq:wp_addition}, which is the traditional form of the addition theorem, involves the derivative of Weierstrass elliptic function, the latter can be eliminated with the use of Weierstrass differential equation \eqref{eq:Weierstrass_equation} resulting in an algebraic relation between $\wp \left( {z} \right)$, $\wp \left( {w} \right)$ and $\wp \left( {z + w} \right)$.

\subsubsection*{Pseudo-addition Theorems for $\zeta$ and $\sigma$}
The functions $\zeta$ and $\sigma$ are not elliptic, and, thus, they are not guaranteed to obey algebraic addition theorems. However, the fact that any elliptic function can be written as ratio of $\sigma$ functions can be used to deduce pseudo-addition theorems for them. Consider the function $\wp \left( z \right) - \wp \left( w \right)$ as a function of $z$. This function, obviously has a second order pole in each cell, congruent to $z=0$. It also obviously has two roots congruent to $z=w$ and $z=-w$. Consequently, equation \eqref{eq:elliptic_terms_sigma} implies that $\wp \left( z \right) - \wp \left( w \right)$ can be written as
\begin{equation*}
\wp \left( z \right) - \wp \left( w \right) = A\frac{{\sigma \left( {z - w} \right)\sigma \left( {z + w} \right)}}{{{\sigma ^2}\left( z \right)}} .
\end{equation*}
Writing down the principal part of the Laurent series of the above relation at the region of $z=0$, we find
\begin{equation*}
\frac{1}{{{z^2}}} = A\frac{{\sigma \left( { - w} \right)\sigma \left( w \right)}}{{{z^2}}} ,
\end{equation*}
implying that
\begin{equation*}
A =  - \frac{1}{{{\sigma ^2}\left( w \right)}} .
\end{equation*}
The above results in the following pseudo-addition theorem for $\sigma$ functions
\begin{equation}
\wp \left( z \right) - \wp \left( w \right) =  - \frac{{\sigma \left( {z - w} \right)\sigma \left( {z + w} \right)}}{{{\sigma ^2}\left( z \right){\sigma ^2}\left( w \right)}} .
\label{eq:sigma_addition}
\end{equation}

Differentiating equation \eqref{eq:sigma_addition} with respect to $z$ and $w$, we arrive of the following relations,
\begin{align*}
\wp '\left( z \right) &=  - \frac{{\sigma \left( {z - w} \right)\sigma \left( {z + w} \right)}}{{{\sigma ^2}\left( z \right){\sigma ^2}\left( w \right)}}\left( {\zeta \left( {z - w} \right) + \zeta \left( {z + w} \right) - 2\zeta \left( z \right)} \right) ,\\
 - \wp '\left( w \right) &=  - \frac{{\sigma \left( {z - w} \right)\sigma \left( {z + w} \right)}}{{{\sigma ^2}\left( z \right){\sigma ^2}\left( w \right)}}\left( { - \zeta \left( {z - w} \right) + \zeta \left( {z + w} \right) - 2\zeta \left( w \right)} \right) ,
\end{align*}
Adding them up, we find,
\begin{equation*}
\wp '\left( z \right) - \wp '\left( w \right) =  - 2\frac{{\sigma \left( {z - w} \right)\sigma \left( {z + w} \right)}}{{{\sigma ^2}\left( z \right){\sigma ^2}\left( w \right)}}\left( {\zeta \left( {z + w} \right) - \zeta \left( z \right) - \zeta \left( w \right)} \right) ,
\end{equation*}
which implies the following pseudo addition theorem for the $\zeta$ function
\begin{equation}
\frac{1}{2}\frac{{\wp '\left( z \right) - \wp '\left( w \right)}}{{\wp \left( z \right) - \wp \left( w \right)}} = \zeta \left( {z + w} \right) - \zeta \left( z \right) - \zeta \left( w \right) .
\label{eq:zeta_addition}
\end{equation}

\newpage
\subsection*{Problems}

\begin{problem}
Use the definitions \eqref{eq:zeta_definition} and \eqref{eq:sigma_definition} to deduce the homogeneity properties of the Weierstrass quasi-periodic functions
\begin{align}
\zeta \left( {z;{g_2},{g_3}} \right) &= \mu \zeta \left( {\mu z;\frac{{{g_2}}}{{{\mu ^4}}},\frac{{{g_3}}}{{{\mu ^6}}}} \right) ,\\
\sigma \left( {z;{g_2},{g_3}} \right) &= \frac{1}{\mu }\sigma \left( {\mu z;\frac{{{g_2}}}{{{\mu ^4}}},\frac{{{g_3}}}{{{\mu ^6}}}} \right).
\end{align}
\label{pr:homogeneity_quasi}
\end{problem}

\begin{problem}
Prove the parity properties of Weierstrass quasi-periodic functions. Namely,
\begin{itemize}
\item Show that the definition \eqref{eq:zeta_definition} together with the defining condition \eqref{eq:zeta_definition_condition} imply that $\zeta$ is an odd function.
\item
Show that the definition \eqref{eq:sigma_definition} together with the defining condition \eqref{eq:sigma_definition_condition} imply that $\sigma$ is an odd function.
\end{itemize}
\label{pr:parity_quasi_periodic}
\end{problem}

\begin{problem}
Prove the relations giving the Weierstrass quasi-periodic functions after a shift of their argument by an arbitrary period in the lattice of the corresponding Weierstrass elliptic function. Namely,
\begin{itemize}
\item Use equation \eqref{eq:zeta_quasi_periodicity_step} to prove \eqref{eq:zeta_quasi_periodicity}.
\item Use equation \eqref{eq:sigma_quasi_periodicity_step} to prove \eqref{eq:sigma_quasi_periodicity}.
\end{itemize}
\label{pr:quasi_periodic_property}
\end{problem}

\begin{problem}
Use the addition theorem for Weierstrass elliptic function to show that
\begin{align}
\wp \left( {z + {\omega _1}} \right) &= {e_1} + \frac{{2e_1^2 + {e_2}{e_3}}}{{\wp \left( z \right) - {e_1}}} ,\\
\wp \left( {z + {\omega _2}} \right) &= {e_3} + \frac{{2e_3^2 + {e_1}{e_2}}}{{\wp \left( z \right) - {e_3}}} ,\\
\wp \left( {z + {\omega _3}} \right) &= {e_2} + \frac{{2e_2^2 + {e_3}{e_1}}}{{\wp \left( z \right) - {e_2}}} .
\end{align}
\label{pr:half_period_shift}
\end{problem}

\begin{problem}
Use the fact that every elliptic function can be written in terms of the $\zeta$ function and its derivatives to deduce the pseudo-addition theorem for the $\zeta$ function.
\label{pr:zeta_addition}
\end{problem}

\begin{problem}
Use the pseudo-addition theorem of the $\zeta$ function to deduce the addition theorem for the $\wp$ function.
\label{pr:wp_addition}
\end{problem}

\begin{problem}
Use the addition theorem of the $\wp$ function to deduce a duplication formula. Then, Use the fact that $\wp \left( 2z \right)$ can be considered an elliptic function with the same periods as $\wp \left( z \right)$ to express it in terms of $\zeta \left( z \right)$ and its derivatives and result in the same duplication formula. For this purpose, you will find the results of problem \ref{pr:half_period_shift} useful.
\label{pr:wp_duplication}
\end{problem}

\newpage

\setcounter{equation}{0}
\section{Lecture 3: Applications in Classical Mechanics}
\label{sec:mechanics}
The first applications of the Weierstrass functions in physics that we will face are problems in classical mechanics with one degree of freedom, where Weierstrass equation \eqref{eq:Weierstrass_equation} emerges as conservation of energy. In such problems, the moduli $g_2$ and $g_3$ are connected to physical quantities, and, thus, they are real. Furthermore, the unknown function, as well as the independent variable in Weierstrass equation represent some physical quantity and they have to be real, too. It follows that we need to understand the real solutions in the real domain of equation \eqref{eq:Weierstrass_equation}, in the case the moduli are also real.

\subsection{Real Solutions of Weierstrass Equation in the Real Domain}
\label{subsec:real_Weierstrass}
\subsubsection*{The Weierstrass Elliptic Function with Real Moduli}
As the general solution of equation \eqref{eq:Weierstrass_equation} is given in terms of the Weierstrass elliptic function, we need to study the special properties of the former with real moduli.

When the parameters $g_2$ and $g_3$ are real, there are two possible cases for the reality of the three roots:
\begin{enumerate}
\item All three roots are real; conventionally we define them such that $e_1 > e_2 > e_3$. In this case, \emph{we may select the fundamental half-periods so that $\omega_1$ is real and $\omega_2$ is purely imaginary}. Then, they are given by the expressions,
\begin{align}
{\omega _1} &= \int_{{e_1}}^{ + \infty } {\frac{{dt}}{{\sqrt {4\left( {t - {e_1}} \right)\left( {t - {e_2}} \right)\left( {t - {e_3}} \right)} }}} , \label{eq:real_period}\\
{\omega _2} &= i\int_{ - \infty }^{{e_3}} {\frac{{dt}}{{\sqrt {4\left( {{e_1} - t} \right)\left( {{e_2} - t} \right)\left( {{e_3} - t} \right)} }}} . \label{eq:imaginary_period}
\end{align}
The above expressions imply that at the limit the two larger roots coincide, the real half-period diverges, whereas at the limit the two smaller roots coincide, the imaginary half-period diverges.
\item There is one real root and two complex ones being complex conjugate to each other; conventionally, we define $e_2$ as the real one and $e_1$ and $e_3$ so that $\textrm{Im} e_1 > 0$. In this case, \emph{it is not possible to select the fundamental half-periods as in the case of the three real roots, but we may select them so that they are complex conjugate to each other}. Then, they are given by the expressions,
\begin{align}
{\omega _1} + {\omega _2} &= \int_{{e_2}}^{ + \infty } {\frac{{dt}}{{\sqrt {4\left( {t - {e_1}} \right)\left( {t - {e_2}} \right)\left( {t - {e_3}} \right)} }}} , \label{eq:complex_period1}\\
{\omega _1} - {\omega _2} &= i\int_{ - \infty }^{{e_2}} {\frac{{dt}}{{\sqrt {4\left( {t - {e_1}} \right)\left( {{e_2} - t} \right)\left( {t - {e_3}} \right)} }}} . \label{eq:complex_period2}
\end{align}
\end{enumerate}
A proof of this statement is presented in the appendix.

\subsubsection*{The Locus of Complex Numbers $z$ for Whom $\wp \left( z \right)$ Is Real}
We return to the investigation for real solutions of equation \eqref{eq:Weierstrass_equation} with real moduli $g_2$ and $g_3$. Since the general solution of the latter is given in terms of Weierstrass elliptic function, this investigation requires the specification of the locus of complex numbers $z$ for whom $\wp \left( z \right)$ is real. A preliminary observation that can be made is the fact that the Weierstrass elliptic function assumes real values on the real and imaginary axes. The definition of the elliptic function $\wp$ \eqref{eq:Weierstrass_definition} implies that
\begin{equation*}
\overline {\wp \left( {z;{\omega _1},{\omega _2}} \right)}  = \wp \left( {\bar z;{{\bar \omega }_1},{{\bar \omega }_2}} \right) .
\end{equation*}
The definitions of the moduli $g_2$ and $g_3$, as given by equations \eqref{eq:g2_definition} and \eqref{eq:g3_definition} imply that
\begin{equation*}
{g_2}\left( {{{\bar \omega }_1},{{\bar \omega }_2}} \right) = \overline {{g_2}\left( {{\omega _1},{\omega _2}} \right)} ,\quad {g_3}\left( {{{\bar \omega }_1},{{\bar \omega }_2}} \right) = \overline {{g_3}\left( {{\omega _1},{\omega _2}} \right)} 
\end{equation*}
and consequently,
\begin{equation*}
\overline {\wp \left( {z;{g_2},{g_3}} \right)}  = \wp \left( {\bar z;{{\bar g}_2},{{\bar g}_3}} \right) .
\end{equation*}
Thus, when the moduli $g_2$ and $g_3$ are real, it holds that
\begin{equation}
\overline {\wp \left( {z;{g_2},{g_3}} \right)}  = \wp \left( {\bar z;{g_2},{g_3}} \right) .
\end{equation}

The above equation and the fact that $\wp$ is even imply that $\wp$ is real on the real and imaginary axes. Let $x \in \mathbb{R}$, then,
\begin{align}
\overline {\wp \left( {x;{g_2},{g_3}} \right)}  &= \wp \left( {\bar x;{g_2},{g_3}} \right) = \wp \left( {x;{g_2},{g_3}} \right) ,\\
\overline {\wp \left( {ix;{g_2},{g_3}} \right)}  &= \wp \left( { - ix;{g_2},{g_3}} \right) = \wp \left( {ix;{g_2},{g_3}} \right) .
\end{align}

\begin{figure}[ht]
\vspace{10pt}
\begin{center}
\begin{picture}(100,40)
\put(8,7.5){\includegraphics[width = 0.4\textwidth]{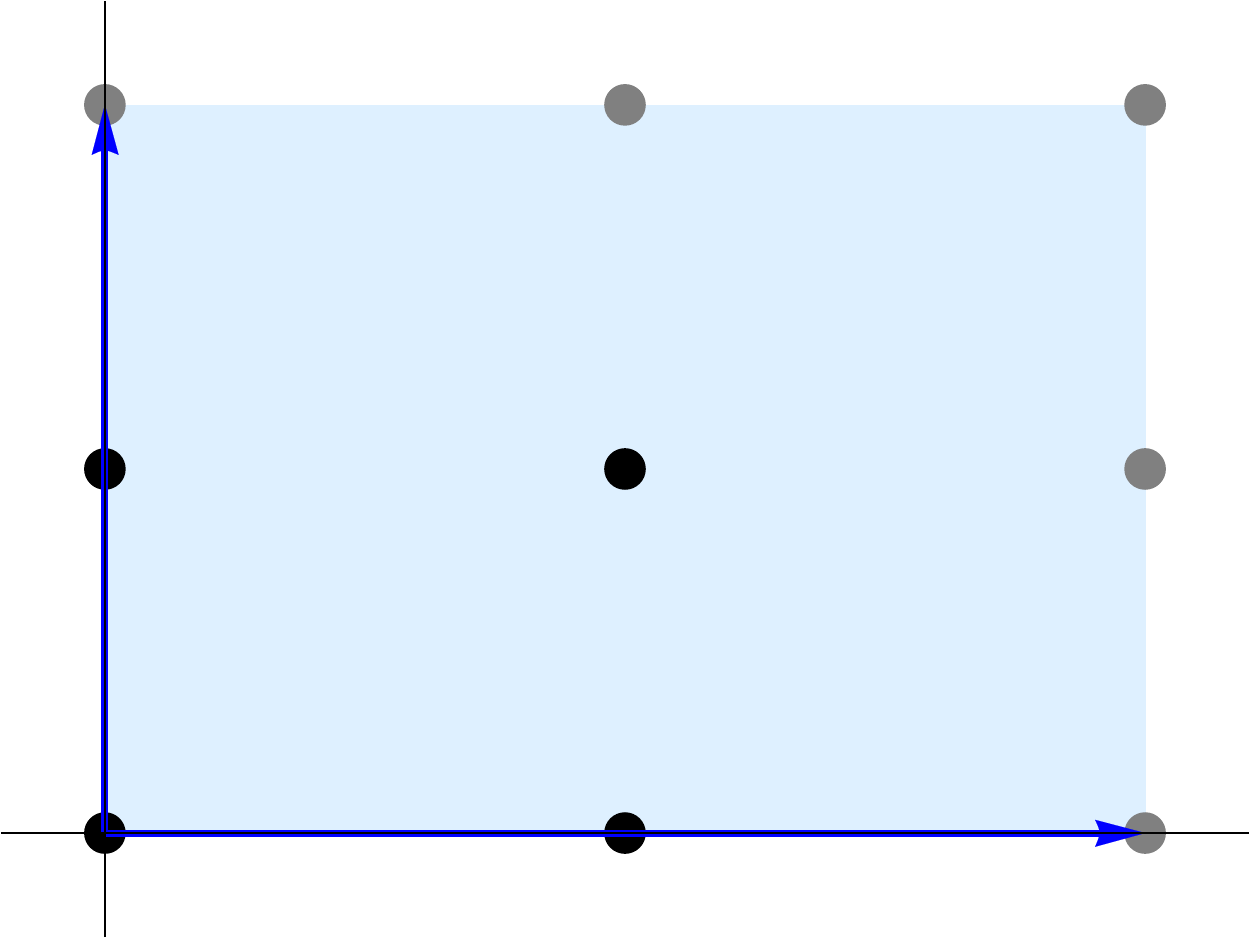}}
\put(52,7.5){\includegraphics[width = 0.4\textwidth]{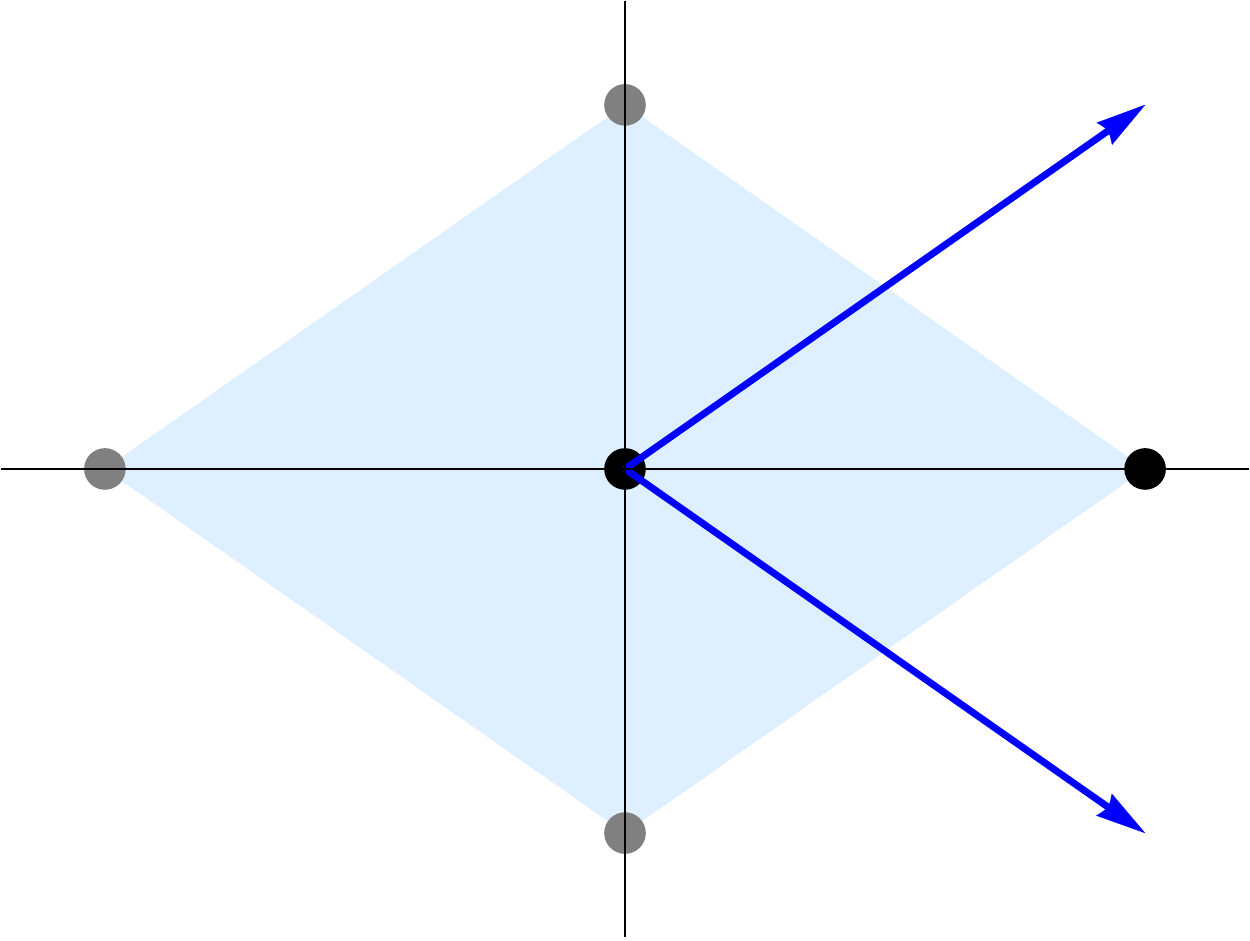}}
\put(48.5,10.25){\color{green}Re$z$}
\put(9,38){\color{red}Im$z$}
\put(92.5,21.75){\color{green}Re$z$}
\put(70,38){\color{red}Im$z$}
\put(34.5,8.5){\color{blue}$2\omega_1$}
\put(7,27,5){\color{blue}$2\omega_2$}
\put(86,8.5){\color{blue}$2\omega_1$}
\put(86,35,5){\color{blue}$2\omega_2$}
\put(89,20.5){\color{green}$e_2$}
\put(72,20.5){\color{green}$+\infty$}
\put(55.5,20.5){\color{green}$e_2$}
\put(69,35){\color{red}$e_2$}
\put(66.5,23.5){\color{red}$-\infty$}
\put(69,12){\color{red}$e_2$}
\put(44.5,8.75){\color{green}$+\infty$}
\put(28,8.75){\color{green}$e_1$}
\put(11.25,8.75){\color{green}$+\infty$}
\put(6,34.5){\color{red}$-\infty$}
\put(8.25,23.25){\color{red}$e_3$}
\put(6,11.75){\color{red}$-\infty$}
\put(28.25,23.5){$e_2$}
\put(17,3){three real roots}
\put(65,3){one real root}
\end{picture}
\end{center}
\vspace{-10pt}
\caption{The values of $\wp$ on the real and imaginary axes}
\vspace{5pt}
\label{fig:real_cell}
\end{figure}
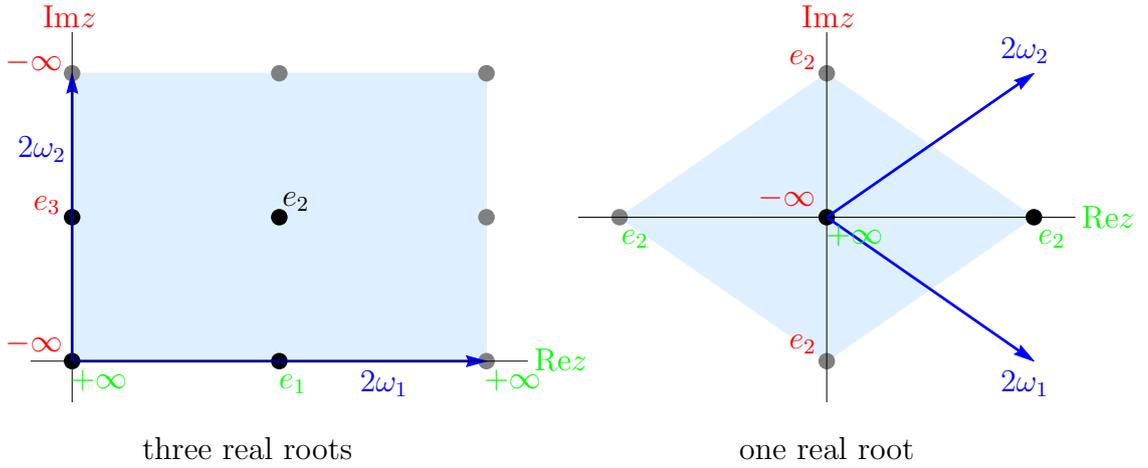

But is $\wp$ real on any other points not congruent to the real or imaginary axes? The answer depends on the reality of the roots $e_1$, $e_2$ and $e_3$. The function $\wp$ is an order two elliptic function, and, thus, it assumes any real value (as well as any complex value) twice in every cell. The only exception to this rule are the three roots $e_i$, which appear only once, since they correspond to double roots of the equation $\wp \left( z \right) =e_i$. In figure \ref{fig:real_cell}, one cell of $\wp$ with real moduli is plotted for either three or one real root.
Dotted points are congruent to a period or a half-period. Grey dots at the boundary of the plotted cell are congruent to a black dot at another point of the boundary, and thus, they are not considered to be part of the cell. This holds for segments that connect grey dots. Moreover, as the poles are second order with Laurent coefficient equal to one, as one approaches a pole from the real axis, $\wp$ tends to $+\infty$, while when one approaches a pole from the imaginary axis, $\wp$ tends to $-\infty$. Finally, as the derivative vanishes only at the half periods, $\wp$ is monotonous at the segments between consequent half-periods and poles.

Having the above into consideration, and following picture \ref{fig:real_cell}, it is clear that
\begin{itemize}
\item when there are three real roots, the function $\wp$ assumes all real values larger than $e_1$ twice in the segment $\left[ 0 , 2 \omega_1 \right]$ on the real axis; each value appears once in $\left[ 0 , \omega_1 \right]$ and once in $\left[ \omega_1 , 2\omega_1 \right]$. Similarly, it takes all real values smaller than $e_3$ twice in the segment $\left[ 0 , 2\omega_2 \right]$ on the imaginary axis.
\item when there is only one real root, $\wp$ takes all real values larger than $e_2$ twice in the segment $\left[ - \omega_1 - \omega_2 , \omega_1 + \omega_2 \right]$. Similarly, it takes all real values smaller than $e_2$ twice in the segment $\left[ \omega_2 - \omega_1 , \omega_1 - \omega_2 \right]$ on the imaginary axis.
\end{itemize}

Therefore in the case of one real root, the function $\wp$ assumes all real values exactly twice in the cell of figure \ref{fig:real_cell} at positions on the real and imaginary axes. Indeed, only $e_2$ appears once, as it appears only at positions congruent to each other. Consequently, the function $\wp$ cannot assume any real value at any other point within the cell, and, thus all positions where $\wp$ is real on the complex plane are congruent to a point either on the real or the imaginary axis.

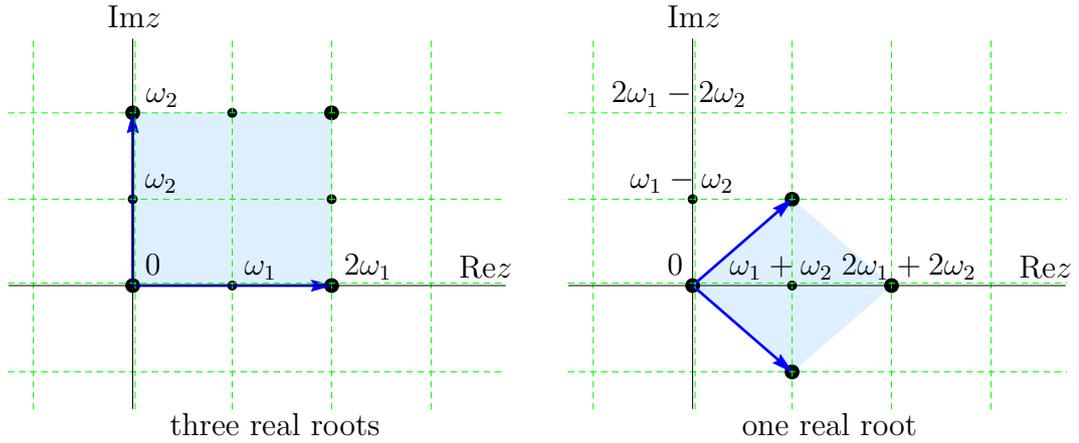
\begin{figure}[ht]
\vspace{10pt}
\begin{center}
\begin{picture}(100,36)
\put(7.5,2){\includegraphics[width = 0.4\textwidth]{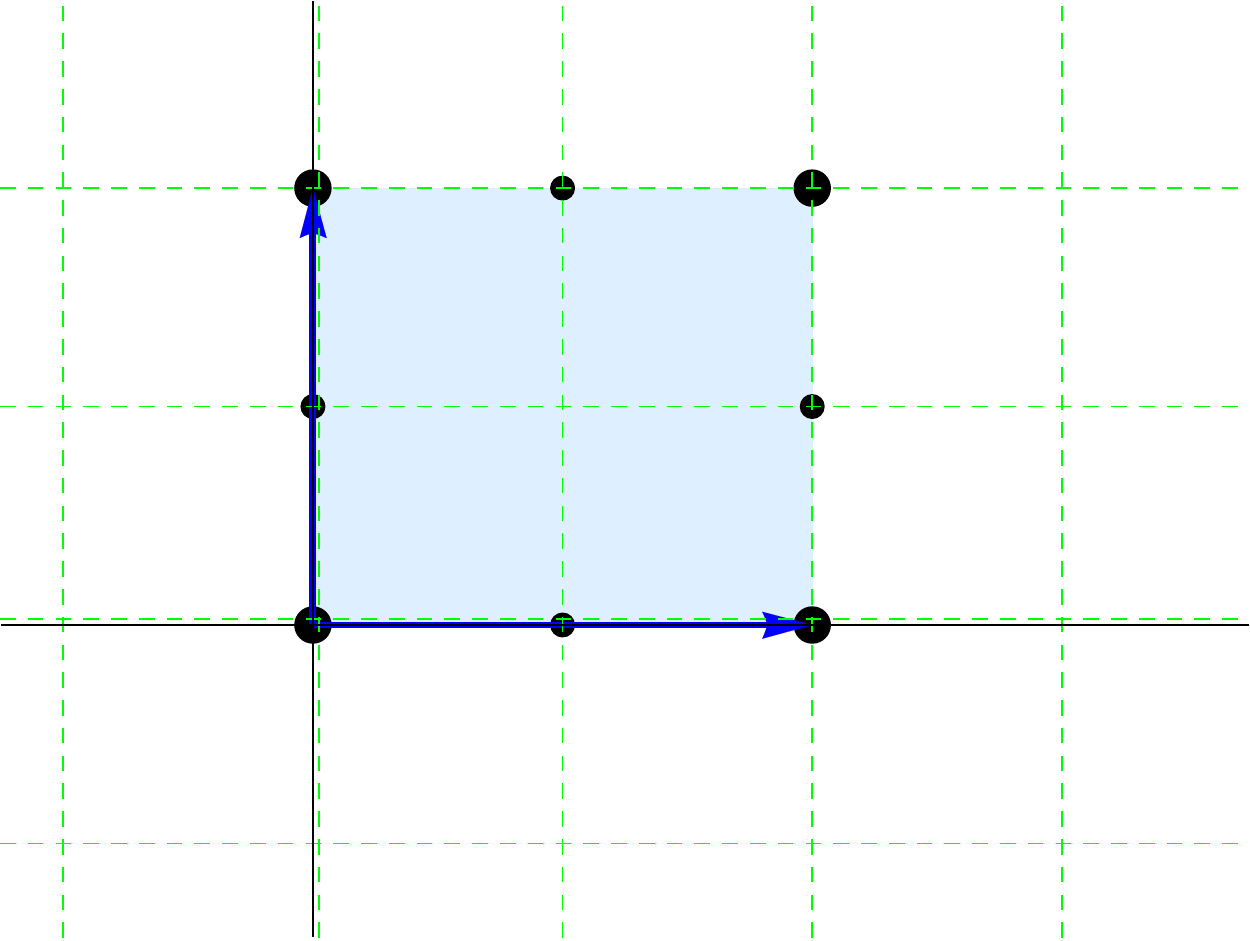}}
\put(52.5,2){\includegraphics[width = 0.4\textwidth]{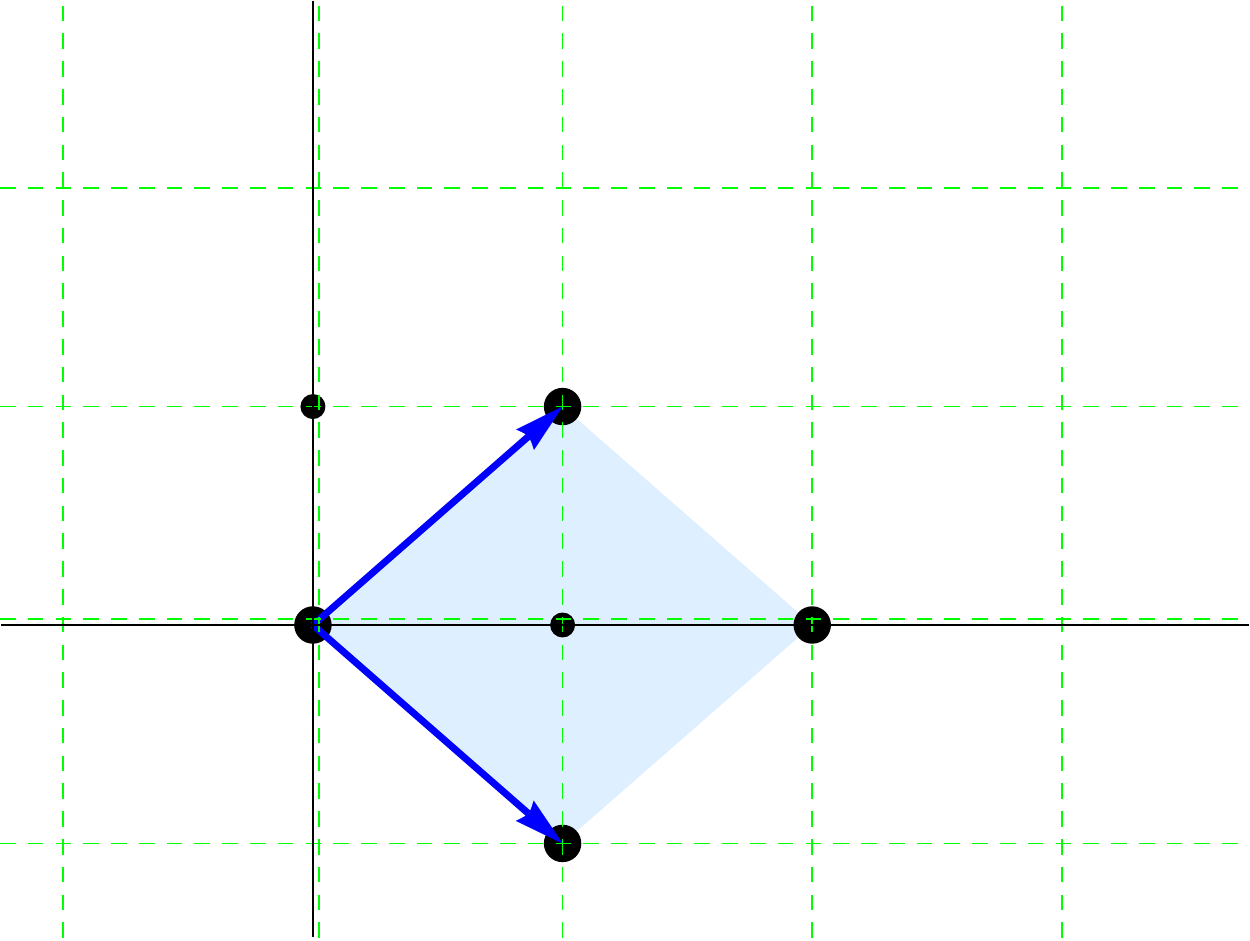}}
\put(43.75,12.75){Re$z$}
\put(15.5,33){Im$z$}
\put(88.75,12.75){Re$z$}
\put(60.5,33){Im$z$}
\put(18.5,13){0}
\put(26.5,13){$\omega_1$}
\put(34.5,13){$2\omega_1$}
\put(18.5,20){$\omega_2$}
\put(18.6,27){$\omega_2$}
\put(60.5,13){0}
\put(65.5,13){$\omega_1 + \omega_2$}
\put(74.5,13){$2\omega_1 + 2\omega_2$}
\put(57.5,20){$\omega_1 - \omega_2$}
\put(56,27){$2\omega_1 - 2\omega_2$}
\put(20.5,0){three real roots}
\put(66.5,0){one real root}
\end{picture}
\end{center}
\vspace{-10pt}
\caption{The locus of complex numbers $z$ for whom $\wp \left( z \right)$ is real}
\vspace{5pt}
\label{fig:preal}
\end{figure}

In the case there are three roots, the function $\wp$ assumes all real values larger than $e_1$ or smaller than $e_3$ exactly twice in the cell at positions on the real and imaginary axes. This means that there are other positions within the cell, where $\wp$ is real and it assumes all real values between $e_3$ and $e_1$. We already know such a point, namely $\omega_3$, where the function $\wp$ assumes the value $e_2$. It is a natural guess that $\wp$ is real on the horizontal and vertical axes passing through $\omega_3$. This is indeed true. Assuming that $x \in \mathbb{R}$,
\begin{align}
\overline {\wp \left( {ix + {\omega _1}} \right)}  &= \wp \left( {-ix + {\omega _1}} \right) = \wp \left( {ix - {\omega _1}} \right) = \wp \left( {ix - {\omega _1} + 2{\omega _1}} \right) = \wp \left( {ix + {\omega _1}} \right), \\
\overline {\wp \left( {x + {\omega _2}} \right)}  &= \wp \left( {x - {\omega _2}} \right) = \wp \left( {x - {\omega _2} + 2{\omega _2}} \right) = \wp \left( {x + {\omega _2}} \right) .
\end{align}
Thus, $\wp$ assumes the values between $e_2$ and $e_1$ twice in the segment $\left[ \omega_1 , \omega_1 + 2\omega_2 \right]$ on the shifted imaginary axis and the values between $e_3$ and $e_2$ exactly twice in the segment $\left[ \omega_2 , 2\omega_1 + \omega_2 \right]$ on the shifted real axis. Thus, all real numbers appear twice on the real, shifted real, imaginary and shifted imaginary axes except for the roots that appear only once. Therefore, the function $\wp$ cannot be real at any other position within the cell. All positions in the complex plane, where $\wp$ assumes real values have to be congruent to point on these four segments. Figure \ref{fig:preal} displays these positions.

\subsubsection*{Real Solutions of Weierstrass Equation in the Real Domain}
It is now simple to find what are the real solutions in the real domain of the equation
\begin{equation}
{\left( {\frac{{dy}}{{dx}}} \right)^2} = 4{y^3} - {g_2}y - {g_3} ,
\label{eq:Weierstrass_eq_real}
\end{equation}
where $g_2$ and $g_3$ are real. We know that in the complex domain, the general solution of this equation is
\begin{equation}
y = \wp \left( x - z_0 \right) ,
\end{equation}
where $z_0 \in \mathbb{C}$. In our problem $x$ and $y$ has to be real. However, $z_0$ is a constant of integration and has no physical meaning. It can assume any complex value, as long as $y$ is real for any real $x$. This is equivalent to selecting any line on the complex plane that is parallel to the real axis. It is evident in figure \ref{fig:preal} that when there is only one real root, all such lines are congruent to the real axis itself. On the contrary, when there are three real roots, there are two options, they are congruent to the real axis, or the real axis shifted by $\omega_2$. Consequently, \emph{the general real solution of \eqref{eq:Weierstrass_eq_real} in the real domain is}
\begin{equation}
y = \wp \left( x - x_0 \right) ,
\label{eq:solution_single}
\end{equation}
\emph{where $x_0 \in \mathbb{R}$, when there is one real root and}
\begin{equation}
y = \wp \left( x - x_0 \right) \quad \mathrm{\emph{or}} \quad y = \wp \left( x - x_0 +\omega_2 \right) ,
\label{eq:solution_three}
\end{equation}
\emph{where $x_0 \in \mathbb{R}$, when there are three real roots; the appropriate choice depends on initial conditions, namely whether the initial value of $y$ lies within $\left[ e_1 , \infty \right)$ or $\left[ e_3 , e_2 \right]$}.

\subsection{Point Particle in a Cubic Potential}
\label{subsec:cubic}
\subsubsection*{Problem Definition}
Let's now consider a point particle moving in one dimension under the influence of a force that is a quadratic function of position. We select the origin of the coordinate system as the position of extremal force and we select units, such that the mass of the particle equals 2 and the coefficient of the quadratic term in the force equals 12. The equation of motion is written as
\begin{equation}
2\ddot x = {F_0} + 12{x^2} .
\end{equation}
This equation can be integrated once to the form of conservation of energy. Fixing the integration constant so that the potential vanishes at $x=0$ we get
\begin{equation}
{{\dot x}^2} + V\left( x \right) = E,\quad V\left( x \right) =  - {F_0}x - 4{x^3} 
\end{equation}
or
\begin{equation}
{{\dot x}^2} = 4{x^3} + {F_0}x + E .
\label{eq:cubic_energy_conservation}
\end{equation}

Obviously, there is no local minimum of the potential when $F_0 > 0$. In this case all motions of the problem are scattering solutions evolving from $+\infty$ to a minimum $x$ and back to $+\infty$. On the contrary, when $F_0 < 0$ there is a local maximum at $x = x_0 \equiv \sqrt{- F_0 / 12}$ and a local minimum at $x = - x_0$, and, thus, a range of values for the energy $E$, namely,
\begin{equation}
\left| E \right| < E_0, \quad E_0 = {\left( { - \frac{{{F_0}}}{3}} \right)^{\frac{3}{2}}} ,
\end{equation}
for which the equation $V \left( x \right) = E$ has three roots and consequently there are two possible kinds of motion. One of them is a scattering solution evolving from $+\infty$ to a minimum $x$ and back to $+\infty$ and the other is an oscillating solution in the region of the local minimum of the potential.
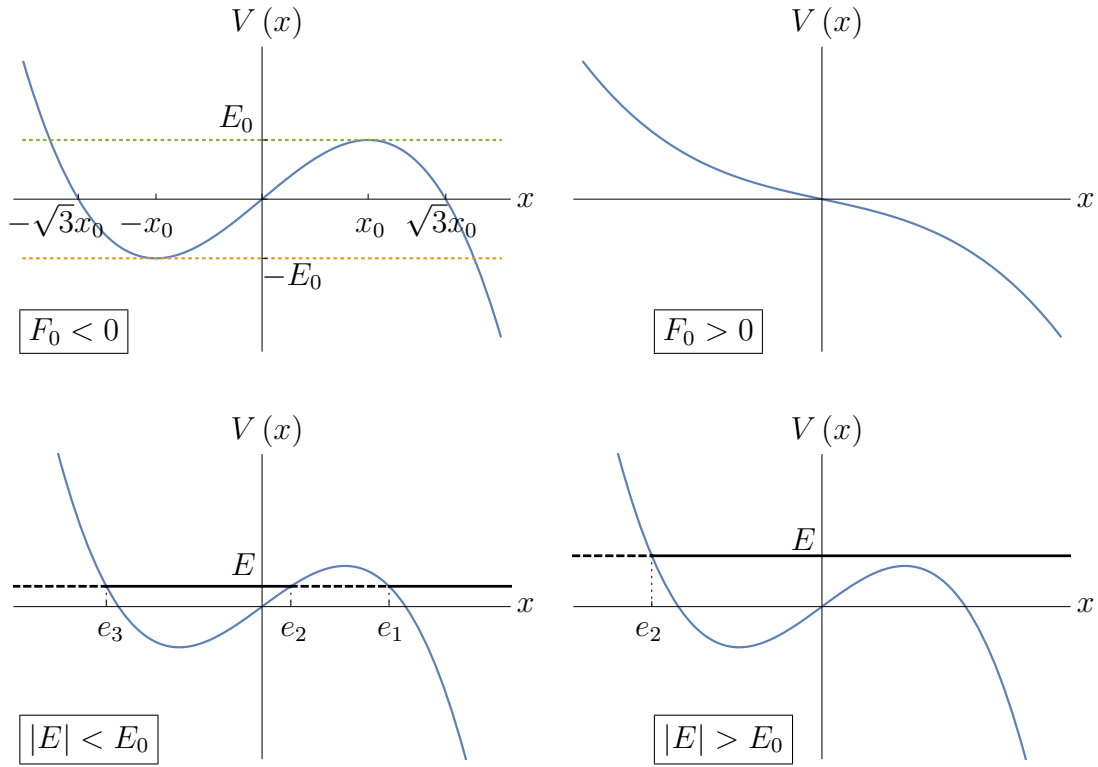
\begin{figure}[ht]
\vspace{10pt}
\begin{center}
\begin{picture}(100,63)
\put(7.5,35){\includegraphics[width = 0.4\textwidth]{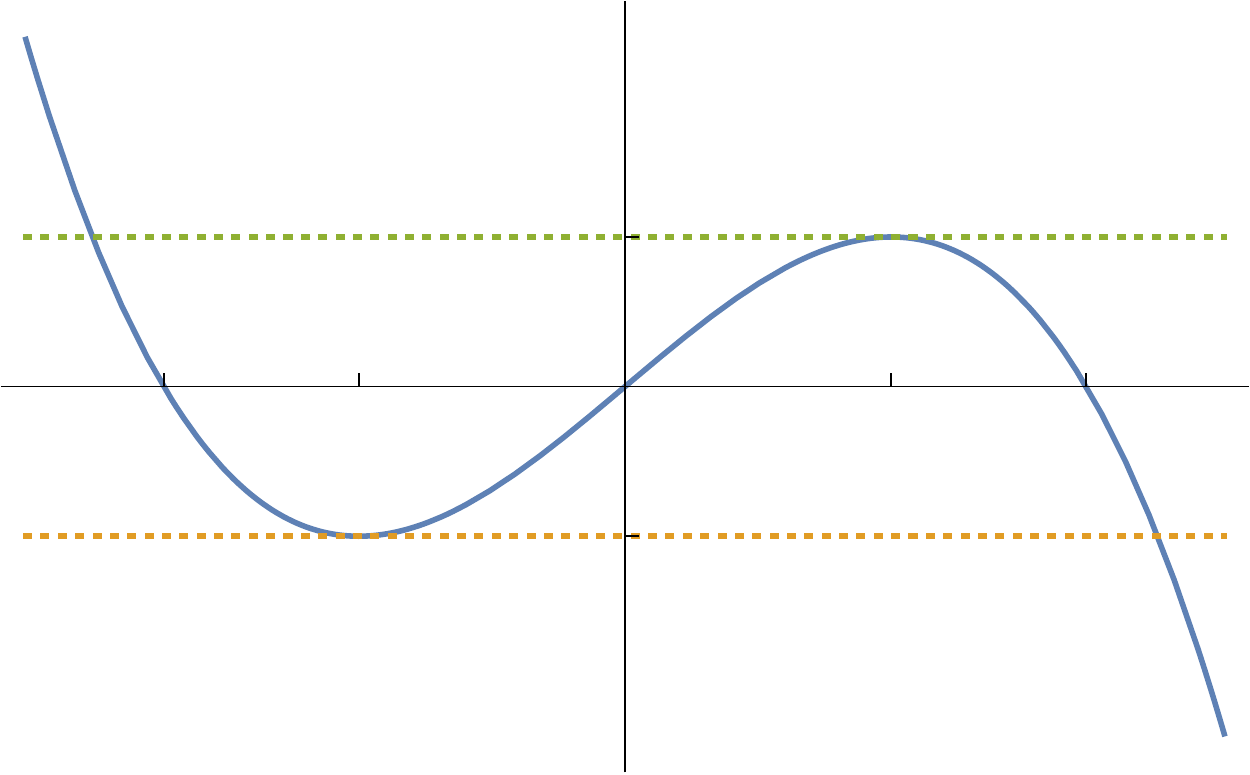}}
\put(52.5,35){\includegraphics[width = 0.4\textwidth]{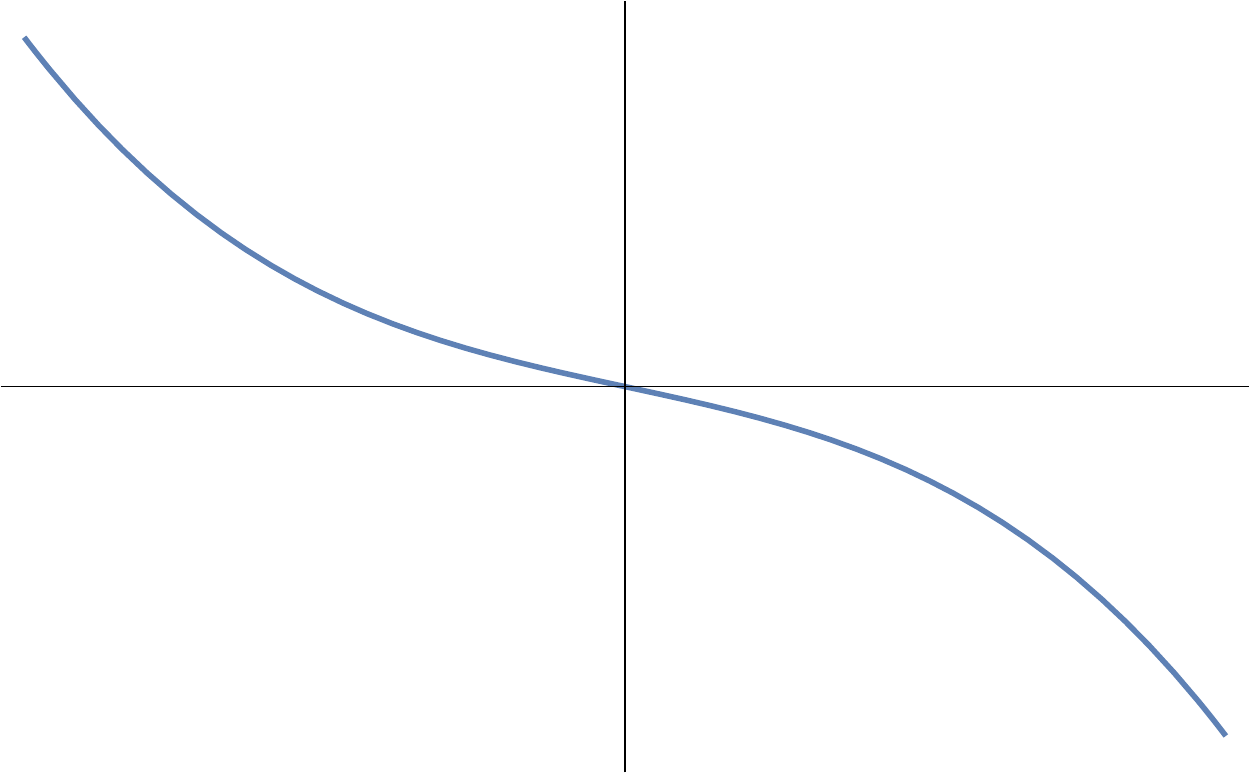}}
\put(7.5,2){\includegraphics[width = 0.4\textwidth]{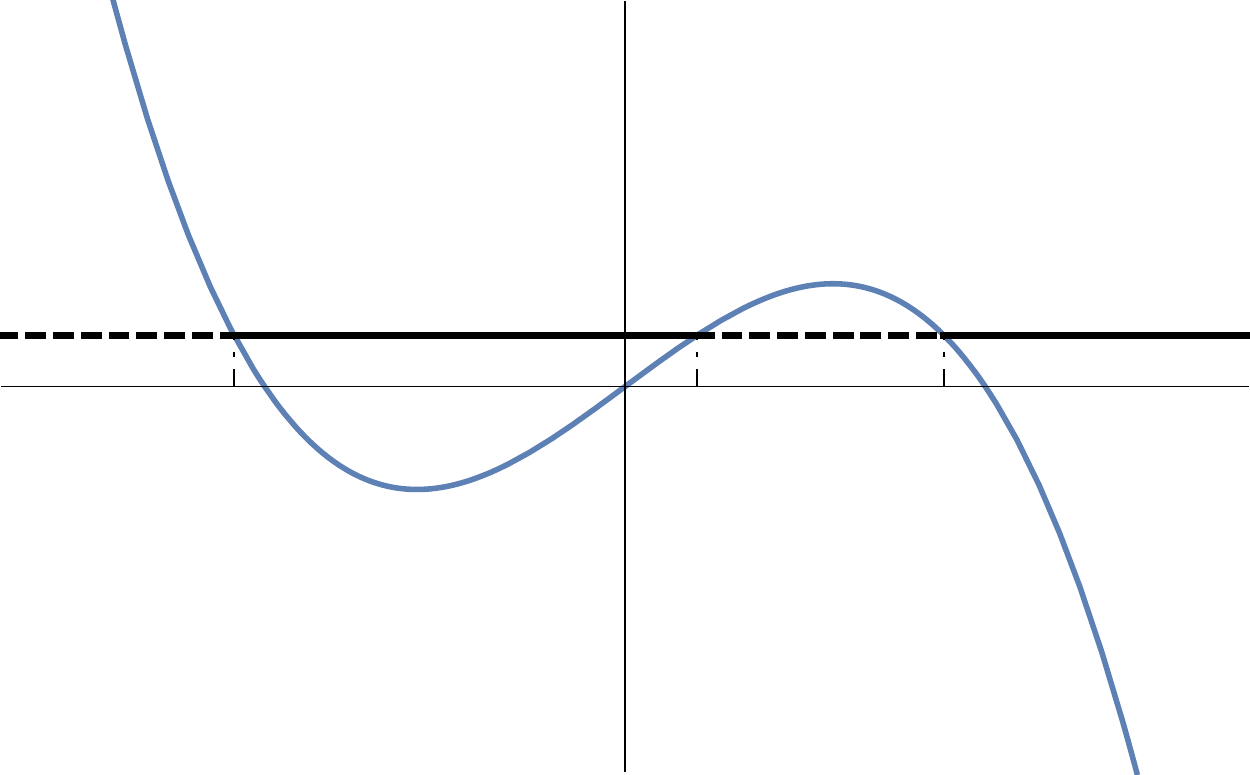}}
\put(52.5,2){\includegraphics[width = 0.4\textwidth]{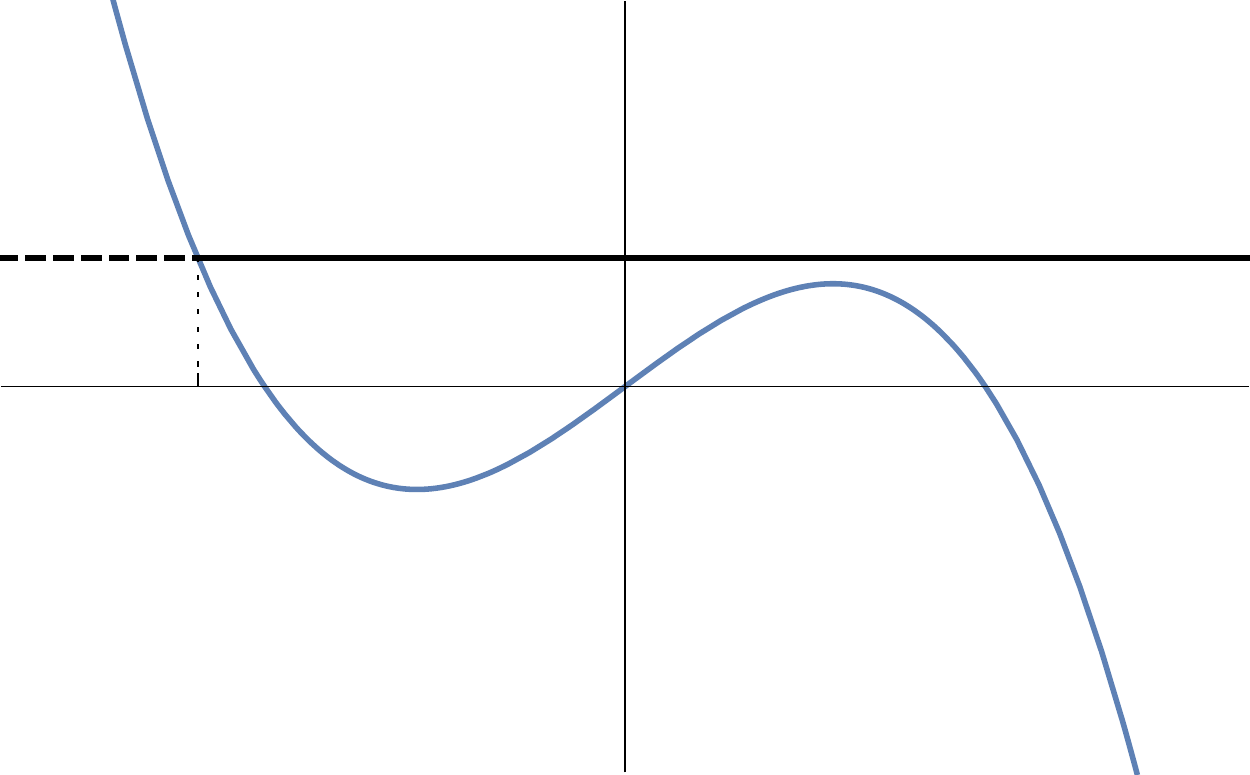}}
\put(25,61){$V \left( x \right)$}
\put(25,28){$V \left( x \right)$}
\put(70,61){$V \left( x \right)$}
\put(70,28){$V \left( x \right)$}
\put(48,46.75){$x$}
\put(93,46.75){$x$}
\put(48,14){$x$}
\put(93,14){$x$}
\put(57.5,12){$e_2$}
\put(36.75,12){$e_1$}
\put(29,12){$e_2$}
\put(14.25,12){$e_3$}
\put(25,17){$E$}
\put(70,19){$E$}
\put(27.5,40.5){$- E_0$}
\put(24,53.25){$E_0$}
\put(35,44.75){$x_0$}
\put(16,44.75){$-x_0$}
\put(39,44.75){$\sqrt{3}x_0$}
\put(7,44.75){$-\sqrt{3}x_0$}
\put(8,3){\framebox{$\left| E \right| < E_0$}}
\put(59,3){\framebox{$\left| E \right| > E_0$}}
\put(8,36){\framebox{$F_0 < 0$}}
\put(59,36){\framebox{$F_0 > 0$}}
\end{picture}
\end{center}
\vspace{-10pt}
\caption{The cubic potential for $F_0 < 0$ (top left) and $F_0 > 0$ top right. In the case $F_0 < 0$, there may exist one (bottom right) or two (bottom left) possible motions depending on the value of the energy.}
\vspace{5pt}
\label{fig:cubic}
\end{figure}

\subsubsection*{Problem Solution}
In the language of classical mechanics, it now becomes obvious why the Weierstrass differential equation has two independent real solutions when there are three real roots and only one when there is one real root. The roots play the role of the extrema of the motion, which are indeed the positions the velocity vanishes, as required for the roots. Furthermore, the solution that always exists is the one defined on the real axis,
\begin{equation}
x = \wp \left( {t - {t_0}; - {F_0}, - E} \right) ,
\end{equation}
which contains the pole, and, thus, it corresponds to the scattering solution. The solution on the shifted real axis,
\begin{equation}
x = \wp \left( {t - {t_0} + \omega_2; - {F_0}, - E} , \right),
\end{equation}
is bounded between $e_3$ and $e_2$ and corresponds to the oscillating solution in the region of the local minimum. Finally, we would like to commend that from the point of view of classical mechanics, it is natural that the Weierstrass elliptic function (or more literally the solution of Weierstrass equation) has order equal to two. In every position there are two possible initial conditions that correspond to the same energy, depending on the direction of the initial velocity. This is mirrored in the fact that the same real value appears in two non-congruent positions in every cell. This does not apply only at the extrema of motion, where the appropriate initial velocity vanishes and indeed these correspond to the roots of the Weierstrass function, which appear only once in every cell.

Finally, let's calculate the ``time of flight'' for scattering solutions and the period of the oscillating solutions. The first is the distance between two consecutive poles on the real axis, which obviously equals
\begin{equation}
T_{\rm {scattering}} = 2 \omega_1.
\end{equation}
Similarly, the period of the oscillating solutions is the distance between two consecutive appearances of the same root. This is also clearly
\begin{equation}
T_{\rm {oscillating}} = 2 \omega_1.
\end{equation}
Therefore, \emph{for the energies that an oscillating solution exists, the ``time of flight'' of the scattering solution and the period of the oscillating solution with the same energy are equal}.

\subsubsection*{The Role of the Imaginary Period}
In the above we found that the Weierstrass elliptic function naturally describes the motion of a point particle in a cubic potential. The real period of Weierstrass function ($ 2\omega_1$ in the case of three real roots and $2 \omega_3$ in the case of one real root) plays the role of the ``time of flight'' or the period of the motion. Is there any physical meaning for the imaginary period?

It is easy to answer this question using the homogeneity transformation \eqref{eq:Weierstras_homogeneity_wp}. This relation with $\mu = i$ implies that
\begin{equation}
\wp \left( {iz;{g_2},{g_3}} \right) =  - \wp \left( {z;{g_2}, - {g_3}} \right) .
\end{equation}
It is a direct consequence that $\wp \left( {iz;{g_2},{g_3}} \right)$ obeys the differential equation,
\begin{equation*}
{\left( {\frac{{d\wp \left( {iz;{g_2}, - {g_3}} \right)}}{{dz}}} \right)^2} =  - 4 {\wp ^3}\left( {iz;{g_2}, - {g_3}} \right) + {g_2}\wp \left( {iz;{g_2}, - {g_3}} \right) - {g_3} .
\end{equation*}
Selecting $g_2 = - F_0$ and $g_3 = E$, we find that the function $\wp \left( {iz; - {F_0}, - E} \right)$ obeys the differential equation,
\begin{equation*}
{\left( {\frac{{d\wp \left( {iz; - {F_0}, - E} \right)}}{{dz}}} \right)^2} =  - 4 {\wp ^3}\left( {iz; - {F_0}, - E} \right) - {F_0}\wp \left( {iz; - {F_0}, - E} \right) - E
\end{equation*}
and thus it is a solution to yet another one dimensional point particle problem, namely,
\begin{equation}
{{\dot y}^2} + \tilde V\left( y \right) =  - E,\quad \tilde V\left( y \right) = {y^3} + {F_0}y =  - V\left( y \right) ,
\end{equation}
which clearly is the problem of motion of a point particle in the inverted potential to that of the initial problem, having the opposite energy.
\begin{figure}[ht]
\vspace{10pt}
\begin{center}
\begin{picture}(100,32)
\put(7.5,2){\includegraphics[width = 0.4\textwidth]{cubic3.pdf}}
\put(52.5,2){\includegraphics[width = 0.4\textwidth]{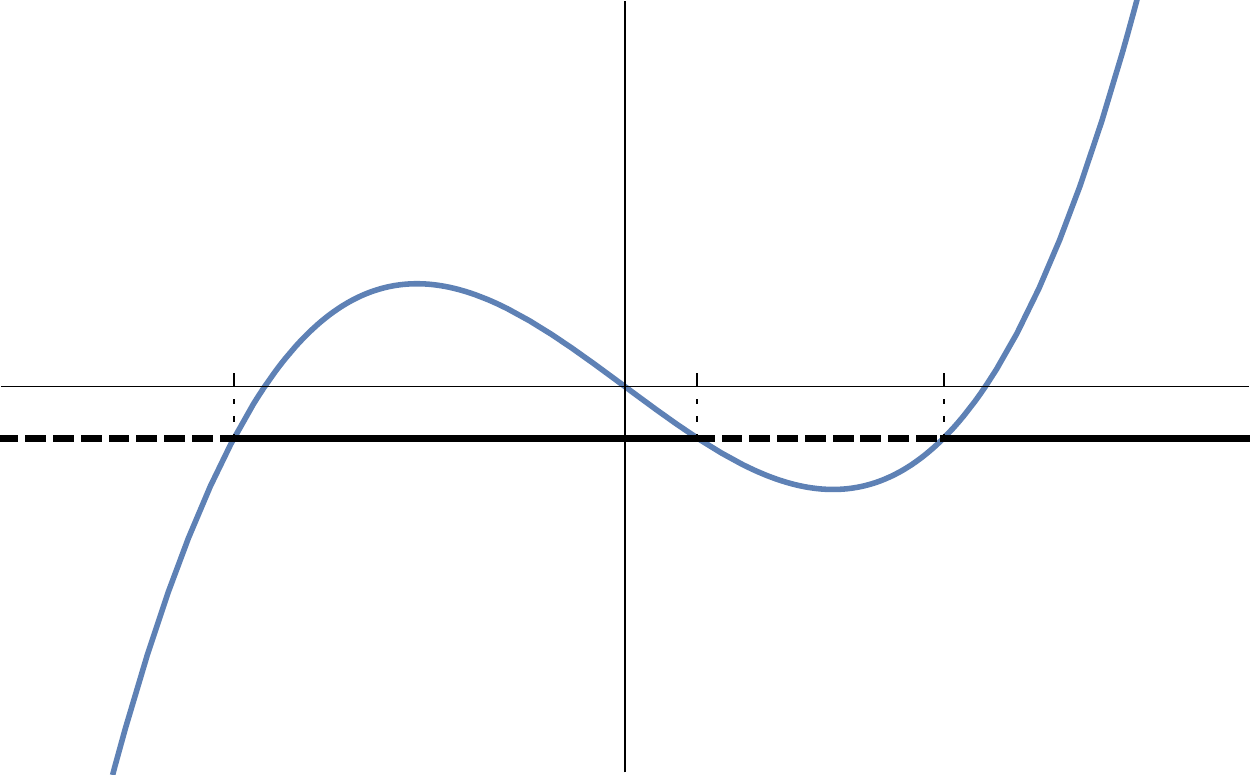}}
\put(25,28){$V \left( x \right)$}
\put(70,28){$\tilde V \left( y \right)$}
\put(48,14){$x$}
\put(93,14){$y$}
\put(36.75,12){$e_1$}
\put(29,12){$e_2$}
\put(14.25,12){$e_3$}
\put(81.75,15.5){$e_1$}
\put(74,15.5){$e_2$}
\put(59.25,15.5){$e_3$}
\put(25,17){$E$}
\put(70,10.5){$E$}
\end{picture}
\end{center}
\vspace{-10pt}
\caption{The original and inverted point particle problems}
\vspace{5pt}
\label{fig:inverted_cubic}
\end{figure}

It is obvious that in this problem, the point particle is moving in the complementary region than that of the initial problem. Furthermore, since the position of the particle has to be real, it turns out that the unbounded motion of the point particle under the influence of the potential $\tilde V$ is given by
\begin{equation}
y = \wp \left( i {\left( t - {t_0} \right) ; - {F_0}, - E} \right) ,
\end{equation}
while if there is a bounded one, it is given by
\begin{equation}
y = \wp \left( i {\left( t - {t_0} \right) + \omega_1; - {F_0}, - E} , \right).
\end{equation}

The physical meaning of the imaginary period is now obvious. The ``time of flight'' of the unbounded motion, as well as the period of the bounded motion in this inverted problem are given by the imaginary period of $\wp \left( {z ; - {F_0}, - E} , \right)$,
\begin{equation}
{\tilde T}_{\rm {scattering}} = {\tilde T}_{\rm {oscillating}} = - 2 i \omega_2.
\end{equation}

\subsection{The Simple Pendulum}
\label{subsec:pendulum}
The Weierstrass elliptic function naturally describes a point particle in a cubic potential, due to the fact that the conservation of energy takes the form of the Weierstrass equation \eqref{eq:Weierstrass_equation}. Its applications though are not limited to this problem. There are several other problems with one degree of freedom that can be transformed to that of a cubic potential with an appropriate coordinate transformation.

\subsubsection*{Problem Definition and Equivalence to Weierstrass Equation}
\begin{figure}[ht]
\vspace{10pt}
\begin{center}
\begin{picture}(53,33)
\put(0,0){\includegraphics[width = 0.5\textwidth]{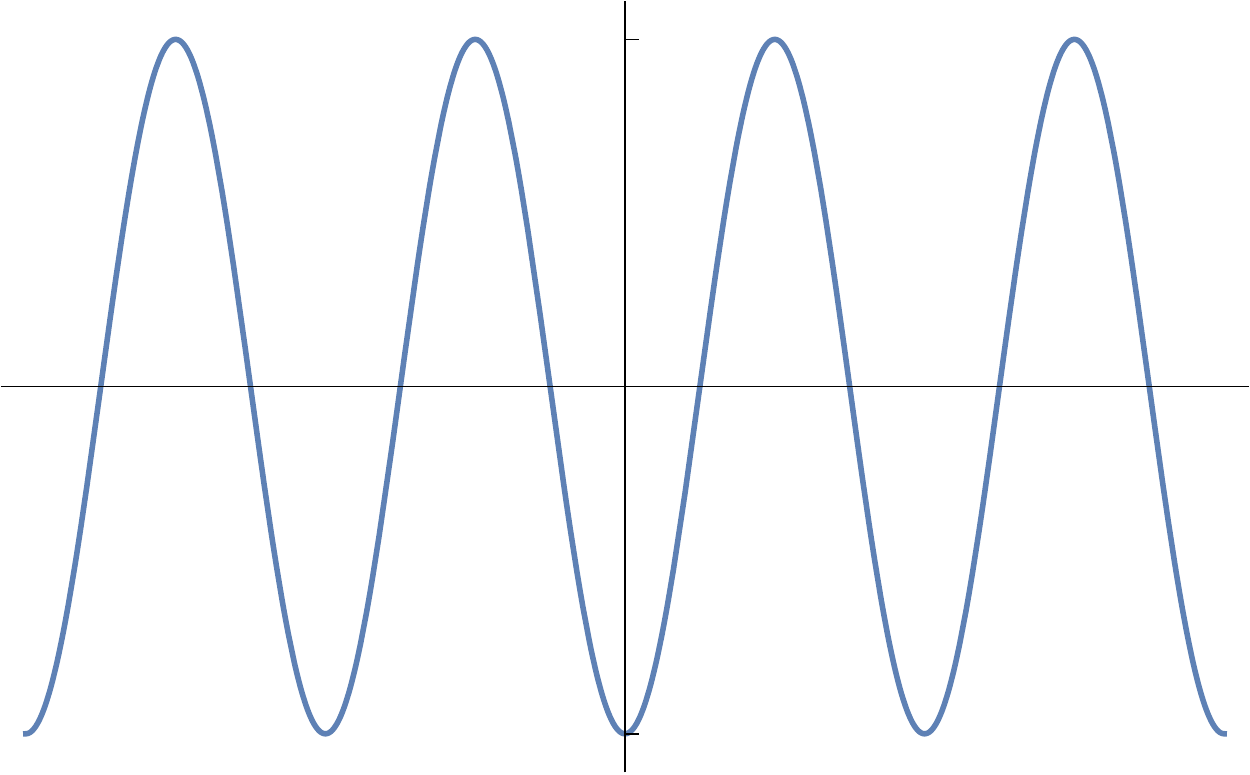}}
\put(19,1){$-\omega^2$}
\put(26,29){$\omega^2$}
\put(50.5,15){$x$}
\put(22.5,32.5){$V \left( x \right)$}
\end{picture}
\end{center}
\vspace{-5pt}
\caption{The simple pendulum potential}
\vspace{5pt}
\label{fig:potential_pendulum}
\end{figure}
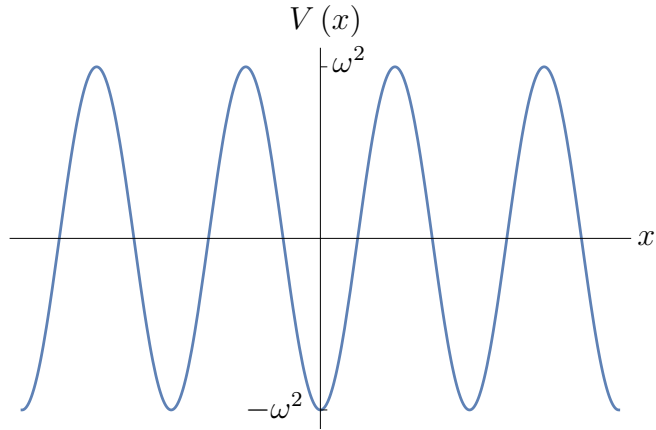

One simple and important problem that can be transformed to a cubic potential problem is the simple pendulum. The equation of motion reads
\begin{equation}
\ddot \theta = - \omega^2 \sin \theta.
\end{equation}
It can be integrated once to take the form of energy conservation,
\begin{equation}
\frac{1}{2} {\dot \theta}^2 + V \left( \theta \right) = E, \quad V \left( \theta \right) = - \omega^2 \cos \theta .
\end{equation}
The potential $V \left( \theta \right)$ is plotted in figure \ref{fig:potential_pendulum}.
The form of the potential indicates that
\begin{itemize}
\item there are no solutions for $E < - \omega^2$,
\item there are oscillating solutions for $\left| E \right| < \omega^2$,
\item there are continuously rotating solutions for $E > \omega^2$.
\end{itemize}

We perform the change of variable
\begin{equation}
- {\omega ^2}\cos \theta  = 2y + \frac{E}{3} .
\label{eq:change_variable_pendulum}
\end{equation}
Then, the conservation of energy takes the form
\begin{equation}
{{\dot y}^2} = 4{y^3} - \left( {\frac{{{E^2}}}{3} + {\omega ^4}} \right)y - \frac{E}{3}\left( {{{\left( {\frac{E}{3}} \right)}^2} - {\omega ^4}} \right) .
\label{eq:Weierstrass_pendulum}
\end{equation}
This is the standard form of Weierstrass differential equation. The solution for $y$ of course should be real, but we should also ensure that
\begin{equation}
\left| 2y + \frac{E}{3} \right| < \omega^2 ,
\end{equation}
so that the change of variable \eqref{eq:change_variable_pendulum} leads to a real $\theta$.

\subsubsection*{Problem Solution and Classification of Solutions}
The roots of the cubic polynomial in the right hand side of equation \eqref{eq:Weierstrass_pendulum} turn out to acquire quite simple expressions,
\begin{equation}
Q \left( y \right) = 4 \left( y - \frac{E}{3} \right) \left( y + \frac{E}{6} - \frac{\omega^2}{2} \right) \left( y + \frac{E}{6} + \frac{\omega^2}{2} \right) .
\end{equation}
The three roots are real for all values of the energy constant $E$. They equal
\begin{equation}
x_1 := \frac{E}{3} , \quad x_2 := - \frac{E}{6} + \frac{\omega^2}{2}, \quad x_3 := - \frac{E}{6} - \frac{\omega^2}{2} .
\label{eq:pendulum_xi}
\end{equation}
We use the notation $x_i$ for the roots as given by equations \eqref{eq:pendulum_xi} to reserve the notation $e_i$ for the roots appropriately ordered.

As there are always three real roots, there are always two independent real solutions of equation \eqref{eq:Weierstrass_pendulum}. They are
\begin{align}
y &= \wp \left( {t - {t_0};{g_2}\left( E \right),{g_3}\left( E \right)} \right) , \label{eq:solution_pendulum_unbounded}\\
y &= \wp \left( {t - {t_0} + {\omega _2};{g_2}\left( E \right),{g_3}\left( E \right)} \right) ,\label{eq:solution_pendulum_bounded}
\end{align}
where
\begin{equation}
{g_2}\left( E \right) = \frac{{{E^2}}}{3} + {\omega ^4},\quad {g_3} = \frac{E}{3}\left( {\left( {{{\frac{E}{3}}}} \right)^2 - {\omega ^4}} \right) .
\end{equation}

\begin{figure}[ht]
\vspace{10pt}
\begin{center}
\begin{picture}(60,35)
\put(0,5){\includegraphics[width = 0.4\textwidth]{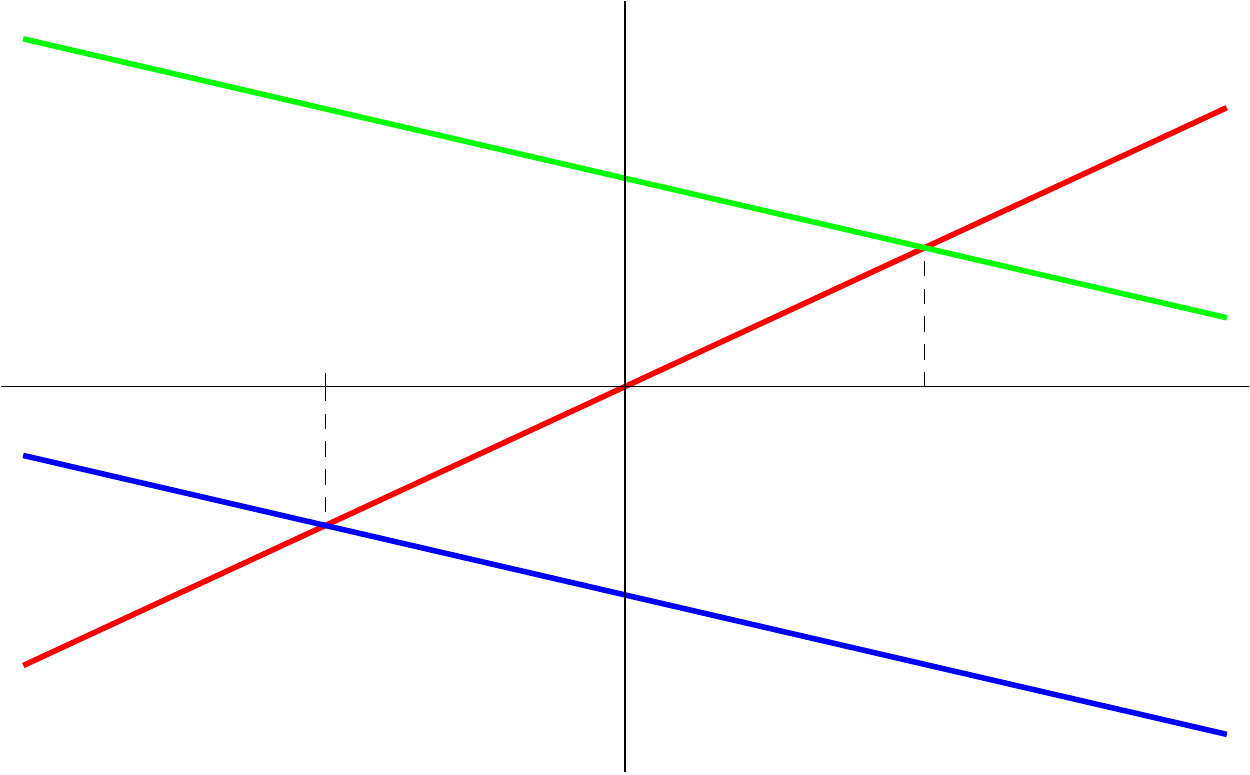}}
\put(44.5,9.5){\includegraphics[height = 0.15\textwidth]{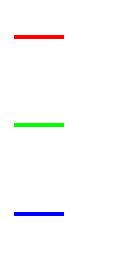}}
\put(7.25,19){$-\omega^2$}
\put(28.25,14.5){$\omega^2$}
\put(49,21.75){$x_1$}
\put(49,16.75){$x_2$}
\put(49,11.75){$x_3$}
\put(19,31){$x_i$}
\put(40,16.75){$E$}
\put(44.5,10.5){\line(0,1){14}}
\put(44.5,10.5){\line(1,0){7.75}}
\put(52.25,10.5){\line(0,1){14}}
\put(44.5,24.5){\line(1,0){7.75}}
\end{picture}
\end{center}
\vspace{-5pt}
\caption{The roots of the cubic polynomial \eqref{eq:Weierstrass_pendulum} as function of the energy constant $E$}
\vspace{5pt}
\label{fig:roots_pendulum}
\end{figure}

The ordering of the three roots $x_i$ depends on the value of the energy constant $E$, as shown in figure \ref{fig:roots_pendulum}.
As the roots $e_i$ have to be appropriately ordered, the identification of $x_i$ with $e_i$ depends on the value of the constant $E$. The appropriate assignments of the three roots are summarized in table \ref{tb:root_assignments_pendulum}.
\begin{table}[ht]
\vspace{10pt}
\begin{center}
\begin{tabular}{ | c || c | }
\hline
 & ordering of roots \\
\hline\hline		
$E < - \omega^2$ & $e_1 = x_2$, $e_2 = x_3$, $e_3 = x_1$ \\
\hline
$\left| E \right| < \omega^2$ & $e_1 = x_2$, $e_2 = x_1$, $e_3 = x_3$ \\
\hline
$E > \omega^2$ & $e_1 = x_1$, $e_2 = x_2$, $e_3 = x_3$ \\
\hline
\end{tabular}
\vspace{3pt}
\caption{The possible orderings of the roots $x_i$}
\label{tb:root_assignments_pendulum}
\end{center}
\end{table}

The unbounded solution \eqref{eq:solution_pendulum_unbounded} ranges from $e_1$ to infinity, whereas the bounded one \eqref{eq:solution_pendulum_bounded} ranges between $e_3$ and $e_2$. Using the appropriate assignment of roots for each energy range, we find that the range of $- \omega^2 \cos \theta = 2 y + E / 3$ is given in table \ref{tb:cos_range}.
\begin{table}[ht]
\vspace{10pt}
\begin{center}
\begin{tabular}{ | c || c | c | }
\hline
 & unbounded range & bounded range \\
\hline\hline		
$E < - \omega^2$ & $\left[ {\omega^2 , + \infty} \right)$ & $\left[ { E , - \omega^2} \right]$ \\
\hline
$\left| E \right| < \omega^2$ & $\left[ {\omega^2 , + \infty} \right)$ & $\left[ {- \omega^2 , E} \right]$ \\
\hline
$E > \omega^2$ & $\left[ {E , + \infty} \right)$ & $\left[ { - \omega^2 , \omega^2 } \right]$ \\
\hline
\end{tabular}
\vspace{3pt}
\caption{The range of $- \omega^2 \cos \theta$}
\label{tb:cos_range}
\end{center}
\end{table}

Table \ref{tb:cos_range} clearly implies that the unbounded solution never corresponds to a real $\theta$. The bounded solution corresponds to a real solution only when $E > - \omega^2$. Thus, as expected by the form of the potential, the pendulum problem has a real solution only when $E > - \omega^2$ and they are given by the bounded solution \eqref{eq:solution_pendulum_bounded}. There is naturally a qualitative change of the form of the solutions at $E = \omega^2$, which is mirrored in the ordering of the roots. The time evolution of $\theta$ is sketched in table \ref{tb:theta_range}.
\begin{table}[ht]
\vspace{10pt}
\begin{center}
\begin{tabular}{ | c || c | c | c || c | c | c | }
\hline
\multirow{2}{*}{} & \multicolumn{3}{|c||}{unbounded} & \multicolumn{3}{|c|}{bounded} \\
\cline{2-7}
{}& $\theta \left( 0 \right)$ & $\theta \left( \omega \right)$ & $\theta \left( 2\omega \right)$ & $\theta \left( 0 \right)$ & $\theta \left( \omega \right)$ & $\theta \left( 2\omega \right)$ \\
\hline\hline
$E < - \omega^2$ & \multicolumn{3}{|c|}{--} & \multicolumn{3}{|c|}{--} \\
\hline
$\left| E \right| < \omega^2$ & \multicolumn{3}{|c|}{--} & $0$ & $\pm \left(\pi - \arccos {\frac{E}{\omega^2}}\right) $ & $0$ \\
\hline
$E > \omega^2$ & \multicolumn{3}{|c|}{--} & $0$ & $\pi$ & $0$ \\
\hline
\end{tabular}
\vspace{3pt}
\caption{The range of the elliptic solutions of the simple pendulum equation}
\label{tb:theta_range}
\end{center}
\end{table}
The period of the oscillatory motions is
\begin{equation}
T_{\rm {oscillating}} = 4 \omega_1.
\end{equation}
while the period (or more literally the quasi-period) of the rotating motions is
\begin{equation}
T_{\rm {rotating}} = 2 \omega_1.
\end{equation}
The difference between the two expressions is due to the change of the topology of the solution. A half-period of the solution corresponds to the transition from the equilibrium position to the maximum displacement position. In the case of oscillatory motion four such segments are required to complete a period, as after two segments the pendulum is back at the equilibrium position but with inverted velocity. On the contrary, in the case of rotating solution, the maximum displacement equals $\pi$, the velocity is never inverted, and, thus, only two half-periods are required.

Since the solution is single valued for $\cos \theta$, in order to find an analytic expression for $\theta$, one has to match appropriate patches, so that the overall solution is everywhere continuous and smooth. It is not difficult to show that selecting initial conditions, so that $\theta \left( 0 \right) = 0$ and $\dot \theta \left( 0 \right) > 0$, the appropriate expression for the angle theta is
\begin{equation}
\theta  = \begin{cases}
{\left( { - 1} \right)^{\left\lfloor {\frac{t}{{2{\omega _1}}}} \right\rfloor }}\arccos \left[ {\frac{1}{{{\omega ^2}}}\left( {2\wp \left( {t + {\omega _2}} \right) + \frac{E}{3}} \right)} \right], & E < \omega^2 , \\
{\left( { - 1} \right)^{\left\lfloor {\frac{t}{{{\omega _1}}}} \right\rfloor }}\arccos \left[ {\frac{1}{{{\omega ^2}}}\left( {2\wp \left( {t + {\omega _2}} \right) + \frac{E}{3}} \right)} \right] + 2\pi \left\lfloor {\frac{t}{{2{\omega _1}}} + \frac{1}{2}} \right\rfloor , & E > \omega^2 .
\end{cases}
\end{equation}
It is left as an exercise for the reader to verify that the above expressions are everywhere continuous and smooth.

\subsection{Point Particle in Hyperbolic Potential}
\label{subsec:hyperbolic}
\subsubsection*{Problems Definition and Equivalence to Weierstrass Equation}

\begin{figure}[ht]
\vspace{10pt}
\begin{center}
\begin{picture}(85,35)
\put(0,0){\includegraphics[width = 0.5\textwidth]{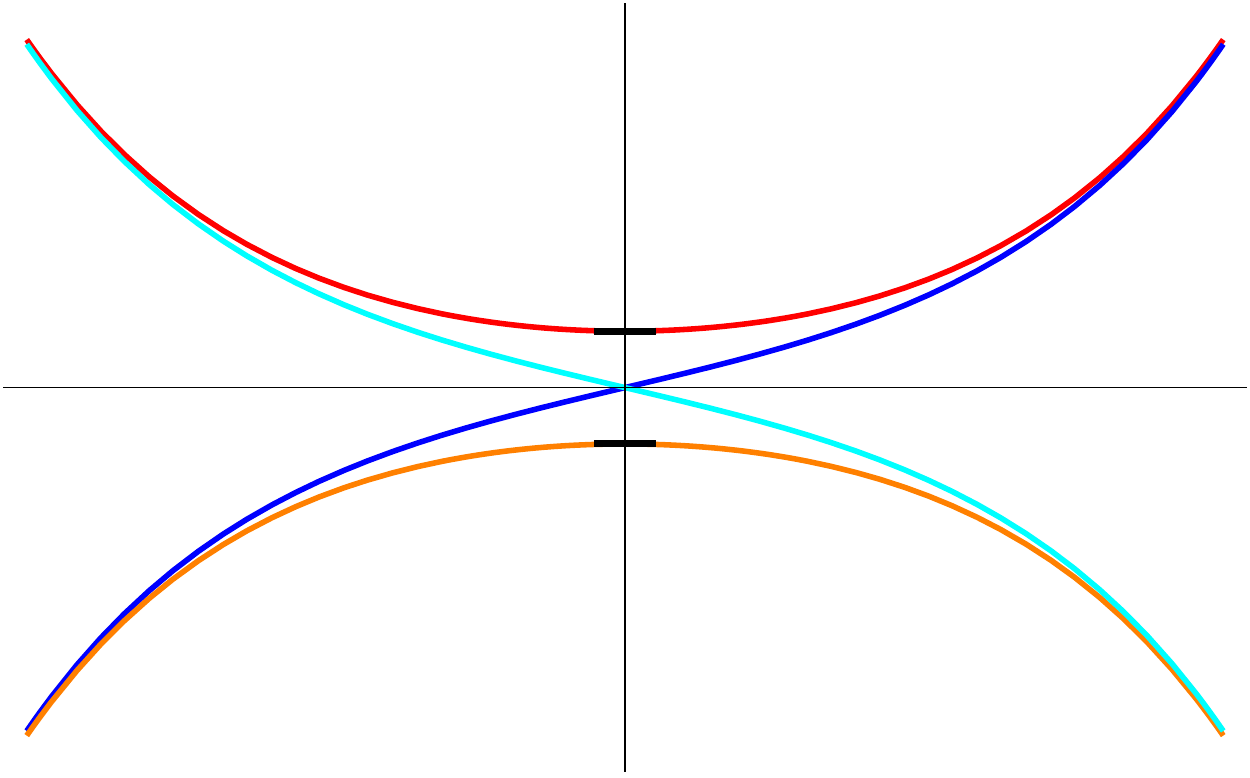}}
\put(54,1){\includegraphics[width = 0.1\textwidth]{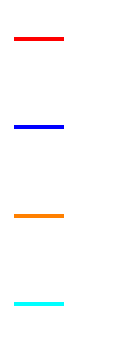}}
\put(19,10){$-m^2$}
\put(26,19){$m^2$}
\put(50.5,15){$x$}
\put(22.5,32.5){$V \left( x \right)$}
\put(60,11.5){$\ddot x = \omega^2 \sinh x$}
\put(60,4.5){$\ddot x = \omega^2 \cosh x$}
\put(60,25.5){$\ddot x = - \omega^2 \sinh x$}
\put(60,18.5){$\ddot x = - \omega^2 \cosh x$}
\put(54.25,3){\line(0,1){25.5}}
\put(54.25,3){\line(1,0){31}}
\put(85.25,3){\line(0,1){25.5}}
\put(54.25,28.5){\line(1,0){31}}
\end{picture}
\end{center}
\vspace{-5pt}
\caption{The potential for the four cases of a hyperbolic force}
\vspace{5pt}
\label{fig:1dpotential}
\end{figure}
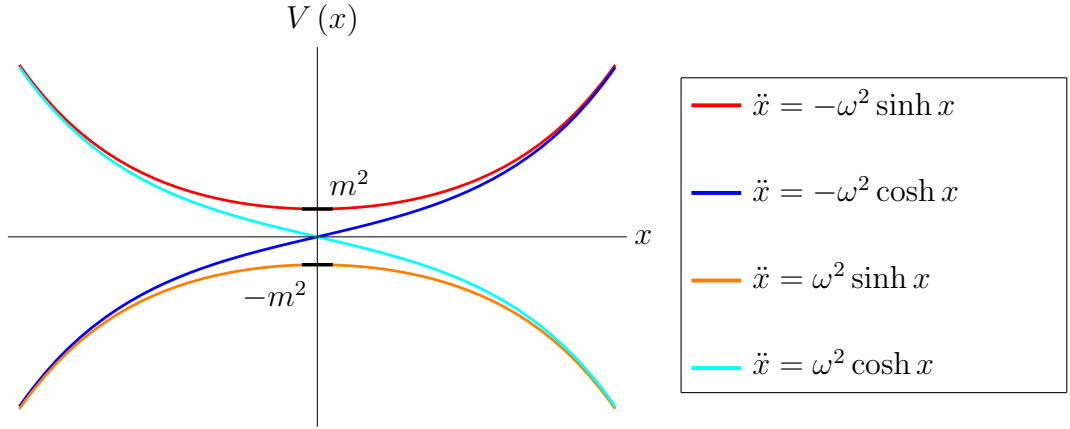

Now let's consider the case of a point particle moving in one dimension under the influence of a hyperbolic force. There are four such possible cases, namely,
\begin{align}
\ddot x &= {\omega^2}\sinh x ,\\
\ddot x &=  - {\omega^2}\sinh x ,\\
\ddot x &= {\omega^2}\cosh x ,\\
\ddot x &=  - {\omega^2}\cosh x .
\end{align}
We will group all those four cases and study them at once writing the equation of motion as
\begin{equation}
\ddot x =  - s{\omega^2}\frac{{{e^x} + t{e^{ - x}}}}{2} .
\label{eq:hyperbolic_eom}
\end{equation}
The parameters $s$ and $t$ take the values $\pm 1$. Appropriate selection of $s$ and $t$ results in any of the four possible hyperbolic forces. Equation \eqref{eq:hyperbolic_eom} can be integrated once to yield
\begin{equation}
\frac{1}{2}{\ddot x}^2 + V \left( x \right) = E , \quad V \left( x \right) = s\frac{{{\omega^2}}}{2}\left( {{e^x } - t{e^{ - x }}} \right).
\label{eq:elliptic_energy_conservation_1}
\end{equation}
The potential energy is plotted in figure \ref{fig:1dpotential} for all the four cases that we are studying.
Considering the form of the potential, we obtain a qualitative picture for the behaviour of the solutions. In the case $s=+1$ and $t=-1$, we expect to find oscillating solutions with energy $E > m^2$ and no solutions for $E < m^2$. In the case $s=-1$ and $t=-1$, we expect to find two different classes of solutions; for $E < - m^2$ we expect to find reflecting scattering solutions since the effective particle does not have enough energy to overcome the potential barrier, whereas for $E > - m^2$, we expect to find transmitting scattering solutions since the particle overcomes the potential barrier. Finally, In the case $t=+1$, we expect to find reflecting scattering solutions for all energies.

Performing the change of variable
\begin{equation}
- s\frac{{{\omega^2}}}{2}{e^x } = 2y - \frac{E}{3} ,
\label{eq:elliptic_y_definition}
\end{equation}
equation \eqref{eq:elliptic_energy_conservation_1} takes the standard Weierstrass form
\begin{equation}
{\dot y}^2 = 4{y^3} - \left( {\frac{1}{3}{E^2} + t\frac{{{\omega^4}}}{4}} \right)y + \frac{E}{3}\left( {\frac{1}{9}{E^2} + t\frac{{{\omega^4}}}{8}} \right) .
\label{eq:elliptic_p_equation}
\end{equation}
The change of variable \eqref{eq:elliptic_y_definition} transforms the problem of the motion of a particle under the influence of a hyperbolic force to yet another one-dimensional problem, describing the motion of a particle with zero energy under the influence of a cubic potential, which has already been studied in section \ref{subsec:cubic}. Real solutions of this equation correspond to real solutions of the initial variable $x$ only when $2y - \frac{E}{3}$ has the same sign as $s$.

\subsubsection*{Four Problems Solved by the Same Expression}
It is interesting to understand how the same equation can be used to describe a variety of solutions that exhibit qualitatively different behaviour, as suggested by the form of the potential in the four cases of hyperbolic forces under study.

Equation \eqref{eq:elliptic_p_equation} is of the standard Weierstrass form \eqref{eq:Weierstrass_equation} with a specific selection for the constants $g_2$ and $g_3$. Equation \eqref{eq:elliptic_p_equation} is solved by
\begin{align}
y &= \wp \left( {t ;{g_2}\left( E, t \right),{g_3}\left( E, t \right)} \right),\\
y &= \wp \left( {t +\omega_2 ;{g_2}\left( E, t \right),{g_3}\left( E, t \right)} \right),
\end{align}
bearing in mind that the second solution is valid only when there are three real roots. The coefficients $g_2$ and $g_3$ are given by
\begin{equation}
{g_2}\left( E, t \right) = \frac{1}{3}{E^2} + t\frac{{{\omega^4}}}{4},\quad {g_3}\left( E, t \right) =  - \frac{E}{3}\left( {\frac{1}{9}{E^2} + t\frac{{{\omega^4}}}{8}} \right) 
\label{eq:elliptic_moduli_energy}
\end{equation}
and the related cubic polynomial is
\begin{equation}
Q \left( x \right) = 4{x^3} - \left( {\frac{1}{3}{E^2} + t\frac{{{\omega^4}}}{4}} \right)x + \frac{E}{3}\left( {\frac{1}{9}{E^2} + t\frac{{{\omega^4}}}{8}} \right) .
\label{eq:elliptic_cubic_equation}
\end{equation}

\begin{figure}[ht]
\vspace{10pt}
\begin{center}
\begin{picture}(100,35)
\put(5,5){\includegraphics[width = 0.4\textwidth]{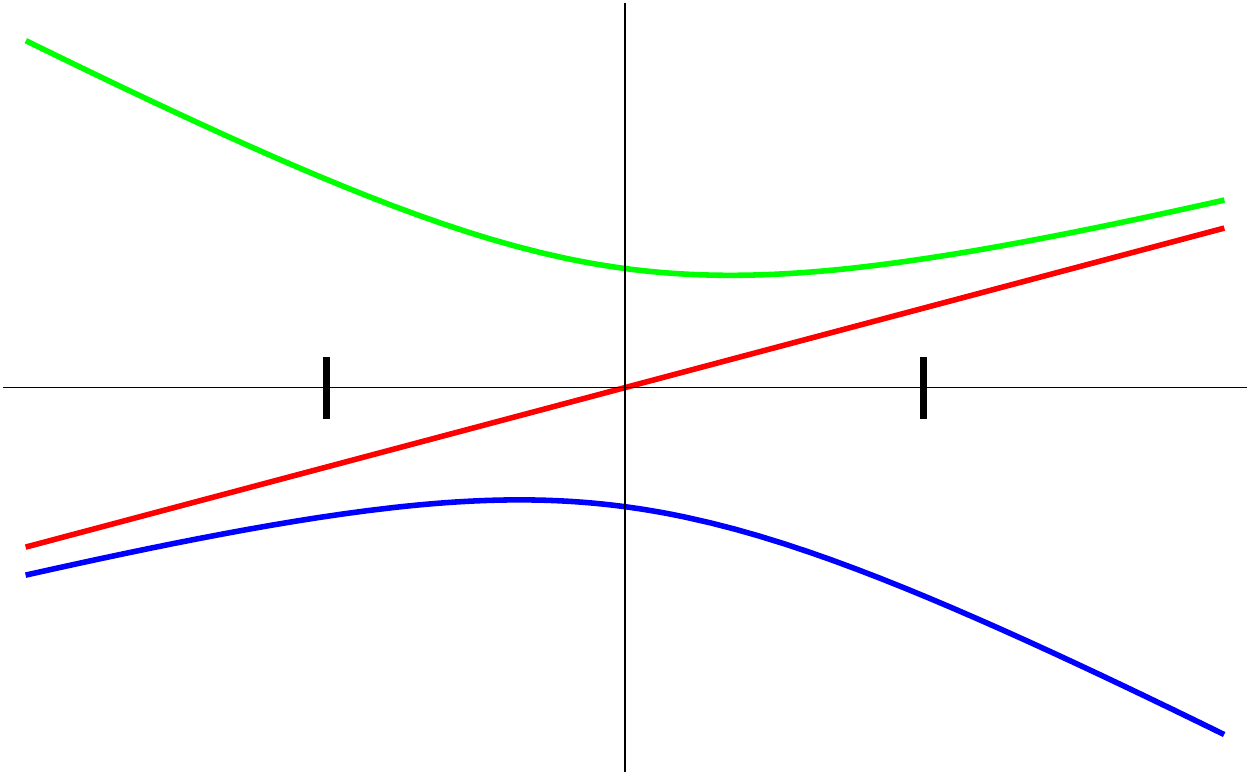}}
\put(55,5){\includegraphics[width = 0.4\textwidth]{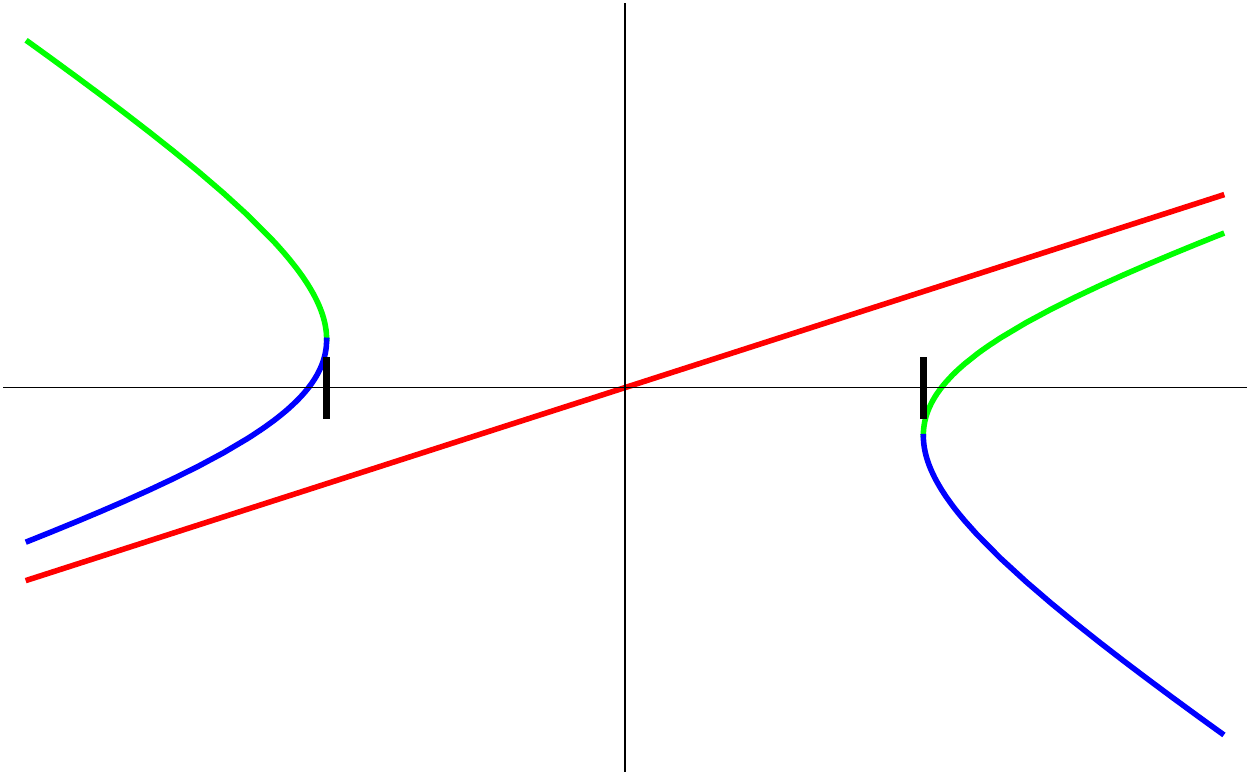}}
\put(46,9.5){\includegraphics[height = 0.15\textwidth]{roots_legend.pdf}}
\put(12.25,19){$-m^2$}
\put(33.55,14.5){$m^2$}
\put(59.5,18){$-m^2$}
\put(85,15){$m^2$}
\put(50.5,21.75){$x_1$}
\put(50.5,16.75){$x_2$}
\put(50.5,11.75){$x_3$}
\put(24,31){$x_i$}
\put(74,31){$x_i$}
\put(43,18){$E$}
\put(93,18){$E$}
\put(20,0){cosh force}
\put(70,0){sinh force}
\put(46,10.5){\line(0,1){14}}
\put(46,10.5){\line(1,0){7.75}}
\put(54,10.5){\line(0,1){14}}
\put(46,24.5){\line(1,0){7.75}}
\end{picture}
\end{center}
\vspace{-5pt}
\caption{The roots of the cubic polynomial \eqref{eq:elliptic_cubic_equation} as function of the energy constant $E$}
\vspace{5pt}
\label{fig:roots}
\end{figure}

The roots of the cubic polynomial are easy to obtain noting that $x = E / 6$ is one of them. Thus, all three roots of $Q \left( x \right)$ are
\begin{equation}
{x_1} = \frac{E}{6},\quad {x_{2,3}} =  - \frac{E}{{12}} \pm \frac{1}{4}\sqrt {{E^2} + t{\omega^4}} .
\label{eq:elliptic_xroots}
\end{equation}
In the following, we use the notation $x_i$ for the roots of $Q \left( x \right)$ as written in equations \eqref{eq:elliptic_xroots} and reserve the notation $e_i$ for the ordered roots of $Q \left( x \right)$ as described in section \ref{subsec:real_Weierstrass}. The roots $x_i$ are plotted as functions of the energy constant $E$ in figure \ref{fig:roots}.

The Weierstrass function allows for a unifying description of the elliptic solutions of both sinh and cosh forces. Different classes of solutions simply correspond to different ordering of the roots $x_i$. Figure \ref{fig:roots} suggests that there are four distinct cases for the ordering of the three roots $x_i$, which are summarized in table \ref{tb:root_assignments}.
\begin{table}[ht]
\vspace{10pt}
\begin{center}
\begin{tabular}{ | c || c | c | }
\hline
 & reality of roots & ordering of roots \\
\hline\hline		
$t = + 1$ & 3 real roots & $e_1 = x_2$, $e_2 = x_1$, $e_3 = x_3$ \\
\hline
$t = - 1$, $E > \omega^2$ & 3 real roots & $e_1 = x_1$, $e_2 = x_2$, $e_3 = x_3$ \\
\hline
$t = - 1$, $\left| E \right| < \omega^2$ & 1 real, 2 complex roots & $e_1 = x_2$, $e_2 = x_1$, $e_3 = x_3$ \\
\hline
$t = - 1$, $E < - \omega^2$ & 3 real roots & $e_1 = x_2$, $e_2 = x_3$, $e_3 = x_1$ \\
\hline
\end{tabular}
\vspace{3pt}
\caption{The possible orderings of the roots $x_i$}
\label{tb:root_assignments}
\end{center}
\end{table}

The unbounded solution ranges from $e_1$ to infinity when there are three real roots and from $e_2$ to infinity when there is only one real root, whereas the bounded solution ranges from $e_3$ to $e_2$. Then, using equation \eqref{eq:elliptic_y_definition}, $- s\frac{{{\omega^2}}}{2}{e^x } = 2 \left( y - x_1 \right)$, we can explore the range of $- s\frac{{{\omega^2}}}{2}{e^x }$ in all cases. The results are summarized in table \ref{tb:v1_range}.
\begin{table}[ht]
\vspace{10pt}
\begin{center}
\begin{tabular}{ | c || c | c | c | }
\hline
 & $-s \frac{\omega^2}{2} e^x$ & unbounded range & bounded range \\
\hline\hline		
$t = + 1$ & $2 \left( y - e_2 \right)$ & $\left[ {2 \left( {e_1 - e_2} \right) , + \infty} \right)$ & $\left[ { - 2 \left( {e_2 - e_3} \right) , 0} \right]$\\
\hline
$t = - 1$, $E > \omega^2$ & $2 \left( y - e_1 \right)$ & $\left[ {0 , + \infty} \right)$ & $\left[ { - 2 \left( {e_1 - e_3} \right) , - 2 \left( {e_1 - e_2} \right)} \right]$ \\
\hline
$t = - 1$, $\left| E \right| < \omega^2$ & $2 \left( y - e_2 \right)$ & $\left[ {0 , + \infty} \right)$ & -- \\
\hline
$t = - 1$, $E < - \omega^2$ & $2 \left( y - e_3 \right)$ & $\left[ {2 \left( {e_1 - e_3} \right) , + \infty} \right)$ & $\left[ { 0 , 2 \left( {e_2 - e_3} \right)} \right]$ \\
\hline
\end{tabular}
\vspace{3pt}
\caption{The solutions for $-s \frac{\omega^2}{2} e^x$ and their range}
\label{tb:v1_range}
\end{center}
\end{table}
In all cases, the sign of $- s\frac{{{\omega^2}}}{2}{e^x }$ does not alternate within its range. Consequently, each solution corresponds to a real solution for exactly one value of the sign $s$.

In table \ref{tb:phi_range}, we sketch the evolution of each solution. In this table $2\Omega$ stands for the real period of $y$, which is equal to $2 \omega_1$ when there are three real roots and $2 \left( \omega_1 + \omega_2 \right)$ when there is only one real root.
\begin{table}[ht]
\vspace{10pt}
\begin{center}
\begin{tabular}{ | c || c | c | c || c | c | c | }
\hline
\multirow{2}{*}{} & \multicolumn{3}{|c||}{unbounded} & \multicolumn{3}{|c|}{bounded} \\
\cline{2-7}
{}& $x \left( 0 \right)$ & $x \left( \Omega \right)$ & $x \left( 2\Omega \right)$ & $x \left( 0 \right)$ & $x \left( \Omega \right)$ & $x \left( 2\Omega \right)$ \\
\hline\hline
\multicolumn{1}{|c||}{} & \multicolumn{6}{|c|}{$s = - 1$}\\
\hline		
$t = + 1$ & $+ \infty$ & $\ln {\frac{4\left( {e_1 - e_2} \right)}{\omega^2}} $ & $+ \infty$ & \multicolumn{3}{|c|}{--}\\
\hline
$t = - 1$, $E > \omega^2$ & $+ \infty$ & $- \infty$ & $+ \infty$ & \multicolumn{3}{|c|}{--} \\
\hline
$t = - 1$, $\left| E \right| < \omega^2$ & $+ \infty$ & $- \infty$ & $+ \infty$ & \multicolumn{3}{|c|}{--} \\
\hline
$t = - 1$, $E < - \omega^2$ & $+ \infty$ & $\ln {\frac{4\left( {e_1 - e_3} \right)}{\omega^2}} $ & $+ \infty$ & $- \infty$ & $\ln {\frac{4\left( {e_2 - e_3} \right)}{m^2}} $ & $- \infty$ \\
\hline\hline
\multicolumn{1}{|c||}{} & \multicolumn{6}{|c|}{$s = + 1$}\\
\hline
$t = + 1$ & \multicolumn{3}{|c||}{--} & $\ln {\frac{4\left( {e_2 - e_3} \right)}{\omega^2}} $ & $- \infty$ & $\ln {\frac{4\left( {e_2 - e_3} \right)}{\omega^2}} $ \\
\hline
$t = - 1$, $E > \omega^2$ & \multicolumn{3}{|c||}{--} & $\ln {\frac{4\left( {e_1 - e_2} \right)}{\omega^2}} $ & $\ln {\frac{4\left( {e_1 - e_3} \right)}{\omega^2}} $ & $\ln {\frac{4\left( {e_1 - e_2} \right)}{\omega^2}} $ \\
\hline
$t = - 1$, $\left| E \right| < \omega^2$ & \multicolumn{3}{|c||}{--} & \multicolumn{3}{|c|}{--} \\
\hline
$t = - 1$, $E < - \omega^2$ & \multicolumn{3}{|c||}{--} & \multicolumn{3}{|c|}{--} \\
\hline
\end{tabular}
\vspace{3pt}
\caption{The extrema of the motion}
\label{tb:phi_range}
\end{center}
\end{table}

All solutions take the following form
\begin{equation}
x \left( {t} \right) = \ln \left[ { - s\frac{2}{{{\omega^2}}}\left( {2\wp \left( {t + z_0 ;{g_2}\left( E \right),{g_3}\left( E \right)} \right) - \frac{E}{3}} \right)} \right] ,\label{eq:elliptic_general_phi1}
\end{equation}
for all choices of the overall sign $s$ and an appropriate choice of the complex integration constant $z_0$. In particular:
\begin{itemize}
\item In the case $s = - 1$ and $t = + 1$, as expected from the form of the potential, only reflecting solutions, coming from and going to the right are found for all energies. In this case there are always three real roots and the solution is given by the unbounded solution. If we select initial conditions, so that the particle is at the minimum position at $t = t_0$, we need to select $z_0 = \omega_1 - t_0$. The ``time of flight'' equals $T = 2 \omega_1$.
\item In the case $s = + 1$ and $t = + 1$, as expected from the form of the potential, only reflecting solutions, coming from and going to the left exist for all energies. In this case there are always three real roots and the solution is given by the bounded solution. Selecting initial conditions, so that the particle is at the maximum position at $t = t_0$, we have to select $z_0 = \omega_2 - t_0$. The ``time of flight'' equals $T = 2 \omega_1$.
\item In the case $s = -1$ and $t = -1$, the form of the potential suggests that there are two possible cases depending on the energy.
\begin{itemize}
\item When $E < - \omega^2$, the particle does not have enough energy to overcome the potential barrier. Therefore, there are reflecting solution coming from and going to either of the two directions. In this case there are three real roots. The particles coming from the right are described from the unbounded solution, while the particles coming from the left are described from the bounded solution. Selecting initial condition, so that the particle  is at the extremal position at $t = t_0$ requires the selection $z_0 = \omega_1 - t_0$ for the particles coming from the right and $z_0 = \omega_1 + \omega_2 - t_0$ for the particles coming from the left. The ``time of flight'' in both cases equals $T = 2 \omega_1$.
\item When $E > - \omega^2$, the particle has enough energy to overcome the potential barrier. Therefore, there are two classes of transmitting solutions coming from either direction. It is interesting though that the reality of the three roots depends on whether the energy is smaller or larger than the critical value $\omega^2$.
\begin{itemize}
\item When $E < \omega^2$, there is only one real root and both left-incoming and right-incoming particles are described by the unbounded solution. The ``time of flight'' in both cases equals $T = \omega_3$. Selecting initial conditions such that the particle at $t = t_0$ lies at the origin, one should select $z_0 = \omega_3 / 2 - t_0$ for a particle coming from the right and $z_0 = 3 \omega_3 / 2 - t_0$ for a particle coming from the left.
\item When $E > \omega^2$, there are three real roots. Apart from this, the situation is similar to the case $E < \omega^2$, with the substitution of $\omega_3$ with $\omega_1$.
\end{itemize}
\end{itemize}
\item Finally, when $s = + 1$ and $t = - 1$, there are oscillatory solutions only when $E > \omega^2$, as expected by the form of the potential. The period of the oscillation equals $T = 2 \omega_1$. Selecting initial conditions so that the particle lies at its minimum position at $t = t_0$ yields $z_0 = - t_0$.
\end{itemize}

\newpage
\subsection*{Problems}
\begin{problem}
Find the energies for which the solution for the cubic potential degenerates to a Weierstrass function with a double root, thus a simply periodic function. Find the special expressions and the periods of the motion at these limits. Verify that in one of these limits, the unique period is equal to the period of small oscillations in the region of the local minimum.
\label{pr:qubic_double_roots}
\end{problem}

\begin{problem}
In section \ref{subsec:cubic}, we showed that while the real period of the Weierstrass elliptic function equals the ``time of flight'' or the period of the motion of a point particle with energy $E$ in a cubic potential, the absolute value of the imaginary period equals the corresponding quantities for the motion of a point particle with energy $-E$ in the inverted potential. However, the inverted potential is identical to the initial one if a coordinate reflection is performed. Thus, the absolute value of the imaginary period should also equal the ``time of flight'' or period of the motion of a point particle with energy $-E$ in the initial potential. Verify this using expressions \eqref{eq:real_period}, \eqref{eq:imaginary_period}, \eqref{eq:complex_period1} and \eqref{eq:complex_period2}
\label{pr:real_imaginary_periods}
\end{problem}

\begin{problem}
In section \ref{subsec:cubic}, we showed that when a bounded oscillatory motion exists in a cubic potential, its period equals the ``time of flight'' of the scattering solution for the same energy. This happens due to a discrete symmetry of the conservation of energy equation that exists when there are three real roots.

More specifically, find how the conservation of energy \eqref{eq:cubic_energy_conservation} is transformed under the change of variables
\begin{equation*}
x \to {e_3} + \frac{{\left( {{e_3} - {e_1}} \right)\left( {{e_3} - {e_2}} \right)}}{{y - {e_3}}} .
\end{equation*}

Then, find where the segments $\left( - \infty , e_3 \right]$, $\left[ e_3 , e_2 \right]$, $\left[ e_2 , e_1 \right]$ and $\left[ e_1 , + \infty \right)$ are mapped through this coordinate transformation.

Finally, show that the above imply that the period of the oscillatory motion and the ``time of flight'' of the scattering solution with the same energy are equal.
\label{pr:discrete_symmetry}
\end{problem}

\begin{problem}
Find the energies for which the solution to the pendulum problem is expressed in terms of the Weierstrass elliptic function with a double root. Find the special expressions for the pendulum motion at these limits and verify that in one of the limits the solution degenerates to the stable equilibrium and simultaneously the period of motion equals the period of the small oscillations, $2 \pi / \omega$.
\label{pr:pendulum_double_roots}
\end{problem}

\begin{problem}
For the motion of a particle in a hyperbolic potential, in the case $s=-1$, $t=-1$ and $E > -\omega^2$, which corresponds to the case of transmitting scattering solutions by a repulsive potential barrier, show that the motion is symmetric around the instant $t=0$, namely show that $x \left( - t \right) = - x \left( t \right)$.

Obviously in the case of reflecting scattering solutions, $E < -\omega^2$, the symmetry of the problem implies that  $x \left( - t \right) = x \left( t \right)$. What has to change in your previous derivation in this case?
\label{pr:hyperbolic_transmission}
\end{problem}

\begin{problem}
For the motion of a particle in a hyperbolic potential, in the case $s=+1$, $t=-1$, which corresponds to oscillatory motion, the bounds of motion in table \ref{tb:phi_range} look asymmetric. However, the fact that the potential is even suggests that they should be symmetric. Verify that they actually are. Furthermore, verify that $x \left( t + T / 2 \right) = - x \left( t \right)$, where $T$ is the period of the oscillation.

Show that at the double root limit of the solution, the solution degenerates to the equilibrium solution and that the period of motion tends to the period of the small oscillations $2 \pi / \omega$.
\label{pr:hyperbolic_oscillations}
\end{problem}

\newpage

\setcounter{equation}{0}
\section{Lecture 4: Applications in Quantum Mechanics}
\label{sec:quantum}

It is well known that particles that move in a periodic potential
\begin{equation}
V \left(x + a \right) = V \left( x \right)
\end{equation}
accept as eigenfunctions Bloch wave solutions of the form
\begin{equation}
\psi \left( x \right) = e^{i k x} u\left( x \right) ,
\label{eq:Bloch_states}
\end{equation}
where $u\left( x \right)$ is periodic with the same period as the potential
\begin{equation}
u \left(x + a \right) = u \left( x \right) .
\end{equation}
The parameter $k$ is a function of the energy. When $k$ is real, these wavefunctions are delta-function normalizable, while when $k$ contains an imaginary part the wavefunctions are exponentially diverging. \emph{This fact leads to the band structure of the periodic potentials.}

Although these are well known facts, it is not simple to find an analytically solvable periodic potential that demonstrates the formation of an non-trivial band structure.

\subsection{The $n = 1$ \Lame Potential}
\label{subsec:lame}

\subsubsection*{The $n = 1$ \Lame Potential and its Solutions}

Let's consider the periodic potential
\begin{equation}
V \left( x \right) = 2 \wp \left( x \right) ,
\end{equation}
where it is assumed that the moduli $g_2$ and $g_3$ are considered to be real. The \Schrodinger equation reads
\begin{equation}
- \frac{{{d^2}y}}{{d{x^2}}} + 2\wp \left( x \right)y = \lambda y .
\label{eq:lame_n1_problem}
\end{equation}

Consider the functions
\begin{equation}
{y_ \pm } \left( {x ; a} \right) = \frac{{\sigma \left( {x \pm a} \right)}}{{\sigma \left( x \right) \sigma \left( \pm a \right)}}{e^{ - \zeta \left( \pm \alpha  \right)x}} .
\label{eq:lame_eigenstates}
\end{equation}
It is easy to verify by direct computation that these functions are both eigenfunctions of the \Schrodinger problem \eqref{eq:lame_n1_problem}. Using the defining property of the Weierstrass $\zeta$ and $\sigma$ functions \eqref{eq:zeta_definition} and \eqref{eq:sigma_definition}, we find
\begin{align*}
\frac{{d{y_ \pm }}}{{dx}} &= \left( {\zeta \left( {x \pm a} \right) - \zeta \left( x \right) \mp \zeta \left( \alpha  \right)} \right){y_ \pm } ,\\
\frac{{{d^2}{y_ \pm }}}{{d{x^2}}} &= \left[ {{{\left( {\zeta \left( {x \pm a} \right) - \zeta \left( x \right) \mp \zeta \left( \alpha  \right)} \right)}^2} - \left( {\wp \left( {x \pm a} \right) - \wp \left( x \right)} \right)} \right]{y_ \pm } .
\end{align*}
Applying the addition theorems of Weierstrass $\wp$ and $\zeta$ functions \eqref{eq:wp_addition} and \eqref{eq:zeta_addition}, the above relations can be simplified to the form
\begin{align}
\frac{{d{y_ \pm }}}{{dx}} &= \frac{1}{2}\frac{{\wp '\left( x \right) \mp \wp '\left( a \right)}}{{\wp \left( x \right) - \wp \left( a \right)}}{y_ \pm } , \label{eq:lame_first_der}\\
\frac{{{d^2}{y_ \pm }}}{{d{x^2}}} &= \left( {2\wp \left( x \right) + \wp \left( a \right)} \right){y_ \pm } . \label{eq:lame_second_der}
\end{align}
The last equation implies that $y_\pm$ are eigenfunctions of problem \eqref{eq:lame_n1_problem}, both corresponding to the eigenvalue
\begin{equation}
\lambda = - \wp \left( a \right) .
\label{eq:lame:eigenvalues}
\end{equation}

As long as the eigenfunction modulus $a$ is not equal to any of the three half-periods, the two $\sigma$ functions appearing to the numerator of $y_\pm$ do not have roots at congruent positions. As such the two wavefunctions are linearly independent and they provide the general solution. When the modulus $a$ equals any of the half-periods though, it turns out that
\begin{equation}
{y_ \pm }\left( {x;{\omega _{1,2,3}}} \right) = \sqrt {\wp \left( x \right) - {e_{1,3,2}}} \, .
\end{equation}
For those eigenvalues, the second linearly independent solution is given by
\begin{equation}
\tilde y\left( {x;{\omega _{1,2,3}}} \right) = \sqrt {\wp \left( x \right) - {e_{1,3,2}}} \left( {\zeta \left( {x + {\omega _{1,2,3}}} \right) + {e_{1,3,2}}x} \right) . 
\label{eq:lame_special}
\end{equation}

\subsubsection*{Reality of the Solutions}

We would like to study whether the eigenfunctions \eqref{eq:lame_eigenstates} are real or not. First, we consider the case of three real roots. In this case, $\wp \left( a \right)$ will assume all real values if $a$ runs in the perimeter of the rectangle with corners located at $0$, $\omega_1$, $\omega_2$ and $\omega_3$. Since the Weierstrass elliptic function is of order two, for every point in the perimeter of this rectangle there is another point in the fundamental period parallelogram, where $\wp$ assumes the same value. Due to $\wp$ being an even function, this point is congruent to the opposite of the initial one. Therefore, the selection of the other point does not correspond to new eigenfunctions, but simply corresponds to the reflection $y_+ \leftrightarrow y_-$. As a result, we divide our analysis to four cases, one for each side of the rectangle with corners at the origin and the half-periods. In the following, $b$ is considered always real.
\begin{enumerate}
\item $a$ lies in the segment $\left[0 , \omega_1 \right]$. Then, $a = b$ and $\wp \left( a \right) > e_1$. In this case, trivially, the eigenfunctions \eqref{eq:lame_eigenstates} are real as,
\begin{equation}
\overline{{{y}_ \pm }\left( {x;b} \right)} = {y_ \pm }\left( {x;\bar b} \right) = {y_ \pm }\left( {x;b} \right) .
\end{equation}
\item $a$ lies in the segment $\left[0 , \omega_2 \right]$. Then, $a = ib$ and $\wp \left( a \right) < e_3$. In this case, trivially, the eigenfunctions \eqref{eq:lame_eigenstates} are complex conjugate to each other as,
\begin{equation}
\overline{{{y}_ \pm }\left( {x;ib} \right)} = {y_ \pm }\left( {x;\overline {ib} } \right) = {y_ \pm }\left( {x; - ib} \right) = {y_ \mp }\left( {x;ib} \right) .
\end{equation}
\item $a$ lies in the segment $\left[\omega_2 , \omega_3 \right]$. Then, $a = \omega_2 + b$ and $e_3 < \wp \left( a \right) < e_2$. In this case,
\begin{equation*}
\overline{{{y}_ \pm }\left( {x;{\omega _2} + b} \right)} = {y_ \pm }\left( {x;\overline {{\omega _2} + b} } \right) = {y_ \pm }\left( {x; - {\omega _2} + b} \right) .
\end{equation*}
We use the quasi-periodicity properties of functions $\zeta$ \eqref{eq:zeta_quasi_periodicity_step} and $\sigma$ \eqref{eq:sigma_quasi_periodicity_step} to find that
\begin{equation*}
\begin{split}
{y_ \pm }\left( {x; - {\omega _2} + b} \right) &= \frac{{\sigma \left( {x \mp {\omega _2} \pm b} \right)}}{{\sigma \left( x \right)\sigma \left( { \mp {\omega _2} \pm b} \right)}}{e^{ \mp \zeta \left( { - {\omega _2} + b} \right)x}}\\
 &= \frac{{ - \sigma \left( {x \pm {\omega _2} \pm b} \right){e^{ \mp 2\zeta \left( {{\omega _2}} \right)\left( {x \pm {\omega _2} \pm b \mp {\omega _2}} \right)}}}}{{ - \sigma \left( x \right)\sigma \left( { \pm {\omega _2} \pm b} \right){e^{ \mp 2\zeta \left( {{\omega _2}} \right)\left( { \pm {\omega _2} \pm b \mp {\omega _2}} \right)}}}}{e^{ \mp \left( {\zeta \left( {{\omega _2} + b} \right) - 2\zeta \left( {{\omega _2}} \right)} \right)x}}\\
 &= \frac{{\sigma \left( {x \pm {\omega _2} \pm b} \right)}}{{\sigma \left( x \right)\sigma \left( { \pm {\omega _2} \pm b} \right)}}{e^{ \mp \zeta \left( {{\omega _2} + b} \right)x}} = {y_ \pm }\left( {x;{\omega _2} + b} \right) ,
\end{split}
\end{equation*}
implying that
\begin{equation}
\overline{{{y}_ \pm }\left( {x;{\omega _2} + b} \right)} = {y_ \pm }\left( {x;{\omega _2} + b} \right) ,
\end{equation}
meaning that in this case the eigenfunctions \eqref{eq:lame_eigenstates} are real.
\item $a$ lies in the segment $\left[\omega_1 , \omega_3 \right]$. Then, $a = \omega_1 + ib$ and $e_2 < \wp \left( a \right) < e_1$. In this case,
\begin{equation*}
\overline{{{y}_ \pm }\left( {x;{\omega _1} + ib} \right)} = {y_ \pm }\left( {x;\overline {{\omega _1} + ib} } \right) = {y_ \pm }\left( {x;{\omega _1} - ib} \right) .
\end{equation*}
As in previous case, we use the quasi-periodicity properties of $\zeta$ and $\sigma$ to find
\begin{equation*}
\begin{split}
{y_ \pm }\left( {x;{\omega _1} - ib} \right) &= \frac{{\sigma \left( {x \pm {\omega _1} \mp ib} \right)}}{{\sigma \left( x \right)\sigma \left( { \pm {\omega _1} \mp ib} \right)}}{e^{ \mp \zeta \left( {{\omega _1} - ib} \right)x}}\\
 &= \frac{{ - \sigma \left( {x \mp {\omega _1} \mp ib} \right){e^{ \pm 2\zeta \left( {{\omega _1}} \right)\left( {x \mp {\omega _1} \mp ib \pm {\omega _1}} \right)}}}}{{ - \sigma \left( x \right)\sigma \left( { \mp {\omega _1} \mp ib} \right){e^{ \pm 2\zeta \left( {{\omega _1}} \right)\left( { \mp {\omega _1} \mp ib \pm {\omega _1}} \right)}}}}{e^{ \mp \left( {\zeta \left( { - {\omega _1} - ib} \right) + 2\zeta \left( {{\omega _1}} \right)} \right)x}}\\
 &= \frac{{\sigma \left( {x \mp {\omega _1} \mp ib} \right)}}{{\sigma \left( x \right)\sigma \left( { \mp {\omega _1} \mp ib} \right)}}{e^{ \pm \zeta \left( {{\omega _1} + ib} \right)x}} = {y_ \mp }\left( {x;{\omega _1} + ib} \right) ,
\end{split}
\end{equation*}
meaning that
\begin{equation}
\overline{{{y}_ \pm }\left( {x;{\omega _1} + ib} \right)} = {y_ \mp }\left( {x;{\omega _1} + ib} \right)
\end{equation}
or in other words, in this case, the eigenfunctions \eqref{eq:lame_eigenstates} are complex conjugate to each other.
\end{enumerate}

In the case of one real root, the situation is much simpler. $\wp \left( a \right)$ will assume all real values if $a$ runs in the union of two segments, one on the real axis with endpoints $0$ and $\omega_1 + \omega_2$ and one in the imaginary axis with endpoints $0$ and $\omega_1 - \omega_2$. Similarly to the case of three real roots, there are more points where $\wp$ assumes real values, but their selection corresponds simply to the reflection $y_+ \leftrightarrow y_-$.
\begin{enumerate}
\item $a$ lies in the segment $\left[0 , \omega_1 + \omega_2 \right]$. Then, $a = b$ and $\wp \left( a \right) > e_2$. This case is identical to the first case above, and, thus, the eigenfunctions \eqref{eq:lame_eigenstates} are real.
\item $a$ lies in the segment $\left[0 , \omega_1 - \omega_2 \right]$. Then, $a = ib$ and $\wp \left( a \right) < e_2$. This case is identical to the second case above, and, thus, the eigenfunctions \eqref{eq:lame_eigenstates} are complex conjugate to each other.
\end{enumerate}

\subsubsection*{The Band Structure of the $n = 1$ \Lame Potential: Three Real Roots}

Comparing to the flat potential, we would expect that when the eigenfunctions \eqref{eq:lame_eigenstates} are complex conjugate to each other, they are delta function normalizable Bloch waves, whereas when the eigenfunctions \eqref{eq:lame_eigenstates} are real, they are exponentially diverging non-normalizable states. However, in order to explicitly show that, we need to find how the eigenfunctions \eqref{eq:lame_eigenstates} transform under a shift of their argument by a period of the potential. For this purpose, we need to write the eigenfunctions \eqref{eq:lame_eigenstates} in the form \eqref{eq:Bloch_states}. Let us consider the case of three real roots. Then, the period of the potential equals $2 \omega_1$ and using the quasi-periodicity property of Weierstrass sigma function,
\begin{equation*}
\frac{{\sigma \left( {x \pm a + 2{\omega _1}} \right)}}{{\sigma \left( {x + 2{\omega _1}} \right)\sigma \left( { \pm a} \right)}} = \frac{{ - {e^{2\zeta \left( {{\omega _1}} \right)\left( {x \pm a + {\omega _1}} \right)}}\sigma \left( {x \pm a} \right)}}{{ - {e^{2\zeta \left( {{\omega _1}} \right)\left( {x + {\omega _1}} \right)}}\sigma \left( x \right)\sigma \left( { \pm a} \right)}} = {e^{ \pm 2a\zeta \left( {{\omega _1}} \right)}}\frac{{\sigma \left( {x \pm a} \right)}}{{\sigma \left( x \right)\sigma \left( { \pm a} \right)}} .
\end{equation*}
Thus, we may write the eigenfunctions \eqref{eq:lame_eigenstates} as
\begin{equation}
{y_ \pm }\left( {x;a} \right) = {u_ \pm }\left( {x;a} \right){e^{ \pm ik\left( a \right)x}} ,
\end{equation}
where
\begin{equation}
{u_ \pm }\left( {x;a} \right) = \frac{{\sigma \left( {x \pm a} \right)}}{{\sigma \left( x \right)\sigma \left( a \right)}}{e^{ \mp \frac{{a\zeta \left( {{\omega _1}} \right)}}{{{\omega _1}}}x}},\quad ik\left( a \right) = \frac{{a\zeta \left( {{\omega _1}} \right) - {\omega _1}\zeta \left( a \right)}}{{{\omega _1}}} 
\end{equation}
and it holds that ${u_ \pm }\left( {x + 2 \omega_1 ; a} \right) = {u_ \pm }\left( {x ; a} \right)$.

In order to discriminate non-normalizable states from Bloch waves, we need to study the function
\begin{equation}
f\left( a \right) = a\zeta \left( {{\omega _1}} \right) - {\omega _1}\zeta \left( a \right) .
\label{eq:lame_phase}
\end{equation}
This is clearly not an elliptic function, but rather a quasi-periodic function, since it obeys,
\begin{align}
f\left( {a + 2{\omega _1}} \right) &= \left( {a + 2{\omega _1}} \right)\zeta \left( {{\omega _1}} \right) - {\omega _1}\left( {\zeta \left( a \right) + 2\zeta \left( {{\omega _1}} \right)} \right) = f\left( a \right) ,\\
f\left( {a + 2{\omega _2}} \right) &= \left( {a + 2{\omega _2}} \right)\zeta \left( {{\omega _1}} \right) - {\omega _1}\left( {\zeta \left( a \right) + 2\zeta \left( {{\omega _2}} \right)} \right) = f\left( a \right) + i\pi .
\end{align}
However, its derivative equals
\begin{equation}
f'\left( a \right) = \zeta \left( {{\omega _1}} \right) + {\omega _1}\wp \left( a \right) ,
\label{eq:lame_phase_derivative}
\end{equation}
which is clearly an order two elliptic function. Therefore, the function $f \left( a \right)$ is stationary exactly twice in each cell. Furthermore, $f'\left( a \right)$ is real, wherever $\wp \left( a \right)$ is real, thus, in the space where the moduli $a$ takes values.

The function $ f \left( a \right)$ takes the following values at the origin and the half-periods,
\begin{align}
\mathop {\lim }\limits_{b \to {0^ + }} f\left( b \right) &=  - \mathop {\lim }\limits_{b \to {0^ + }} \frac{{{\omega _1}}}{b} =  - \infty ,\\
\mathop {\lim }\limits_{b \to {0^ + }} f\left( {ib} \right) &=  - \mathop {\lim }\limits_{b \to {0^ + }} \frac{{{\omega _1}}}{{ib}} =  + i\infty ,\\
f\left( {{\omega _1}} \right) &= {\omega _1}\zeta \left( {{\omega _1}} \right) - {\omega _1}\zeta \left( {{\omega _1}} \right) = 0,\\
f\left( {{\omega _2}} \right) &= {\omega _2}\zeta \left( {{\omega _1}} \right) - {\omega _1}\zeta \left( {{\omega _2}} \right) = i\frac{\pi }{2},\\
f\left( {{\omega _3}} \right) &= \left( {{\omega _1} + {\omega _2}} \right)\zeta \left( {{\omega _1}} \right) - {\omega _1}\zeta \left( {{\omega _1} + {\omega _2}} \right) = i\frac{\pi }{2}.
\end{align}

Since the derivative of $f \left( a \right)$ is real at the perimeter of the rectangle with corners the origin and the three half-periods, it follows that in the sides $\left[0 , \omega_1 \right]$ and $\left[\omega_2 , \omega_3 \right]$ only the real part of $f \left( a \right)$ varies, whereas in the sides $\left[0 , \omega_2 \right]$ and $\left[\omega_1 , \omega_3 \right]$ only the imaginary part of $f \left( a \right)$ varies.

Since the real part of $f \left( a \right)$ is identical at $a = \omega_2$ and $a = \omega_3$ (it vanishes), the mean value theorem implies that there is a point in the segment $\left[\omega_2 , \omega_3 \right]$ where the derivative of $f \left( a \right)$ vanishes. In every cell, there is another
point, where the derivative vanishes, which is congruent to the opposite of the above point, and, thus, it is not congruent to any point of the perimeter of the rectangle with corners at the origin and the half-periods. Since the derivative of $f \left( a \right)$ is an order two elliptic function, there is no other point in a cell, and, thus, in the aforementioned rectangle, where $f \left( a \right)$ is stationary.

The above, combined with the values of $f \left( a \right)$ at the origin and the half-periods, imply that
\begin{enumerate}
\item At the segment $\left[0 , \omega_1 \right]$, $f \left( a \right)$ is everywhere real. It is nowhere stationary in this segment and therefore, it varies monotonically from $- \infty$ at the origin to $0$ at $\omega_1$. It follows that it vanishes nowhere except at $a = \omega_1$.
\item At the segment $\left[\omega_1 , \omega_3 \right]$, $f \left( a \right)$ is everywhere purely imaginary. It is nowhere stationary in this segment and therefore, it varies monotonically from $0$ at $\omega_1$ to $i \pi / 2$ at $\omega_3$.
\item At the segment $\left[\omega_2 , \omega_3 \right]$, $f \left( a \right)$ has an imaginary part equal to $i \pi / 2$. The real part vanishes at the endpoints of the segment, it reaches a minimum value at the stationary point of $f \left( a \right)$ and it vanishes nowhere expect of the endpoints, since the derivative of $f \left( a \right)$ vanishes only once.
\item At the segment $\left[0 , \omega_2 \right]$, $f \left( a \right)$ is everywhere purely imaginary. It is nowhere stationary in this segment and therefore, it varies monotonically from $+ i \infty$ at the origin to $i \pi / 2$ at $\omega_2$.
\end{enumerate}

It follows that $k \left( a \right)$ is purely imaginary, as required for Bloch waves, in the segments $\left[0 , \omega_2 \right]$ and $\left[\omega_1 , \omega_3 \right]$ and nowhere else. Thus, \emph{the band structure of the $n = 1$ \Lame potential, in the case of three real roots, contains a finite ``valence'' band between the energies $-e_1$ and $-e_2$ and an infinite ``conduction'' band for energies larger than $-e_3$.}

\subsubsection*{The Band Structure of the $n = 1$ \Lame Potential: One Real Root}

In the case of one real root, the period of the potential equals $2 \omega_1 + 2 \omega_2$. We write the eigenfunctions \eqref{eq:lame_eigenstates} as
\begin{equation}
{y_ \pm }\left( {x;a} \right) = {u_ \pm }\left( {x;a} \right){e^{ \pm ik\left( a \right)x}} ,
\end{equation}
where
\begin{equation}
{u_ \pm }\left( {x;a} \right) = \frac{{\sigma \left( {x \pm a} \right)}}{{\sigma \left( x \right)\sigma \left( a \right)}}{e^{ \mp \frac{{a\zeta \left( {{\omega _1} + {\omega _2}} \right)}}{{{\omega _1} + {\omega _2}}}x}},\quad ik\left( a \right) = \frac{{a\zeta \left( {{\omega _1} + {\omega _2}} \right) - \left( {{\omega _1} + {\omega _2}} \right)\zeta \left( a \right)}}{{{\omega _1} + {\omega _2}}} 
\end{equation}
and ${u_ \pm }\left( {x + 2 \omega_1 +2 \omega_2 ; a} \right) = {u_ \pm }\left( {x ; a} \right)$.

We now have to consider the function
\begin{equation}
f\left( a \right) = a \zeta \left( {{\omega _1} + {\omega _2}} \right) - \left( {{\omega _1} + {\omega _2}} \right)\zeta \left( a \right) .
\end{equation}
Similarly to the case of three real roots, the function $f\left( a \right)$ is quasi-periodic
\begin{align}
f\left( {a + 2{\omega _1}} \right) = \left( {a + 2{\omega _1}} \right)\zeta \left( {{\omega _1} + {\omega _2}} \right) - \left( {{\omega _1} + {\omega _2}} \right)\left( {\zeta \left( a \right) + 2\zeta \left( {{\omega _1}} \right)} \right) = f\left( a \right) - i\pi ,\\
f\left( {a + 2{\omega _2}} \right) = \left( {a + 2{\omega _2}} \right)\zeta \left( {{\omega _1} + {\omega _2}} \right) - \left( {{\omega _1} + {\omega _2}} \right)\left( {\zeta \left( a \right) + 2\zeta \left( {{\omega _2}} \right)} \right) = f\left( a \right) + i\pi 
\end{align}
and its derivative equals
\begin{equation}
f'\left( a \right) = \zeta \left( {{\omega _1} + {\omega _2}} \right) + \left( {{\omega _1} + {\omega _2}} \right)\wp \left( a \right) ,
\end{equation}
which is an order two elliptic function, implying that $f\left( a \right)$ is stationary exactly once in the union of the segments $\left[ 0 , \omega_1 + \omega_2 \right]$ and $\left[ 0 , \omega_1 - \omega_2 \right]$. Furthermore, the derivative $f'\left( a \right)$ is real wherever $\wp \left( a \right)$ is real and therefore only the real part of $f\left( a \right)$ varies on the segment $\left[ 0 , \omega_1 + \omega_2 \right]$, whereas only the imaginary part of $f\left( a \right)$ varies on the segment $\left[ 0 , \omega_1 - \omega_2 \right]$. We can easily verify that
\begin{align}
\mathop {\lim }\limits_{b \to {0^ + }} f\left( b \right) &=  - \mathop {\lim }\limits_{b \to {0^ + }} \frac{{{\omega _1} + {\omega _2}}}{b} =  - \infty ,\\
\mathop {\lim }\limits_{b \to {0^ + }} f\left( {ib} \right) &=  - \mathop {\lim }\limits_{b \to {0^ + }} \frac{{{\omega _1} + {\omega _2}}}{{ib}} =  + i\infty ,\\
f\left( {{\omega _1} + {\omega _2}} \right) &= \left( {{\omega _1} + {\omega _2}} \right)\zeta \left( {{\omega _1} + {\omega _2}} \right) - \left( {{\omega _1} + {\omega _2}} \right)\zeta \left( {{\omega _1} + {\omega _2}} \right) = 0,\\
f\left( {{\omega _1} - {\omega _2}} \right) &= \left( {{\omega _1} - {\omega _2}} \right)\zeta \left( {{\omega _1} + {\omega _2}} \right) - \left( {{\omega _1} + {\omega _2}} \right)\zeta \left( {{\omega _1} - {\omega _2}} \right) =  - i\pi ,
\end{align}
and, thus, $f\left( a \right)$ is real on $\left[ 0 , \omega_1 + \omega_2 \right]$ and purely imaginary on $\left[ 0 , \omega_1 - \omega_2 \right]$.

The only question to be answered is whether the stationary point of $f\left( a \right)$ belongs in $\left[ 0 , \omega_1 + \omega_2 \right]$ or $\left[ 0 , \omega_1 - \omega_2 \right]$. In the first case, there is another point on $\left[ 0 , \omega_1 + \omega_2 \right]$, where $f\left( a \right)$ vanishes, and, thus, it can be considered purely imaginary, while in the second case there is not. It turns out that the stationary point lies in $\left[ 0 , \omega_1 - \omega_2 \right]$ and therefore $f\left( a \right)$ is purely imaginary everywhere on $\left[ 0 , \omega_1 - \omega_2 \right]$ and nowhere else. Thus, \emph{the band structure of the $n = 1$ \Lame potential, in the case of one real roots, contains only an infinite ``conduction'' band for energies larger than $-e_2$.}

\subsection{The Bounded $n = 1$ \Lame Potential}
\label{subsec:lame_bounded}

\subsubsection*{The Bounded $n = 1$ \Lame Potential and its Band Structure}

If three real roots exist, the whole process of finding the eigenstates and the band structure can be repeated for the potential $V = 2 \wp \left( x + \omega_2 \right)$. Unlike the potential of the previous section, this one is bounded. In a trivial way, the functions
\begin{equation*}
{\psi _ \pm }\left( {x;a} \right) = {y_ \pm }\left( {x + {\omega _2};a} \right) = \frac{{\sigma \left( {x + {\omega _2} \pm a} \right)}}{{\sigma \left( {x + {\omega _2}} \right)\sigma \left( a \right)}}{e^{ \mp \zeta \left( a \right)\left( {x + {\omega _2}} \right)}}
\end{equation*}
obey
\begin{equation}
\frac{{{d^2}{\psi _ \pm }\left( {x;a} \right)}}{{d{x^2}}} = \left( {2\wp \left( {x + {\omega _2}} \right) + \wp \left( a \right)} \right){\psi _ \pm }\left( {x;a} \right) ,
\end{equation}
and, thus, they are eigenfunctions of the bounded $n = 1$ \Lame problem, both with an eigenfunction equal to
\begin{equation}
\lambda = \wp \left( a \right) .
\end{equation}

However, this asymmetric insertion of $\omega_2$ has deprived the eigenfunctions from their nice reality properties. For example, for real $a$,
\begin{equation}
\begin{split}
{{\bar \psi }_ + }\left( {x;a} \right) &= \frac{{\sigma \left( {x - {\omega _2} + a} \right)}}{{\sigma \left( {x - {\omega _2}} \right)\sigma \left( a \right)}}{e^{ - \zeta \left( a \right)\left( {x - {\omega _2}} \right)}}\\
 &= \frac{{ - \sigma \left( {x + {\omega _2} + a} \right){e^{ - 2\zeta \left( {{\omega _2}} \right)\left( {x + {\omega _2} + a - {\omega _2}} \right)}}}}{{ - \sigma \left( {x + {\omega _2}} \right)\sigma \left( a \right){e^{ - 2\zeta \left( {{\omega _2}} \right)\left( {x + {\omega _2} - {\omega _2}} \right)}}}}{e^{ - \zeta \left( a \right)\left( {x + {\omega _2}} \right)}}{e^{2\zeta \left( a \right){\omega _2}}}\\
 &= {e^{2\left( {{\omega _2}\zeta \left( a \right) - a\zeta \left( {{\omega _2}} \right)} \right)}}{\psi _ + }\left( {x;a} \right) ,
\end{split}
\end{equation}
implying that ${\psi _ + }\left( {x;a} \right)$ is not real, but it has to be rotated by a constant complex phase in order to become so. It is left as an exercise to the audience to show that the eigenfunctions
\begin{equation}
{\psi_ \pm }\left( {x;a} \right) = \frac{{\sigma \left( {x + {\omega _2} \pm a} \right)\sigma \left( {{\omega _2}} \right)}}{{\sigma \left( {x + {\omega _2}} \right)\sigma \left( {{\omega _2} \pm a} \right)}}{e^{ - \zeta \left( { \pm a} \right)x}} ,
\label{eq:lame_eigenstates_shifted}
\end{equation}
have reality properties similar to those of the eigenfunctions of the unbounded \Lame potential \eqref{eq:lame_eigenstates}. It is also left to the audience to verify that \emph{the band structure of the bounded potential is identical to the band structure of the unbounded one.}

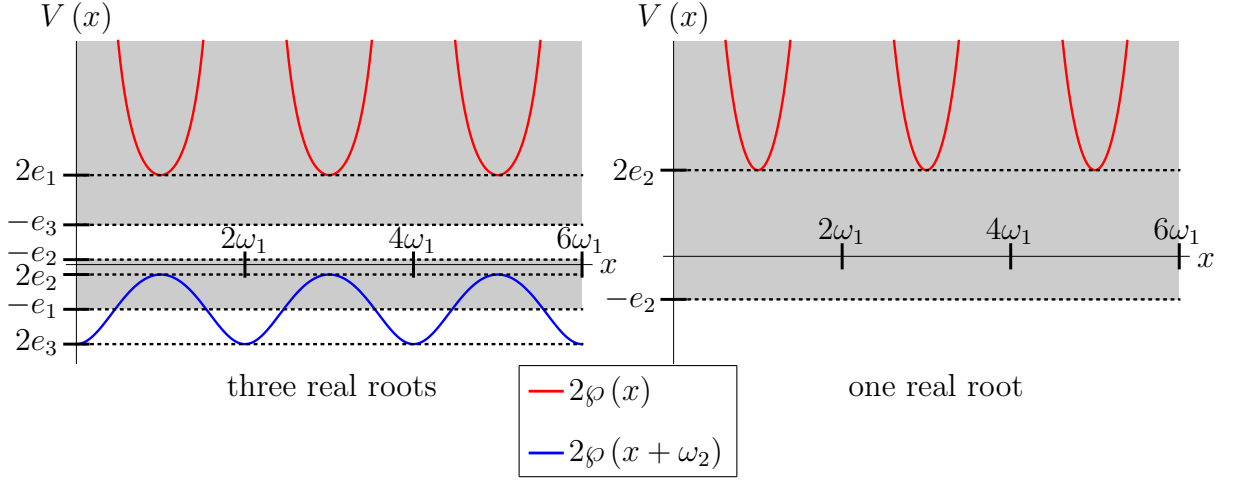
\begin{figure}[ht]
\vspace{10pt}
\begin{center}
\begin{picture}(100,40)
\put(4.5,10){\includegraphics[width = 0.43\textwidth]{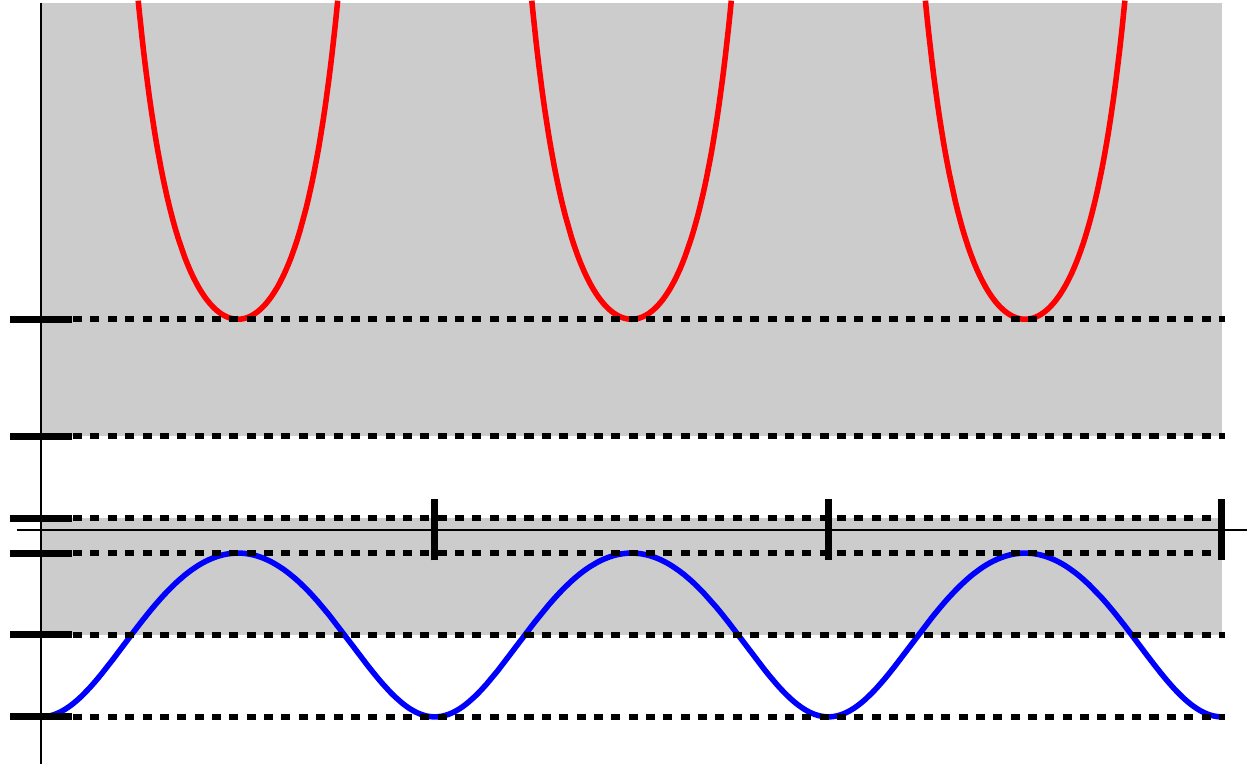}}
\put(52.5,10){\includegraphics[width = 0.43\textwidth]{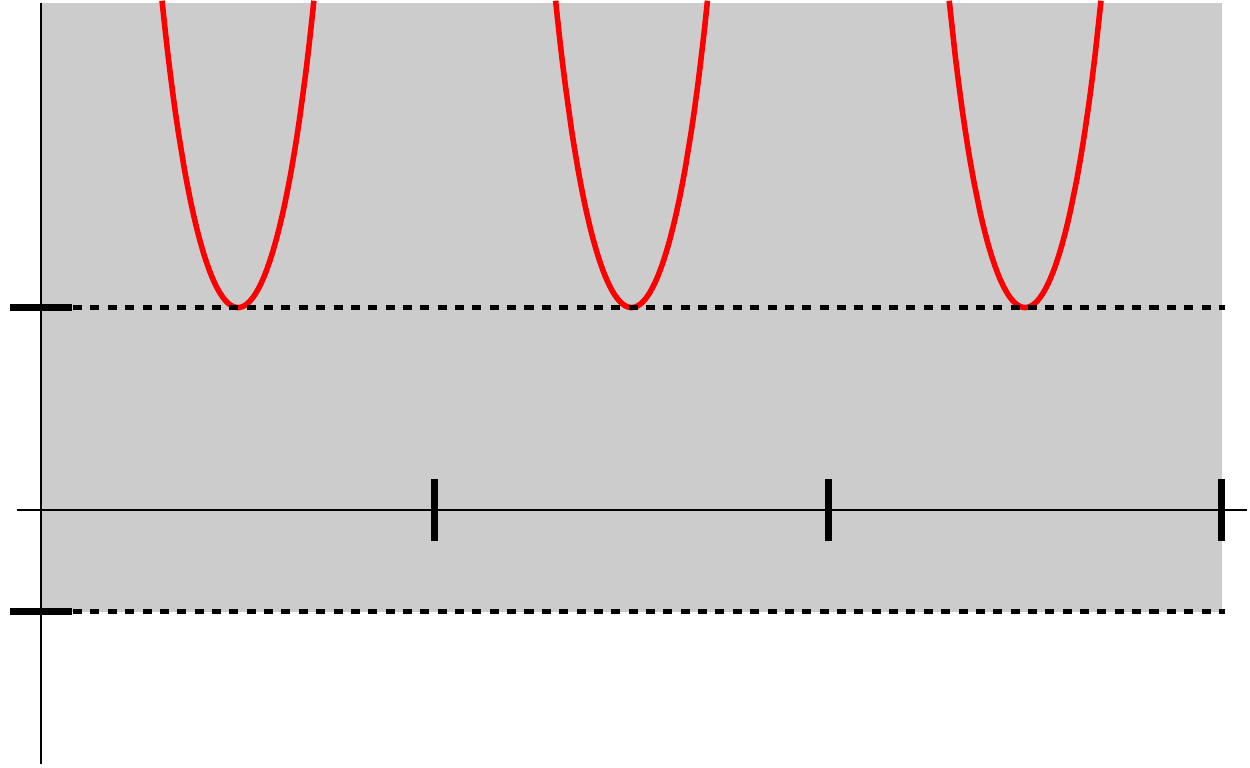}}
\put(41.5,0){\includegraphics[height = 0.1\textwidth]{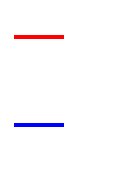}}
\put(1,25){$2e_1$}
\put(1,16.5){$2e_2$}
\put(1,11.25){$2e_3$}
\put(0.25,14.25){$-e_1$}
\put(0.25,18.5){$-e_2$}
\put(0.25,21){$-e_3$}
\put(17.5,19.5){$2\omega_1$}
\put(31,19.5){$4\omega_1$}
\put(44.5,19.5){$6\omega_1$}
\put(49,25){$2e_2$}
\put(48.35,14.75){$-e_2$}
\put(65.5,20.25){$2\omega_1$}
\put(79,20.25){$4\omega_1$}
\put(92.5,20.25){$6\omega_1$}
\put(3,37.5){$V \left( x \right)$}
\put(51,37.5){$V \left( x \right)$}
\put(48,17.5){$x$}
\put(96,18){$x$}
\put(18,7.5){three real roots}
\put(68,7.5){one real root}
\put(45.5,7.25){$2\wp \left( x \right)$}
\put(45.5,2.25){$2\wp \left( x + \omega_2 \right)$}
\put(41.5,1){\line(0,1){9}}
\put(41.5,1){\line(1,0){17.5}}
\put(59,1){\line(0,1){9}}
\put(41.5,10){\line(1,0){17.5}}
\end{picture}
\end{center}
\vspace{-10pt}
\caption{The band structure of the $n=1$ \Lame potential}
\vspace{5pt}
\label{fig:bands}
\end{figure}

It is quite interesting that the potentials $V = 2 \wp \left( x \right)$ and $V = 2 \wp \left( x + \omega_2 \right)$ have the same band structure. The two potentials are quite dissimilar functions, the unbounded one having second order poles, whereas the bounded one being smooth, as shown in figure \ref{fig:bands}.

\subsubsection*{Connection Between the Bounded and Unbounded $n = 1$ \Lame Problems}

The fact that the two potentials are isospectral is not a coincidence. Assume the function
\begin{equation}
W\left( x \right) = \frac{{\wp '\left( x \right)}}{{2\left( {\wp \left( x \right) - {e_3}} \right)}} ,
\label{eq:lame_superpotential}
\end{equation}
which we will call the ``superpotential'' and the operators
\begin{equation}
A = \frac{d}{{dx}} + W\left( x \right),\quad {A^\dag } =  - \frac{d}{{dx}} + W\left( x \right) ,
\end{equation}
which we will call ``annihilation'' and ``creation'' operators respectively. Then,
\begin{align}
{A^\dag }A =  - \frac{{{d^2}}}{{d{x^2}}} + {V_1}\left( x \right),\quad {V_1}\left( x \right) = {W^2}\left( x \right) - W'\left( x \right) ,\\
A{A^\dag } =  - \frac{{{d^2}}}{{d{x^2}}} + {V_2}\left( x \right),\quad {V_2}\left( x \right) = {W^2}\left( x \right) + W'\left( x \right) .
\end{align}

It is a matter of algebra to show that for the specific superpotential given by \eqref{eq:lame_superpotential}, it turns out that
\begin{equation}
{V_1}\left( x \right) = 2\wp \left( {x + {\omega _2}} \right) + {e_3} , \quad {V_2}\left( x \right) = 2\wp \left( x \right) + {e_3} .
\end{equation}
In other words,
\begin{equation}
{A^\dag }A = {{\tilde H}} + {e_3},\quad A{A^\dag } = {H} + {e_3} ,
\end{equation}
where ${H}$ is the Hamiltonian of the unbounded $n = 1$ \Lame problem and ${{\tilde H}_{n = 1}}$ is the Hamiltonian of the bounded $n = 1$ \Lame problem.

Now consider an eigenfunction of the unbounded problem with energy $E$,
\begin{equation}
{H}y = \left( {A{A^\dag } - {e_3}} \right)y = Ey
\end{equation}
Then, the function $\tilde y = {A^\dag }y$ is an eigenfunction of the bounded problem with the same energy, since
\begin{equation}
{{\tilde H}}\tilde y = \left( {{A^\dag }A - {e_3}} \right){A^\dag }y = {A^\dag }\left( {A{A^\dag } - {e_3}} \right)y = {A^\dag }Hy = {A^\dag }Ey = E\tilde y .
\end{equation}

Action with the operator ${A^\dag }$ cannot change a Bloch wave to an exponentially diverging function or the other way around. As a result, the two Hamiltonians are isospectral. It is left as an exercise to the audience to show that indeed
\begin{equation}
{A^\dag } y_\pm \left( x ; a \right) = c \psi_\pm \left( x ; a \right) ,
\end{equation}
where $y_\pm$ are given by equation \eqref{eq:lame_eigenstates}, $\psi_\pm$ are given by equation \eqref{eq:lame_eigenstates_shifted} and $c$ is a constant.

%

\subsection{Epilogue}
\label{subsec:epilogue}

The presentation of the $n = 1$ \Lame problem concludes this set of lectures on applications of Weierstrass elliptic and related functions in Physics. The applications of elliptic functions in not constrained in the content of these lectures.

Many more problems in classical mechanics can be analytically solved in terms of elliptic functions. We may refer to the point particle in a quartic potential, the spherical pendulum and the symmetric top as three such examples. Furthermore, in quantum mechanics, potentials of the form
\begin{equation}
V \left( x \right) = n \left( n + 1 \right) \wp \left( x \right) ,
\end{equation}
where $n \in \mathbb{Z}$ can be analytically solved in terms of elliptic functions and present amazing features. Such potentials have a richer band structure containing up to $n$ finite gaps in their spectrum.

The \Lame equation also appears in other problems of classical Physics, whenever one expresses the Laplace operator in elliptical coordinates. Actually, this is the historical reason for the study of this class of linear differential equations. For example, elliptical functions will appear if one studies the heat diffusion on the surface of a ellipsoid by rotation, such as the earth surface.

Furthermore, elliptic functions find applications in many fields of more modern physics. Several solutions of very interesting integrable systems, such as the sine-Gordon, the sinh-Gordon or the Korteweg-de Vries (KdV) equation can be analytically expressed in terms of elliptic functions. Such solutions can be further used to construct analytic string solutions in symmetric spaces, such as dS and AdS spaces, as well as minimal surfaces in hyperbolic spaces. Elliptic functions also emerge naturally when one calculates string scattering amplitudes for world-sheets that have the topology of a torus.

\subsection*{Acknowledgements}
I would like to thank the organizer of these lectures E. Papantonopoulos and the School of Applied Mathematics and Physical Sciences of the National Technical University of Athens for the hospitality.

The research of G.P. is funded by the ``Post-doc\-toral researchers support'' action of the operational programme ``human resources development, education and long life learning, 2014-2020'', with priority axes 6, 8 and 9, implemented by the Greek State Scholarship Foundation and co-funded by the European Social Fund - ESF and National Resources of Greece.

\newpage
\subsection*{Problems}
\begin{problem}
Show that the special solution \eqref{eq:lame_special} is indeed an eigenfunction of the $n=1$ \Lame problem, corresponding to an eigenvalue equal to one of the three roots of the respective Weierstrass elliptic function.
\label{pr:lame_special}
\end{problem}

\begin{problem}
Show that the eigenfunctions \eqref{eq:lame_eigenstates_shifted} of the bounded $n = 1$ \Lame potential have the same reality properties as the eigenfunctions \eqref{eq:lame_eigenstates} of the unbounded one.
\label{pr:lame_bounded_reality}
\end{problem}

\begin{problem}
Find how the eigenfunctions \eqref{eq:lame_eigenstates_shifted} of the bounded $n = 1$ \Lame potential transform under a shift of their argument by the period of the potential. Once you have accomplish that, deduce the band structure of the bounded $n = 1$ \Lame potential.
\label{pr:lame_bounded_bands}
\end{problem}

\begin{problem}
Using the addition theorem of the Weierstrass $\sigma$ function \eqref{eq:sigma_addition}, show that the eigenstates $y_\pm$, as given by equation \eqref{eq:lame_eigenstates} obey the following ```normalization'' properties:
\begin{align}
{y_ + }{y_ - } &= \wp \left( x \right) - \wp \left( a \right) , \label{eq:lame_eigen_property_1}\\
{y_ + }'{y_ - } - {y_ + }{y_ - }' &= - \wp '\left( a \right) . \label{eq:lame_eigen_property_2}
\end{align}

How are these properties modified for the eigenfunctions of the bounded $n = 1$ \Lame potential?
\label{pr:lame_normalization}
\end{problem}

\begin{problem}
Show that the action of the creation operator $A^\dag$ on the eigenstates of the unbounded $n = 1$ \Lame problem \eqref{eq:lame_eigenstates} results to the eigenstates of the bounded $n = 1$ \Lame problem \eqref{eq:lame_eigenstates_shifted}, namely
\begin{equation}
{A^\dag } y_\pm \left( x ; a \right) = c \psi_\pm \left( x ; a \right) 
\end{equation}
How does the coefficient $c$ compares with the results of problem \ref{pr:lame_normalization}?
\label{pr:lame_creation}
\end{problem}


\newpage

\appendix

\renewcommand{\thesection}{\Alph{section}}
\renewcommand{\thesubsection}{\Alph{section}.\arabic{subsection}}
\renewcommand{\theequation}{\Alph{section}.\arabic{equation}}

\setcounter{equation}{0}
\section{The Half-Periods for Real Moduli}
\label{sec:proof_periods}
It is obvious that formulas \eqref{eq:g2_definition} and \eqref{eq:g3_definition} put certain restrictions on the fundamental periods of Weierstrass elliptic function, if the moduli are real. Unfortunately, the problem of inverting \eqref{eq:g2_definition} and \eqref{eq:g3_definition} is quite difficult.

For this purpose, we will apply the integral formula for Weierstrass elliptic function \eqref{eq:integral_formula_sim}, for the three half periods. In all cases, we select the integration path along the real axes.

Following section \ref{subsec:real_Weierstrass}, when the moduli $g_2$ and $g_3$ are real, the roots of the cubic polynomial are either all real or there is one real and two complex ones, the latter being complex conjugate to each other. In the case there are three real roots, we find
\begin{align}
{\omega _1} \sim \int_{{e_1}}^\infty  {\frac{1}{{\sqrt {4 \left( t - e_1 \right) \left( t - e_2 \right) \left( t - e_3 \right)} }}dt} \equiv x, \label{eq:integral_root1}\\
{\omega _2} \sim \int_{{e_3}}^{ - \infty } {\frac{1}{{\sqrt {4\left( t - e_1 \right) \left( t - e_2 \right) \left( t - e_3 \right)} }}dt} \equiv y, \label{eq:integral_root3}
\end{align}
whereas when there is one real and two complex roots
\begin{align}
{\omega _3} \sim \int_{{e_2}}^\infty  {\frac{1}{{\sqrt {4\left( t - e_2 \right) \left( \left( z - {\mathop{\rm Re}\nolimits} e_1 \right) ^2 + \left( {\mathop{\rm Im}\nolimits} e_1 \right) ^2 \right)} }}dt} \equiv x' , \label{eq:integral_root2a}\\
{\omega _3} \sim \int_{{e_2}}^{ - \infty } {\frac{1}{{\sqrt {4\left( t - e_2 \right) \left( \left( z - {\mathop{\rm Re}\nolimits} e_1 \right) ^2 + \left( {\mathop{\rm Im}\nolimits} e_1 \right) ^2 \right)} }}dt} \equiv y' . \label{eq:integral_root2b}
\end{align}
The integrated function is real everywhere in the range of integration for the first integral in both cases, whereas it is imaginary everywhere in the range of integration for the second integral. Therefore,
\begin{equation}
x,x',iy,iy' \in \mathbb{R} .
\end{equation}
The above imply that when there are three real roots the half period $\omega_1$ is congruent to a real number and the half-period $\omega_2$ is congruent to an purely imaginary number,
\begin{align}
x &= \left( {2{m_1} + 1} \right){\omega _1} + 2{n_1}{\omega _2} ,\\
y &= 2{m_2}{\omega _1} + \left( {2{n_2} + 1} \right){\omega _2} ,
\end{align}
whereas when there is only one real root the half period $\omega_3 = \omega_1 + \omega_2$ is congruent to both a real and a purely imaginary number,
\begin{align}
x' &= \left( {2{k_1} + 1} \right){\omega _1} + \left( {2{l_1} + 1} \right){\omega _2} ,\\
y' &= \left( {2{k_2} + 1} \right){\omega _1} + \left( {2{l_2} + 1} \right){\omega _2} ,
\end{align}
where $m_1 , n_1 , m_2 , n_2 \in \mathbb{Z}$.

The integrated quantity in \eqref{eq:integral_root1} never changes sign in the whole integration range. Consequently, as the integration limit approaches from infinity to $e_1$, the integral monotonically changes from $0$ to its final value. This means that $x$ is not just a point on the real axis that is congruent to the half-period $\omega_1$, but there is no other such position on the real axis between $0$ and $x$. This implies that $2m_1 + 1$ and $2n_1$ are relatively prime. In a similar manner, the above statement holds for $x'$, as well as for $y$ and $y'$ on the imaginary axis, and thus the pairs $\left\{2m_2 , 2n_2 + 1 \right\}$, $\left\{2k_1 + 1 , 2l_1 + 1 \right\}$ and $\left\{2k_2 +1 , 2l_2 + 1 \right\}$ are also pairs of relatively prime numbers,
\begin{align*}
\gcd \left( {2{m_1} + 1,2{n_1}} \right) = \gcd \left( {2{m_2},2{n_2} + 1} \right) &= 1 \\
\gcd \left( {2{k_1} + 1,2{l_1} + 1} \right) = \gcd \left( {2{k_2} + 1,2{l_2} + 1} \right) &= 1 .
\end{align*}

We may redefine the periods $\omega_1$ and $\omega_2$ with a modular transformation, as described in section \ref{subsec:elliptic},
\begin{align*}
{\omega _1} &= a{\omega _1}' + b{\omega _2}' ,\\
{\omega _2} &= c{\omega _1}' + d{\omega _2}' ,
\end{align*}
where $ad - bc = 1$. Let us select $b =  - 2{n_1}$ and $d = 2{m_1} + 1$ in the case of three real roots and $b =  - 2{l_1} - 1$ and $d = 2{k_1} + 1$ in the case of one real root. We find,
\begin{align*}
x &= \left( {ad - bc} \right){\omega _1}' ,\\
y &= \left( {2{m_2}a + \left( {2{n_2} + 1} \right)c} \right){\omega _1}' + \left( {2{m_2}b + \left( {2{n_2} + 1} \right)d} \right){\omega _2}' ,\\
x' &= \left( {ad - bc} \right){\omega _1}' ,\\
y' &= \left( {\left( {2{k_2} + 1} \right)a + \left( {2{l_2} + 1} \right)c} \right){\omega _1}' + \left( {\left( {2{k_2} + 1} \right)b + \left( {2{l_2} + 1} \right)d} \right){\omega _2}' .
\end{align*}
We managed to eliminate ${\omega_2}'$ in the expressions for $x$ and $x'$. However, we should ask whether there are modular transformations for the specific selections of $b$ and $d$ made above. In other words, are there integer solutions for $a$ and $c$ to the equation
\begin{equation}
ad - bc = 1 
\label{eq:diophantine_modular}
\end{equation}
for the specific selections of $b$ and $d$ made above? This equation is a linear Diophantine equation of the form $ax + by = c$ and it is known that such equations have integer solutions, as long as $c$ is a multiple of the greatest common divisor of $a$ and $b$. In both cases $b$ and $d$ are relatively prime, and, thus, their greatest common divisor is equal to one. Therefore, in both cases, equation \eqref{eq:diophantine_modular} does have solutions. Furthermore, The parity of the selected $b$ and $d$ implies that in the case of three real roots, $a$ will be odd, while in the case of one real root $a$ and $c$ have to be of opposite parity. The above imply that
\begin{align}
x &= {\omega _1}' ,\\
y &= {m_2}'{\omega _1}' + \left( {2{n_2}' + 1} \right){\omega _2}' ,\\
x' &= {\omega _1}' ,\\
y' &= \left( {2{k_2}' + 1} \right){\omega _1}' + {2{l_2}'}{\omega _2}' ,
\end{align}
where $m_2 , n_2 , k_2 , l_2 \in \mathbb{Z}$. In both cases, the new lattice has been formed so that the basic period parallelogram has one side parallel to the real axis.

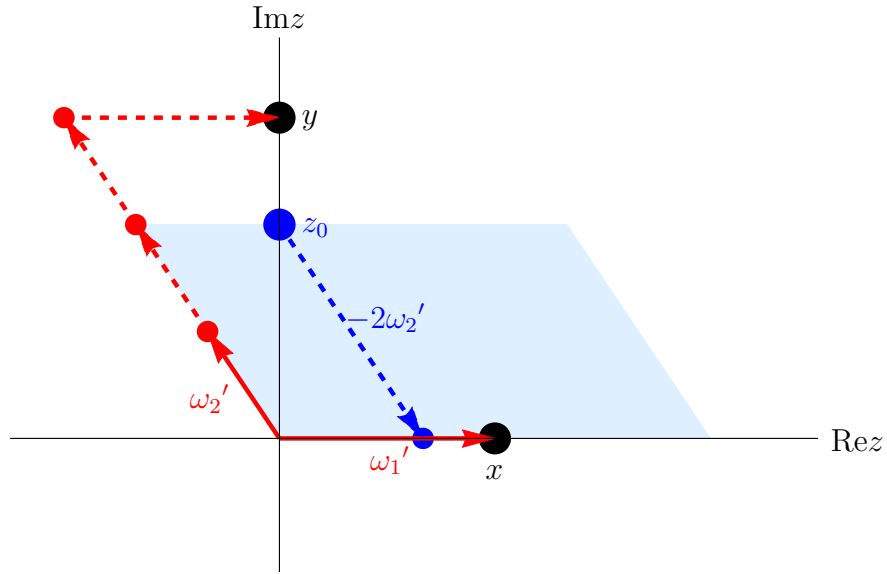
\begin{figure}[ht]
\vspace{10pt}
\begin{center}
\begin{picture}(85,48)
\put(10,2){\includegraphics[width = 0.65\textwidth]{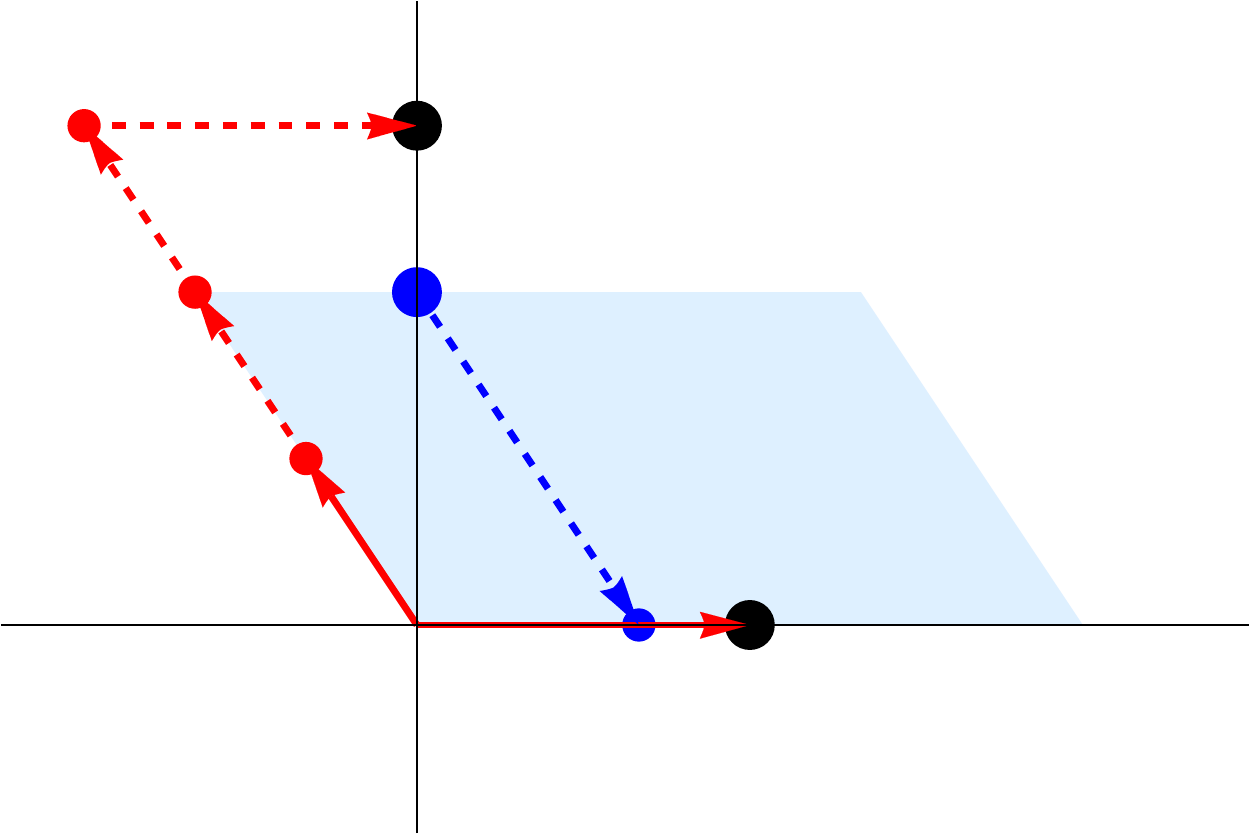}}
\put(76,12){Re$z$}
\put(29.5,46.25){Im$z$}
\put(39,10.5){\color{red}${\omega_1}'$}
\put(24.5,15,5){\color{red}${\omega_2}'$}
\put(48.25,9.75){\color{black}$x$}
\put(33.5,38.5){\color{black}$y$}
\put(33.5,29.75){\color{blue}$z_0$}
\put(37,22){\color{blue}${- 2\omega_2}'$}
\end{picture}
\end{center}
\vspace{-10pt}
\caption{An example of the inconsistency that appears when ${n_2}' \neq 0$. In this example it is assumed that ${m_2}' = {n_2}' = 1$. It is evident that $z_0$ is congruent to a point on the real axis, thus, $\wp \left( z_0 \right) \in \left[ e_1 , + \infty \right)$. At the same time $z_0$ lies in the imaginary axis between 0 and $y$, therefore $\wp \left( z_0 \right) \in \left( - \infty , e_3 \right]$. Obviously, since $e_1 > e_3$, these statements cannot be both true.}
\vspace{5pt}
\label{fig:consistency}
\end{figure}

Let us now focus in the case of three real roots. As we commented above, the Weierstrass function ranges between $+ \infty$ and $e_1$ between the origin and the real half-period $x$. The fact that $\wp$ is even means that it actually ranges between $+ \infty$ and $e_1$ in the whole period between $-x$ and $x$ and consequently in the whole real axis. Periodicity implies that this holds for any shifted axis by any multiple of $2 \omega_2$. If $\left| 2 {n_2}' + 1 \right| > 2$, then the segment of the imaginary axis from $0$ to $y$ crosses such a line, thus there is point on this segment where $\wp$ takes a value larger or at most equal to $e_1$ (see figure \ref{fig:consistency}).
But we have already stated that Weierstrass function is real on the imaginary axis and changes monotonically from $-\infty$ at $z=0$ to $e_3$ at $z = y$. Since $e_3 < e_1$, This is not possible. Therefore the only consistent possibility is ${n_2}' = 0$ (or ${n_2}' = -1$) and
\begin{align*}
x &= {\omega _1}' ,\\
y &= {m_2}'{\omega _1}' + {\omega _2}' .
\end{align*}
A further modular transformation of the form
\begin{align*}
{\omega _1}' &= {\omega _1}'' , \\
{\omega _2}' &= {m_2}'{\omega _1}'' - {\omega _2}'' ,
\end{align*}
results in
\begin{align}
x &= {\omega _1}'' ,\\
y &= {\omega _2}'' ,
\end{align}
meaning that when there are three real roots, the two fundamental half-periods can be selected so that one of them is real and the other purely imaginary. The selection ${n_2}' = -1$ simply results in $y = -{\omega _2}''$. If such a selection is performed, then their values are equal to $x$ and $y$, which are given by the integral formulas \eqref{eq:integral_root1} and \eqref{eq:integral_root3}.

In a similar manner, in the case of one real root, monotonicity of the Weierstrass elliptic function on the imaginary axis between 0 and $y'$ implies that ${l_2 }' = 1$ (or ${l_2 }' = -1$) and
\begin{align*}
x &= {\omega _1}' ,\\
y &= \left( {2{k_2}' + 1} \right) {\omega _1}' + 2 {\omega _2}' .
\end{align*}
In this case, it is easy to show that there is no modular transformation that would preserve the reality of $\omega_1$ simultaneously setting $\omega_2$ to an imaginary value. Such a transformation would necessarily be of the form
\begin{align*}
{\omega _1}' &= {\omega _1}'' , \\
{\omega _2}' &= c {\omega _1}'' + {\omega _2}'' 
\end{align*}
and it would transform the half periods to the form,
\begin{align*}
x &= {\omega _1}'' ,\\
y &= \left( {2{k_2}' + 2 c + 1} \right) {\omega _1}'' + 2 {\omega _2}'' .
\end{align*}
Clearly, the coefficient of ${\omega _1}''$ cannot be set to zero by such a transformation. On the contrary, in this case, one may perform the transformation
\begin{align*}
{\omega _1}' &= {\omega _1}'' + {\omega _2}'' , \\
{\omega _2}' &= - \left( {k_2}' + 1 \right){\omega _1}'' - {k_2}' {\omega _2}'' ,
\end{align*}
to find
\begin{align}
x &= {\omega _1}'' + {\omega _2}'',\\
y &= - {\omega _1}'' + {\omega _2}'',
\end{align}
meaning that $\omega_1$ and $\omega_2$ can be defined so that they are complex conjugate to each other. If such a selection is performed, then the fundamental half-periods are given by $\left( x' \pm y' \right) / 2$, where $x'$ and $y'$ are given by the integral formulas \eqref{eq:integral_root2a} and \eqref{eq:integral_root2b} respectively.

\newpage

\setcounter{equation}{0}
\section{Problem Solutions}
\label{sec:solutions}
\subsubsection*{Problems on Weierstrass Elliptic Function}
\textbf{Problem \ref{pr:reality_omega_1} Solution}

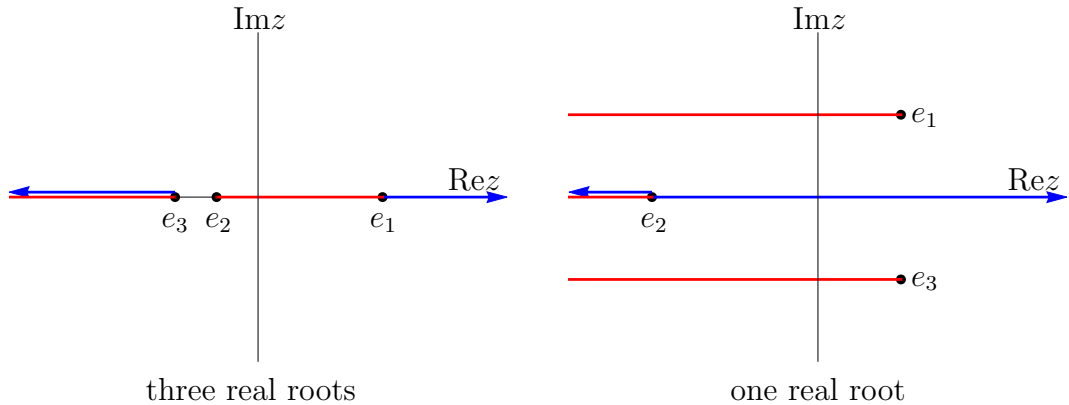
\begin{figure}[ht]
\vspace{10pt}
\begin{center}
\begin{picture}(100,32)
\put(7.5,3){\includegraphics[width = 0.4\textwidth]{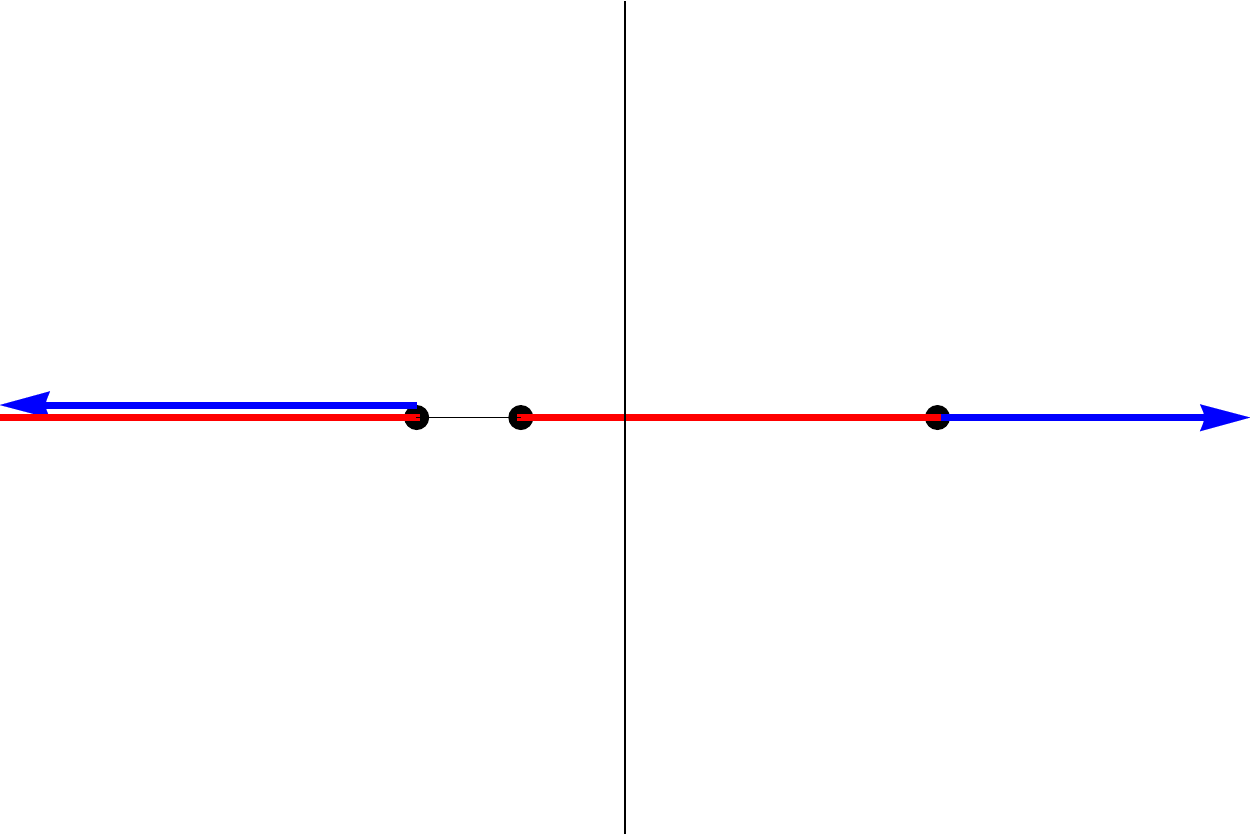}}
\put(52.5,3){\includegraphics[width = 0.4\textwidth]{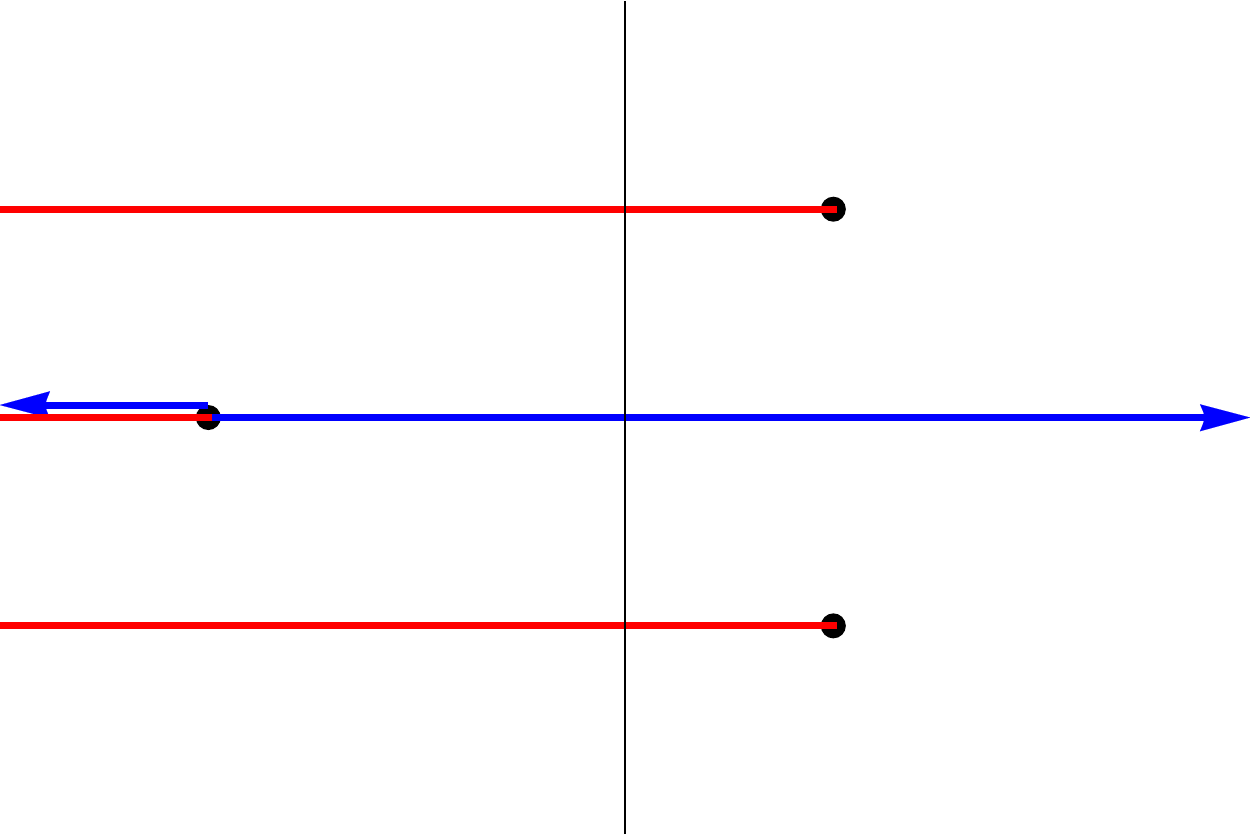}}
\put(42.75,17){Re$z$}
\put(25.5,30){Im$z$}
\put(87.75,17){Re$z$}
\put(70.5,30){Im$z$}
\put(19.75,14){$e_3$}
\put(23.25,14){$e_2$}
\put(36.5,14){$e_1$}
\put(58.25,14){$e_2$}
\put(80,22.5){$e_1$}
\put(80,9.25){$e_3$}
\put(18.5,0){three real roots}
\put(65.5,0){one real root}
\end{picture}
\end{center}
\vspace{-10pt}
\caption{The branch cuts and the integration paths used}
\vspace{5pt}
\label{fig:integration_path}
\end{figure}

We may write the integral formula \eqref{eq:integral_formula_sim} in terms of the roots of the cubic polynomial as,
\begin{equation}
\pm z \sim \int_{\wp \left( z \right)}^\infty  {\frac{1}{{\sqrt {4{t^3} - {g_2}t - {g_3}} }}dt}  = \int_{\wp \left( z \right)}^\infty  {\frac{1}{{\sqrt {4\left( {t - {e_1}} \right)\left( {t - {e_2}} \right)\left( {t - {e_3}} \right)} }}dt} .
\label{eq:integral_formula_roots}
\end{equation}
It is obvious that the integrated function has branch cuts on the complex plane with endpoints the three roots.  We may select the branch cuts as the red lines in figure \ref{fig:integration_path}.

Assume there are three real roots. We may now apply the integral formula \eqref{eq:integral_formula_roots} using the right blue path of figure \ref{fig:integration_path} as the integration path. Let $\omega_1$ be the half-period corresponding to the largest root. Then, we get
\begin{equation}
{\omega _1} \sim \int_{{e_1}}^{ + \infty } {\frac{1}{{\sqrt {4\left( {t - {e_1}} \right)\left( {t - {e_2}} \right)\left( {t - {e_3}} \right)} }}dt} .
\end{equation} 
Clearly, the integrated function is everywhere real (and also positive). It vanishes at infinity as $t^{-3/2}$ and it diverges at the left boundary of the integration as $\left( t - e_1 \right)^{-1/2}$, therefore the integral converges. Thus, we showed that the half-period associated with the largest root is congruent to a real number.

Similarly in the case of the three real roots, we may apply the integral formula \eqref{eq:integral_formula_roots} using the left blue path of figure \ref{fig:integration_path} as the integration path. Let $\omega_2$ be the half-period corresponding to the largest root. Then, we get
\begin{equation}
\begin{split}
{\omega _2} &\sim \int_{{e_3}}^{ - \infty } {\frac{1}{{\sqrt {4\left( {t - {e_1}} \right)\left( {t - {e_2}} \right)\left( {t - {e_3}} \right)} }}dt} \\
 &= i\int_{ - \infty }^{{e_3}} {\frac{1}{{\sqrt {4\left( {{e_1} - t} \right)\left( {{e_2} - t} \right)\left( {{e_3} - t} \right)} }}dt} .
\end{split}
\end{equation}
The integrated function is everywhere real and the integral converges for the same reasons as in previous case. Thus, the half-period associated with the smallest root is congruent to a purely imaginary number.

The situation is similar in the case of one real root. The only difference is that we may use the two different integration paths for the same root. Let $\omega_3$ be the half-period corresponding to the largest root. For the right blue path of figure \ref{fig:integration_path}, we get
\begin{equation}
\begin{split}
{\omega _3} &\sim \int_{{e_2}}^{ + \infty } {\frac{1}{{\sqrt {4\left( {t - {e_1}} \right)\left( {t - {e_2}} \right)\left( {t - {e_3}} \right)} }}dt} \\
 &= \int_{{e_2}}^{ + \infty } {\frac{1}{{\sqrt {4\left[ {{{\left( {t - {\mathop{\rm Re}\nolimits} {e_1}} \right)}^2} + {{\left( {{\mathop{\rm Im}\nolimits} {e_1}} \right)}^2}} \right]\left( {t - {e_2}} \right)} }}dt} ,
\end{split}
\end{equation}
whereas for the left blue path of figure \ref{fig:integration_path} we get
\begin{equation}
\begin{split}
{\omega _3} &\sim \int_{{e_2}}^{ - \infty } {\frac{1}{{\sqrt {4\left[ {{{\left( {t - {\mathop{\rm Re}\nolimits} {e_1}} \right)}^2} + {{\left( {{\mathop{\rm Im}\nolimits} {e_1}} \right)}^2}} \right]\left( {t - {e_2}} \right)} }}dt} \\
 &= i\int_{ - \infty }^{{e_2}} {\frac{1}{{\sqrt {4\left[ {{{\left( {t - {\mathop{\rm Re}\nolimits} {e_1}} \right)}^2} + {{\left( {{\mathop{\rm Im}\nolimits} {e_1}} \right)}^2}} \right]\left( {{e_2} - t} \right)} }}dt} .
\end{split}
\end{equation}
Both integrated functions are real and both integrals converge, implying that the half-period associated with the real root is congruent to both a real and a purely imaginary number.

\textbf{Problem \ref{pr:degenerate cases} Solution}

Suppose that the two larger roots coincide $e_1 = e_2 \equiv e_0$. Then, they are necessarily positive, as the three roots have to sum up to zero, implying that $e_3 = - 2 e_0$. In this case the Weierstrass differential equation reads,
\begin{equation*}
{\left( {\frac{{dy}}{{dz}}} \right)^2} = 4{\left( {y - {e_0}} \right)^2}\left( {y + 2{e_0}} \right) .
\end{equation*}
We perform the change of variable $y = {e_0} + {3{e_0}}/{f^2} $ and then the Weierstrass equation assumes the form,
\begin{equation*}
f{'^2} = 3{e_0}\left( {1 + {f^2}} \right).
\end{equation*}
This one can be easily integrated to yield
\begin{equation*}
f = \pm \sinh \left( {\sqrt {3{e_0}} z + c} \right) .
\end{equation*}
Returning to the initial variable $y$, we find,
\begin{equation*}
y = {e_0} + \frac{{3{e_0}}}{{{{\sinh }^2}\left( {\sqrt {3{e_0}} z + c} \right)}} .
\end{equation*}
We know that by definition, the Weierstrass elliptic function should have a second order pole at $z=0$. This clearly sets the value of $c$ to $i n \pi$, $n \in \mathbb{Z}$ and finally we have
\begin{equation}
\wp \left( z ; 12e_0^2 , - 8 e_0^3 \right) = {e_0} + \frac{{3{e_0}}}{{{{\sinh }^2}\left( {\sqrt {3{e_0}} z} \right)}} .
\end{equation}
In this case clearly the Weierstrass elliptic function has degenerated to a simply periodic function. The only period is purely imaginary and it assumes the value $2\omega_2 = i \pi / \sqrt{3 e_0}$. 

Similarly, if the two smaller roots coincide, we have $e_2 = e_3 = - e_0$ and $e_1 = 2 e_0$, where $e_0 > 0$. The Weierstrass differential equation reads,
\begin{equation*}
{\left( {\frac{{dy}}{{dz}}} \right)^2} = 4{\left( {y + {e_0}} \right)^2}\left( {y - 2{e_0}} \right) .
\end{equation*}
We perform the change of variable, $y =  - {e_0} + {3{e_0}}/{f^2} $ and then the Weierstrass equation takes the form,
\begin{equation*}
f{'^2} = 3{e_0}\left( {1 - {f^2}} \right) .
\end{equation*}
This can be easily integrated to yield
\begin{equation*}
f = \pm \sin \left( {\sqrt {3{e_0}} z + c} \right) .
\end{equation*}
Returning to the initial variable $y$, we find,
\begin{equation*}
y =  - {e_0} + \frac{{3{e_0}}}{{{{\sin }^2}\left( {\sqrt {3{e_0}} z + c} \right)}} .
\end{equation*}
Finally, requesting that the Weierstrass elliptic function has a second order pole at $z = 0$ sets the constant $c$ to $n \pi$, $n \in \mathbb{Z}$ and we get,
\begin{equation}
\wp \left( z ; 12e_0^2 , 8 e_0^3 \right) =  - {e_0} + \frac{{3{e_0}}}{{{{\sin }^2}\left( {\sqrt {3{e_0}} z} \right)}} .
\end{equation}
Once again, in this double root limit, the Weierstrass elliptic function has degenerated to a simply periodic function. Its only period is real and assumes the value $2\omega_1 = \pi / \sqrt{3 e_0}$.

Finally, if all three roots coincide, they must be equal to zero, since their sum should vanish. In this case the Weierstrass differential equation reads,
\begin{equation*}
{\left( {\frac{{dy}}{{dz}}} \right)^2} = 4{y^3} .
\end{equation*}
This can trivially be integrated to acquire
\begin{equation*}
y = \frac{1}{{{{\left( {z + c} \right)}^2}}} .
\end{equation*}
Once again, we demand that the Weierstrass ellitpic function has a double pole at $z = 0$. Therefore $c = 0$ and
\begin{equation}
\wp \left( z ; 0 , 0 \right) = \frac{1}{z^2} .
\end{equation}

To sum up, we found that whenever two roots coincide, the Weierstrass elliptic function degenerates to a function that is not doubly but simply periodic. If all three roots coincide, the Weierstrass elliptic function degenerates to a function that is not periodic at all.
\newline\newline
\textbf{Problem \ref{pr:homogeneity_wp} Solution}

By definition \eqref{eq:Weierstrass_definition} of the Weierstrass elliptic function, we have
\begin{multline*}
\frac{1}{{{\mu ^2}}}\wp \left( {z;{\omega _1},{\omega _2}} \right) \\
= \frac{1}{{{{\left( {\mu z} \right)}^2}}} + \sum\limits_{\left\{ {m,n} \right\} \ne \left\{ {0,0} \right\}} {\left( {\frac{1}{{{{\left( {\mu z + 2m\mu {\omega _1} + 2n\mu {\omega _2}} \right)}^2}}} - \frac{1}{{{{\left( {2m\mu {\omega _1} + 2n\mu {\omega _2}} \right)}^2}}}} \right)} ,
\end{multline*}
or
\begin{equation}
\frac{1}{{{\mu ^2}}}\wp \left( {z;{\omega _1},{\omega _2}} \right) = \wp \left( {\mu z;\mu {\omega _1},\mu {\omega _2}} \right) .
\end{equation}
Then, using the definition of the moduli $g_2$ and $g_3$ \eqref{eq:g2_definition} and \eqref{eq:g3_definition}, we get
\begin{align*}
{g_2}\left( {\mu {\omega _1},\mu {\omega _2}} \right) &= 60\sum\limits_{\left\{ {m,n} \right\} \ne \left\{ {0,0} \right\}} {\frac{1}{{{{\left( {2m\mu {\omega _1} + 2n\mu {\omega _2}} \right)}^4}}}}  = \frac{1}{{{\mu ^4}}}{g_2}\left( {{\omega _1},{\omega _2}} \right) ,\\
{g_3}\left( {\mu {\omega _1},\mu {\omega _2}} \right) &= 140\sum\limits_{\left\{ {m,n} \right\} \ne \left\{ {0,0} \right\}} {\frac{1}{{{{\left( {2m\mu {\omega _1} + 2n\mu {\omega _2}} \right)}^6}}}}  = \frac{1}{{{\mu ^6}}}{g_3}\left( {{\omega _1},{\omega _2}} \right) .
\end{align*}
The latter imply
\begin{equation}
\wp \left( {z;{g_2},{g_3}} \right) = {\mu ^2}\wp \left( {\mu z;\frac{{{g_2}}}{{{\mu ^4}}},\frac{{{g_3}}}{{{\mu ^6}}}} \right) ,
\end{equation}
which is the desired homogeneity property of the Weierstrass elliptic function.
\newpage
\subsubsection*{Problems on Weierstrass Quasi-periodic Functions}
\textbf{Problem \ref{pr:homogeneity_quasi} Solution}

Integrating the homogeneity property of Weierstrass elliptic function \eqref{eq:Weierstras_homogeneity_wp} and using the definition \eqref{eq:zeta_definition} of the function $\zeta$, we get
\begin{equation*}
\zeta \left( {z;{g_2},{g_3}} \right) = \mu \zeta \left( {\mu z;\frac{{{g_2}}}{{{\mu ^4}}},\frac{{{g_3}}}{{{\mu ^6}}}} \right) + c .
\end{equation*}
The defining condition \eqref{eq:zeta_definition_condition}, or in other words the fact that $\zeta$ is an odd function, implies that $c=0$, and, thus,
\begin{equation}
\zeta \left( {z;{g_2},{g_3}} \right) = \mu \zeta \left( {\mu z;\frac{{{g_2}}}{{{\mu ^4}}},\frac{{{g_3}}}{{{\mu ^6}}}} \right) .
\end{equation}

Integrating once more and using the definition \eqref{eq:sigma_definition} we get
\begin{equation*}
\sigma \left( {z;{g_2},{g_3}} \right) = c\sigma \left( {\mu z;\frac{{{g_2}}}{{{\mu ^4}}},\frac{{{g_3}}}{{{\mu ^6}}}} \right) .
\end{equation*}
The defining condition \eqref{eq:sigma_definition_condition} implies that $c = 1 / \mu$. Therefore,
\begin{equation}
\sigma \left( {z;{g_2},{g_3}} \right) = \frac{1}{\mu} \sigma \left( {\mu z;\frac{{{g_2}}}{{{\mu ^4}}},\frac{{{g_3}}}{{{\mu ^6}}}} \right) .
\end{equation}
\newline\newline
\textbf{Problem \ref{pr:parity_quasi_periodic} Solution}

The defining equations \eqref{eq:zeta_definition} and \eqref{eq:zeta_definition_condition} imply that
\begin{equation}
\zeta \left( z \right) = \frac{1}{z} - \int_0^z {\left( {\wp \left( w \right) - \frac{1}{{{w^2}}}} \right)dw} .
\end{equation}
Therefore,
\begin{multline*}
\zeta \left( { - z} \right) =  - \frac{1}{z} - \int_0^{ - z} {\left( {\wp \left( w \right) - \frac{1}{{{w^2}}}} \right)dw}  \\
=  - \frac{1}{z} + \int_0^z {\left( {\wp \left( { - w'} \right) - \frac{1}{{{{\left( { - w'} \right)}^2}}}} \right)dw'}  =  - \frac{1}{z} + \int_0^z {\left( {\wp \left( w' \right) - \frac{1}{{{{w'}^2}}}} \right)dw'} ,
\end{multline*}
where we made the change of the integration variable $w = - w'$ and we used the fact that $\wp$ is an even function. The latter implies that
\begin{equation}
\zeta \left( { - z} \right) = - \zeta \left( z \right) ,
\end{equation}
meaning that $\zeta$ is an odd function.

Similarly, equations \eqref{eq:sigma_definition} and \eqref{eq:sigma_definition_condition} imply
\begin{equation}
\sigma \left( z \right) = z{e^{\int_0^z {\left( {\zeta \left( w \right) - \frac{1}{w}} \right)dw} }} .
\end{equation}
Therefore,
\begin{equation*}
\sigma \left( { - z} \right) =  - z{e^{\int_0^{ - z} {\left( {\zeta \left( w \right) - \frac{1}{w}} \right)dw} }} =  - z{e^{ - \int_0^z {\left( {\zeta \left( { - w'} \right) - \frac{1}{{ - w'}}} \right)dw'} }} =  - z{e^{\int_0^z {\left( {\zeta \left( w' \right) - \frac{1}{w'}} \right)dw'} }} 
\end{equation*}
or else
\begin{equation}
\sigma \left( { - z} \right) = - \sigma \left( z \right) ,
\end{equation}
meaning that $\sigma$ is an odd function.
\newline\newline
\textbf{Problem \ref{pr:quasi_periodic_property} Solution}

First, we will prove the quasi-periodicity property \begin{equation}
\zeta \left( {z + 2m{\omega _1} + 2n{\omega _2}} \right) = \zeta \left( z \right) + 2m\zeta \left( {{\omega _1}} \right) + 2n\zeta \left( {{\omega _2}} \right)
\label{eq:zeta_quasi__gr_0}
\end{equation}
for $n = 0$, namely,
\begin{equation}
\zeta \left( {z + 2m{\omega _1}} \right) = \zeta \left( z \right) + 2m\zeta \left( {{\omega _1}} \right) .
\label{eq:zeta_quasi_n0}
\end{equation}
This obviously holds for $m = 0$. Assuming that it holds for a given value $m$, we will show that it does hold for $m + 1$,
\begin{multline*}
\zeta \left( {z + 2\left( {m + 1} \right){\omega _1}} \right) = \zeta \left( {z + 2m{\omega _1}} \right) + 2\zeta \left( {{\omega _1}} \right)\\
 = \zeta \left( z \right) + 2m\zeta \left( {{\omega _1}} \right) + 2\zeta \left( {{\omega _1}} \right) = \zeta \left( z \right) + 2\left( {m + 1} \right)\zeta \left( {{\omega _1}} \right) .
\end{multline*}
Similarly, one can show that \eqref{eq:zeta_quasi_n0} holds for $m - 1$. Thus, inductively, equation \eqref{eq:zeta_quasi_n0} holds for any $m$.
Therefore, the quasi-periodicity property \eqref{eq:zeta_quasi__gr_0} holds for $n = 0$. Assuming that it holds for a given value $n$, we will show that it holds for $n + 1$,
\begin{multline*}
\zeta \left( {z + 2m{\omega _1} + 2\left( {n + 1} \right){\omega _2}} \right) = \zeta \left( {z + 2m{\omega _1} + 2n{\omega _2}} \right) + 2\zeta \left( {{\omega _2}} \right)\\
 = \zeta \left( z \right) + 2m\zeta \left( {{\omega _1}} \right) + 2n\zeta \left( {{\omega _2}} \right) + 2\zeta \left( {{\omega _2}} \right) = \zeta \left( z \right) + 2m\zeta \left( {{\omega _1}} \right) + 2\left( {n + 1} \right)\zeta \left( {{\omega _2}} \right) .
\end{multline*}
Similarly, one can show that \eqref{eq:zeta_quasi__gr_0}  holds for $n - 1$. Therefore, inductively, we have shown that \eqref{eq:zeta_quasi__gr_0} holds for any $m$ and $n$.

In a similar manner, we proceed  to prove the quasi-periodicity property
\begin{equation}
\sigma \left( {z + 2m{\omega _1} + 2n{\omega _2}} \right) {\left( { - 1} \right)^{m + n + mn}}{e^{\left( {2m\zeta \left( {{\omega _1}} \right) + 2n\zeta \left( {{\omega _2}} \right)} \right)\left( {z + m{\omega _1} + n{\omega _2}} \right)}}\sigma \left( z \right)
\label{eq:sigma_quasi_n_gr_0}
\end{equation}
for $n = 0$, namely,
\begin{equation}
\sigma \left( {z + 2m{\omega _1}} \right) = {\left( { - 1} \right)^m}{e^{2m\zeta \left( {{\omega _1}} \right)\left( {z + m{\omega _1}} \right)}}\sigma \left( z \right) .
\label{eq:sigma_quasi_n0}
\end{equation}
This property obviously holds for $m = 0$. Assuming that it holds for a given value of $m$, we will show that it does hold for $m + 1$,
\begin{multline*}
\sigma \left( {z + 2\left( {m + 1} \right){\omega _1}} \right) =  - {e^{2\zeta \left( {{\omega _1}} \right)\left( {z + 2m{\omega _1} + m{\omega _1}} \right)}}\sigma \left( {z + 2m{\omega _1}} \right)\\
 =  - {e^{2\zeta \left( {{\omega _1}} \right)\left( {z + \left( {2m + 1} \right){\omega _1}} \right)}}{\left( { - 1} \right)^m}{e^{2m\zeta \left( {{\omega _1}} \right)\left( {z + m{\omega _1}} \right)}}\sigma \left( z \right)\\
 = {\left( { - 1} \right)^{m + 1}}{e^{2\zeta \left( {{\omega _1}} \right)\left( {\left( {m + 1} \right)z + \left( {{m^2} + 2m + 1} \right){\omega _1}} \right)}}\sigma \left( z \right)\\
 = {\left( { - 1} \right)^{\left( m + 1 \right)}}{e^{2\left( {m + 1} \right)\zeta \left( {{\omega _1}} \right)\left( {z + \left( {m + 1} \right){\omega _1}} \right)}}\sigma \left( z \right) .
\end{multline*}
Similarly one can show that \eqref{eq:sigma_quasi_n0} holds for $m - 1$. Therefore, inductively, property \eqref{eq:sigma_quasi_n0} holds for any $m$. This implies that the quasi periodicity property \eqref{eq:sigma_quasi_n_gr_0} holds for $n = 0$. Assuming that it does hold for a given value of $n$, we will show that it does hold for $n + 1$,
\begin{multline*}
\sigma \left( {z + 2m{\omega _1} + 2\left( {n + 1} \right){\omega _2}} \right) =  - {e^{2\zeta \left( {{\omega _2}} \right)\left( {z + 2m{\omega _1} + 2n{\omega _2} + {\omega _2}} \right)}}\sigma \left( {z + 2m{\omega _1} + 2n{\omega _2}} \right)\\
 =  - {e^{2\zeta \left( {{\omega _2}} \right)\left( {z + 2m{\omega _1} + 2n{\omega _2} + {\omega _2}} \right)}}{\left( { - 1} \right)^{m + n + mn}}{e^{\left( {2m\zeta \left( {{\omega _1}} \right) + 2n\zeta \left( {{\omega _2}} \right)} \right)\left( {z + m{\omega _1} + n{\omega _2}} \right)}}\sigma \left( z \right)\\
 = {\left( { - 1} \right)^{m + n + 1 + mn}}{e^{2m\zeta \left( {{\omega _1}} \right)\left( {z + m{\omega _1} + n{\omega _2}} \right)}}{e^{2\zeta \left( {{\omega _2}} \right)\left( {\left( {n + 1} \right)z + \left( {mn + 2m} \right){\omega _1} + \left( {{n^2} + 2n + 1} \right){\omega _2}} \right)}}\sigma \left( z \right)\\
 = {\left( { - 1} \right)^{m + n + 1 + mn}}{e^{2m\zeta \left( {{\omega _1}} \right)\left( {z + m{\omega _1} + \left( {n + 1} \right){\omega _2}} \right)}}{e^{2\left( {n + 1} \right)\zeta \left( {{\omega _2}} \right)\left( {z + m{\omega _1} + \left( {n + 1} \right){\omega _2}} \right)}}\\
\qquad\qquad\qquad\qquad\qquad\qquad\qquad\qquad\qquad\qquad\qquad\qquad \times {e^{m\left( {2\zeta \left( {{\omega _2}} \right){\omega _1} - 2\zeta \left( {{\omega _1}} \right){\omega _2}} \right)}}\sigma \left( z \right)\\
 = {\left( { - 1} \right)^{m + n + 1 + mn}}{e^{\left( {2m\zeta \left( {{\omega _1}} \right) + 2\left( {n + 1} \right)\zeta \left( {{\omega _2}} \right)} \right)\left( {z + m{\omega _1} + \left( {n + 1} \right){\omega _2}} \right)}}{e^{ - i\pi m}}\sigma \left( z \right)\\
 = {\left( { - 1} \right)^{m + \left( n + 1 \right) + m\left( {n + 1} \right)}}{e^{\left( {2m\zeta \left( {{\omega _1}} \right) + 2\left( {n + 1} \right)\zeta \left( {{\omega _2}} \right)} \right)\left( {z + m{\omega _1} + \left( {n + 1} \right){\omega _2}} \right)}}\sigma \left( z \right),
\end{multline*}
where in the last step we used property \eqref{eq:z1_z2_relation}. Similarly, one can show that \eqref{eq:sigma_quasi_n_gr_0} holds for $n - 1$, and, thus, for any $m$ and $n$.
\newline\newline
\textbf{Problem \ref{pr:half_period_shift} Solution}

The addition formula \eqref{eq:wp_addition} implies that
\begin{equation*}
\wp \left( {z + {\omega _1}} \right) =  - \wp \left( z \right) - \wp \left( {{\omega _1}} \right) + {\left( {\frac{1}{2}\frac{{\wp '\left( z \right) - \wp '\left( {{\omega _1}} \right)}}{{\wp \left( z \right) - \wp \left( {{\omega _1}} \right)}}} \right)^2} .
\end{equation*}
Since $\wp \left( \omega_1 \right) = e_1$ and $\wp$ is stationary at all half-periods, it follows that
\begin{equation*}
\wp \left( {z + {\omega _1}} \right) =  - \wp \left( z \right) - {e_1} + {\left( {\frac{1}{2}\frac{{\wp '\left( z \right)}}{{\wp \left( z \right) - {e_1}}}} \right)^2} .
\end{equation*}
Using Weierstrass differential equation \eqref{eq:Weierstrass_equation}, we get
\begin{equation*}
\wp \left( {z + {\omega _1}} \right) =  - \wp \left( z \right) - {e_1} + \frac{{\left( {\wp \left( z \right) - {e_2}} \right)\left( {\wp \left( z \right) - {e_3}} \right)}}{{\wp \left( z \right) - {e_1}}} .
\end{equation*}
Finally, using the fact that the three roots sum to zero we find
\begin{equation}
\wp \left( {z + {\omega _1}} \right) = \frac{{{e_1}\wp \left( z \right) + e_1^2 + {e_2}{e_3}}}{{\wp \left( z \right) - {e_1}}} = {e_1} + \frac{{2e_1^2 + {e_2}{e_3}}}{{\wp \left( z \right) - {e_1}}} .
\end{equation}
Similarly, it can be shown that
\begin{equation}
\wp \left( {z + {\omega _2}} \right) = {e_3} + \frac{{2e_3^2 + {e_1}{e_2}}}{{\wp \left( z \right) - {e_3}}},\quad \wp \left( {z + {\omega _3}} \right) = {e_2} + \frac{{2e_2^2 + {e_3}{e_1}}}{{\wp \left( z \right) - {e_2}}} .
\end{equation}
\newline\newline
\textbf{Problem \ref{pr:zeta_addition} Solution}

We will write the elliptic function $\frac{1}{2}\frac{{\wp '\left( z \right) - \wp '\left( w \right)}}{{\wp \left( z \right) - \wp \left( w \right)}}$, being considered a function of $z$, in terms of the Weierstrass function $\zeta$ and its derivatives, taking advantage of the techniques analysed in section \ref{subsec:elliptic_terms_Weierstrass}. In order to do, so we need an irreducible set of poles of this function and the pricipal part of its Laurent series at each pole. We have already shown in section \ref{subsec:elliptic} that given a meromorphic function $f \left( z \right)$, the function $f' / f$ has only first order poles at the positions $f$ has a root or a pole with residues equal to the multiplicity of the root and the opposite of the multiplicity of the pole respectively. We will apply that for the function ${\wp \left( z \right) - \wp \left( w \right)}$. The latter has obviously only a second order pole in a position congruent to $z = 0$ in each cell. The function ${\wp \left( z \right) - \wp \left( w \right)}$ obviously vanishes at $z = w$ and $z = - w$, which in general they are not congruent. Therefore, since it is a second order elliptic function, in each cell it has only two first order roots congruent to $z = w$ and $z = - w$. Consequently, an irreducible set of poles of the function $\frac{{\wp '\left( z \right)}}{{\wp \left( z \right) - \wp \left( w \right)}}$ are $z = 0$, $z = w$ and $z = - w$ and the corresponding Laurent series read,
\begin{align*}
\frac{{\wp '\left( z \right)}}{{\wp \left( z \right) - \wp \left( w \right)}} &=  - \frac{2}{z} + O\left( {{z^0}} \right) ,\\
\frac{{\wp '\left( z \right)}}{{\wp \left( z \right) - \wp \left( w \right)}} &= \frac{1}{{z - w}} + O\left( {{z^0}} \right) ,\\
\frac{{\wp '\left( z \right)}}{{\wp \left( z \right) - \wp \left( w \right)}} &= \frac{1}{{z + w}} + O\left( {{z^0}} \right) .
\end{align*}
Similarly, the function $ - \frac{{\wp '\left( w \right)}}{{\wp \left( z \right) - \wp \left( w \right)}}$ has poles only at the locations where the function ${\wp \left( z \right) - \wp \left( w \right)}$ has roots. We have already shown that an irreducible set of roots of the latter is $z = w$ and $z = - w$. The corresponding Laurent series read,
\begin{align*}
 - \frac{{\wp '\left( w \right)}}{{\wp \left( z \right) - \wp \left( w \right)}} &=  - \frac{{\wp '\left( w \right)}}{{\wp '\left( w \right)\left( {z - w} \right)}} + O\left( {{z^0}} \right) =  - \frac{1}{{z - w}} + O\left( {{z^0}} \right) , \\
 - \frac{{\wp '\left( w \right)}}{{\wp \left( z \right) - \wp \left( w \right)}} &=  - \frac{{\wp '\left( w \right)}}{{\wp '\left( { - w} \right)\left( {z + w} \right)}} + O\left( {{z^0}} \right) = \frac{1}{{z + w}} + O\left( {{z^0}} \right) .
\end{align*}
Combining the above information, it turns out that the function $\frac{1}{2}\frac{{\wp '\left( z \right) - \wp '\left( w \right)}}{{\wp \left( z \right) - \wp \left( w \right)}}$ has poles only at $z = 0$ and $z = - w$ and the corresponding Laurent series read,
\begin{align*}
\frac{1}{2}\frac{{\wp '\left( z \right) - \wp '\left( w \right)}}{{\wp \left( z \right) - \wp \left( w \right)}} &=  - \frac{1}{z} + O\left( {{z^0}} \right) , \\
\frac{1}{2}\frac{{\wp '\left( z \right) - \wp '\left( w \right)}}{{\wp \left( z \right) - \wp \left( w \right)}} &= \frac{1}{{z + w}} + O\left( {{z^0}} \right) .
\end{align*}

The above is the necessary information in order to write the elliptic function $\frac{1}{2}\frac{{\wp '\left( z \right) - \wp '\left( w \right)}}{{\wp \left( z \right) - \wp \left( w \right)}}$ in terms of function $\zeta$ and its derivatives. Applying formula \eqref{eq:elliptic_terms_zeta}, we get
\begin{equation*}
\frac{1}{2}\frac{{\wp '\left( z \right) - \wp '\left( w \right)}}{{\wp \left( z \right) - \wp \left( w \right)}} = \zeta \left( {z + w} \right) - \zeta \left( z \right) + c\left( w \right) .
\end{equation*}
Interchanging $z$ and $w$, we get in a similar manner
\begin{equation*}
\frac{1}{2}\frac{{\wp '\left( z \right) - \wp '\left( w \right)}}{{\wp \left( z \right) - \wp \left( w \right)}} = \zeta \left( {z + w} \right) - \zeta \left( w \right) + c\left( z \right) ,
\end{equation*}
which implies that
\begin{equation*}
\frac{1}{2}\frac{{\wp '\left( z \right) - \wp '\left( w \right)}}{{\wp \left( z \right) - \wp \left( w \right)}} = \zeta \left( {z + w} \right) - \zeta \left( z \right) - \zeta \left( w \right) + c .
\end{equation*}
Finally, the fact that $\wp'$ and $\zeta$ are odd functions, whereas $\wp$ is an even function implies that $c = 0$, which results in the desired pseudo-addition formula
\begin{equation}
\frac{1}{2}\frac{{\wp '\left( z \right) - \wp '\left( w \right)}}{{\wp \left( z \right) - \wp \left( w \right)}} = \zeta \left( {z + w} \right) - \zeta \left( z \right) - \zeta \left( w \right) .
\end{equation}
\newline\newline
\textbf{Problem \ref{pr:wp_addition} Solution}

We differentiate the pseudo-addition theorem for Weierstrass $\zeta$ function \eqref{eq:zeta_addition} with respect to $z$ and $w$. We get
\begin{align*}
 - \wp \left( {z + w} \right) + \wp \left( z \right) &= \frac{1}{2}\frac{{\wp ''\left( z \right)\left( {\wp \left( z \right) - \wp \left( w \right)} \right) - \wp '\left( z \right)\left( {\wp '\left( z \right) - \wp '\left( w \right)} \right)}}{{{{\left( {\wp \left( z \right) - \wp \left( w \right)} \right)}^2}}} , \\
 - \wp \left( {z + w} \right) + \wp \left( w \right) &=  - \frac{1}{2}\frac{{\wp ''\left( w \right)\left( {\wp \left( z \right) - \wp \left( w \right)} \right) - \wp '\left( w \right)\left( {\wp '\left( z \right) - \wp '\left( w \right)} \right)}}{{{{\left( {\wp \left( z \right) - \wp \left( w \right)} \right)}^2}}} .
\end{align*}
Adding the above two equations yields
\begin{equation*}
\begin{split}
 - 2\wp \left( {z + w} \right) + \wp \left( z \right) + \wp \left( w \right) &= \frac{1}{2}\frac{{\left( {\wp ''\left( z \right) - \wp ''\left( w \right)} \right)\left( {\wp \left( z \right) - \wp \left( w \right)} \right) - {{\left( {\wp '\left( z \right) - \wp '\left( w \right)} \right)}^2}}}{{{{\left( {\wp \left( z \right) - \wp \left( w \right)} \right)}^2}}}\\
 &= \frac{1}{2}\frac{{\wp ''\left( z \right) - \wp ''\left( w \right)}}{{\wp \left( z \right) - \wp \left( w \right)}} - \frac{1}{2}{\left( {\frac{{\wp '\left( z \right) - \wp '\left( w \right)}}{{\wp \left( z \right) - \wp \left( w \right)}}} \right)^2} .
\end{split}
\end{equation*}

The second derivative of Weierstrass elliptic function can easily be expressed in terms of the Weierstrass elliptic function itself. Differentiating the Weierstrass differential equation, we get
\begin{equation}
\wp '' = 6{\wp ^2} - \frac{{{g_2}}}{2} ,
\end{equation}
which implies that
\begin{equation*}
\wp ''\left( z \right) - \wp ''\left( w \right) = 6\left( {{\wp ^2}\left( z \right) - {\wp ^2}\left( w \right)} \right) .
\end{equation*}
Therefore, we conclude that
\begin{equation*}
- 2\wp \left( {z + w} \right) + \wp \left( z \right) + \wp \left( w \right) = 3\frac{{{\wp ^2}\left( z \right) - {\wp ^2}\left( w \right)}}{{\wp \left( z \right) - \wp \left( w \right)}} - \frac{1}{2}{\left( {\frac{{\wp '\left( z \right) - \wp '\left( w \right)}}{{\wp \left( z \right) - \wp \left( w \right)}}} \right)^2}
\end{equation*}
and finally,
\begin{equation}
\wp \left( {z + w} \right) =  - \wp \left( z \right) - \wp \left( w \right) - \frac{1}{4}{\left( {\frac{{\wp '\left( z \right) - \wp '\left( w \right)}}{{\wp \left( z \right) - \wp \left( w \right)}}} \right)^2} ,
\end{equation}
which the addition theorem for the Weierstrass elliptic function \eqref{eq:wp_addition}.
\newline\newline
\textbf{Problem \ref{pr:wp_duplication} Solution}

The addition formula for Weierstrass elliptic function states that
\begin{equation*}
\wp \left( {z + w} \right) =  - \wp \left( z \right) - \wp \left( w \right) + {\left( {\frac{1}{2}\frac{{\wp '\left( z \right) - \wp '\left( w \right)}}{{\wp \left( z \right) - \wp \left( w \right)}}} \right)^2} .
\end{equation*}
The duplication formula requires taking the limit $w \to z$. At this limit the fraction in the right hand side of the addition formula becomes indeterminate. It is simple though to use Hospital's rule to find
\begin{equation}
\wp \left( {2z} \right) =  - 2\wp \left( z \right) + {\left( {\frac{1}{2}\frac{{\wp ''\left( z \right)}}{{\wp '\left( z \right)}}} \right)^2} .
\end{equation}

One can eliminate the second derivative of $\wp$. For this purpose we have to differentiate the Weierstrass equation to yield
\begin{equation}
\wp ''\left( z \right) = 6{\wp ^2}\left( z \right) - \frac{{{g_2}}}{2} .
\end{equation}
Then, the duplication formula assumes the form
\begin{equation*}
\wp \left( {2z} \right) =  - 2\wp \left( z \right) + \frac{{{{\left( {6{\wp ^2}\left( z \right) - \frac{{{g_2}}}{2}} \right)}^2}}}{{4\left( {4{\wp ^3}\left( z \right) - {g_2}\wp \left( z \right) - {g_3}} \right)}}
\end{equation*}
or
\begin{equation}
\wp \left( {2z} \right) = \frac{1}{4}\wp \left( z \right) + \frac{{3{g_2}{\wp ^2}\left( z \right) + 9{g_3}\wp \left( z \right) + \frac{{g_2^2}}{4}}}{{4\wp {'^2}\left( z \right)}} .
\end{equation}

Let's now derive the duplication formula expressing $\wp \left( {2z} \right)$ in terms of $\zeta \left( {z} \right)$ and its derivatives. Let $2 \omega_1$ and $2 \omega_2$ be the two fundamental periods of $\wp \left( {z} \right)$. Then the function $\wp \left( {2z} \right)$ is an elliptic function with fundamental periods equal to $\omega_1$ and $\omega_2$. This obviously means that it is also an elliptic function with periods $2 \omega_1$ and $2 \omega_2$, but in the parallelogram defined by the latter, there exist four fundamental period parallelograms of $\wp \left( {2z} \right)$. Therefore, the function $\wp \left( {2z} \right)$, as an elliptic function with periods $2 \omega_1$ and $2 \omega_2$ is an order 8 elliptic function with four double poles at positions congruent to $z=0$, $z=\omega_1$, $z=\omega_2$ and, $z=\omega_3$. The principal part of the Laurent series of $\wp \left( {2z} \right)$ at each of these poles reads
\begin{equation*}
\wp \left( {2z} \right) = \frac{1}{4} \frac{1}{\left( z - z_0 \right)^2} + \mathcal{O} \left( \left( z - z_0 \right)^2 \right).
\end{equation*}
We have the necessary data to express $\wp \left( {2z} \right)$ in terms of the function $\zeta \left( {z} \right)$ and its derivatives. Formula \eqref{eq:elliptic_terms_zeta} implies that
\begin{multline*}
\wp \left( {2z} \right) = C - \frac{1}{4}\zeta '\left( z \right) - \frac{1}{4}\zeta '\left( {z - {\omega _1}} \right) - \frac{1}{4}\zeta '\left( {z - {\omega _2}} \right) - \frac{1}{4}\zeta '\left( {z - {\omega _1} - {\omega _2}} \right)\\
 = C + \frac{1}{4}\wp \left( z \right) + \frac{1}{4}\wp \left( {z - {\omega _1}} \right) + \frac{1}{4}\wp \left( {z - {\omega _2}} \right) + \frac{1}{4}\wp \left( {z - {\omega _3}} \right) .
\end{multline*}
Finding the Laurent series of the above equation at the region of $z=0$, we get
\begin{equation*}
\frac{1}{{4{z^4}}} + \mathcal{O}\left( {{z^2}} \right) = C + \frac{1}{{4{z^4}}} + {e_1} + {e_2} + {e_3} + \mathcal{O}\left( {{z^2}} \right) ,
\end{equation*}
implying that $C = 0$, since the three roots ${e_1}$, ${e_2}$ and ${e_3}$ sum to zero. Using the results of problem \ref{pr:half_period_shift}, we get
\begin{equation*}
\wp \left( {2z} \right) = \frac{1}{4}\wp \left( z \right) + \frac{1}{4}\left( {{e_1} + \frac{{2e_1^2 + {e_2}{e_3}}}{{\wp \left( z \right) - {e_1}}}} \right) + \frac{1}{4}\left( {{e_3} + \frac{{2e_3^2 + {e_1}{e_2}}}{{\wp \left( z \right) - {e_3}}}} \right) + \frac{1}{4}\left( {{e_2} + \frac{{2e_2^2 + {e_3}{e_1}}}{{\wp \left( z \right) - {e_2}}}} \right) .
\end{equation*}
After some algebra and using the fact that the three roots sum to zero, we get
\begin{multline*}
\wp \left( {2z} \right) = \frac{1}{4}\wp \left( z \right) + \frac{{\left( {2\left( {e_1^2 + e_2^2 + e_3^2} \right) + {e_1}{e_2} + {e_2}{e_3} + {e_3}{e_1}} \right){\wp ^2}\left( z \right)}}{{\wp {'^2}\left( z \right)}}\\
 + \frac{{\left( {2\left( {e_1^3 + e_2^3 + e_3^3} \right) + 3{e_1}{e_2}{e_3}} \right)\wp \left( z \right)}}{{\wp {'^2}\left( z \right)}} + \frac{{e_1^2e_2^2 + e_2^2e_3^2 + e_3^2e_1^2}}{{\wp {'^2}\left( z \right)}} .
\end{multline*}

The fact that the three roots sum to zero can be used to calculate the coefficients of the above expression. We find,
\begin{equation*}
{\left( {{e_1} + {e_2} + {e_3}} \right)^2} = \left( {e_1^2 + e_2^2 + e_3^2} \right) + 2\left( {{e_1}{e_2} + {e_2}{e_3} + {e_3}{e_1}} \right) = 0 ,
\end{equation*}
which with the help of equation \eqref{eq:g2_roots} implies
\begin{equation}
e_1^2 + e_2^2 + e_3^2 = \frac{{{g_2}}}{2} .
\end{equation}
Furthermore,
\begin{multline*}
\left( {e_1^2 + e_2^2 + e_3^2} \right)\left( {{e_1} + {e_2} + {e_3}} \right) \\
= e_1^3 + e_2^3 + e_3^3 + \left( {e_1^2{e_2} + e_2^2{e_3} + e_3^2{e_1} + {e_1}e_2^2 + {e_2}e_3^2 + {e_3}e_1^2} \right) = 0,
\end{multline*}
\begin{multline*}
{\left( {{e_1} + {e_2} + {e_3}} \right)^3} \\
= \left( {e_1^3 + e_2^3 + e_3^3} \right) + 3\left( {e_1^2{e_2} + e_2^2{e_3} + e_3^2{e_1} + {e_1}e_2^2 + {e_2}e_3^2 + {e_3}e_1^2} \right) + 6{e_1}{e_2}{e_3} = 0.
\end{multline*}
Taking into account equation \eqref{eq:g3_roots}, the last two equations imply that
\begin{equation}
e_1^3 + e_2^3 + e_3^3 = 3{e_1}{e_2}{e_3} = \frac{{3{g_3}}}{4} .
\end{equation}
Finally,
\begin{equation*}
{\left( {{e_1}{e_2} + {e_2}{e_3} + {e_3}{e_1}} \right)^2} = e_1^2e_2^2 + e_2^2e_3^2 + e_3^2e_1^2 + 2{e_1}{e_2}{e_3}\left( {{e_1} + {e_2} + {e_3}} \right) ,
\end{equation*}
implying that
\begin{equation}
e_1^2e_2^2 + e_2^2e_3^2 + e_3^2e_1^2 = {\left( {{e_1}{e_2} + {e_2}{e_3} + {e_3}{e_1}} \right)^2} = \frac{{g_2^2}}{{16}} .
\end{equation}

Putting everything together, we get
\begin{equation}
\wp \left( {2z} \right) = \frac{1}{4}\wp \left( z \right) + \frac{{3{g_2}{\wp ^2}\left( z \right) + 9{g_3}\wp \left( z \right) + \frac{{g_2^2}}{4}}}{{4\wp {'^2}\left( z \right)}} ,
\end{equation}
which is the desired duplication formula.
\newpage
\subsubsection*{Problems on Classical Mechanics Applications}
\textbf{Problem \ref{pr:qubic_double_roots} Solution}

When the cubic polynomial has a double root $e_0$, the third root necessarily equals $- 2 e_0$, as the three roots sum to zero. Therefore, the cubic polynomial in such a case equals,
\begin{equation}
Q\left( x \right) = 4{\left( {x - {e_0}} \right)^2}\left( {x + 2{e_0}} \right) = 4{x^3} - 12e_0^2x + 8e_0^3 .
\end{equation}
Thus, a cubic polynomial with a double root obeys
\begin{equation}
{g_2} = 12e_0^2,\quad {g_3} =  - 8e_0^3 ,
\end{equation}
or
\begin{equation}
{g_2} > 0,\quad g_2^3 - 27g_3^2 = 0 .
\end{equation}
When $g_3 > 0$, the double root is negative, and, thus, the two smaller roots coincide, whereas when $g_3 < 0$, the double root is positive meaning that the two larger roots coincide.

In the case of the motion of a particle in a cubic potential, the solution is given by
\begin{align}
x &= \wp \left( {t - {t_0}; - {F_0}, - E} \right) ,\\
x &= \wp \left( {t - {t_0} + {\omega _2}; - {F_0}, - E} \right) ,
\end{align}
where the second solution exists only when there are three real roots. The moduli take the values $g_2 = - F_0$ and $g_3 = - E$ and therefore a double root exists when
\begin{equation}
{F_0} < 0,\quad E =  \pm {\left( { - \frac{{{F_0}}}{3}} \right)^{\frac{3}{2}}} .
\end{equation}
These values correspond to the case the potential that possesses a local minimum and the energies equal the limiting values for the existence of a bounded motion. 

When $E = + {\left( { - F_0 / 3} \right)^{3 / 2}}$, the double root equals ${e_0} = {\left( { - F_0 / 12} \right)^{1 / 2}} > 0$. Therefore, the real period diverges, while the imaginary one assumes the value
\begin{equation}
2{\omega _2} = i\pi {\left( { - \frac{{3{F_0}}}{4}} \right)^{ - \frac{1}{4}}} .
\end{equation}
Formula \eqref{eq:two_large_roots_coincide} suggests that in this case,
\begin{equation}
\wp \left( {z; - {F_0}, - {{\left( { - \frac{{{F_0}}}{3}} \right)}^{\frac{3}{2}}}} \right) = {\left( { - \frac{{{F_0}}}{{12}}} \right)^{\frac{1}{2}}} + \frac{{{{\left( { - \frac{{3{F_0}}}{4}} \right)}^{\frac{1}{2}}}}}{{{{\sinh }^2}\left( {{{\left( { - \frac{{3{F_0}}}{4}} \right)}^{\frac{1}{4}}}z} \right)}} ,
\end{equation}
implying that the unbounded and bounded motions degenerate to
\begin{align}
x &= {\left( { - \frac{{{F_0}}}{{12}}} \right)^{\frac{1}{2}}} + \frac{{{{\left( { - \frac{{3{F_0}}}{4}} \right)}^{\frac{1}{2}}}}}{{{{\sinh }^2}\left( {{{\left( { - \frac{{3{F_0}}}{4}} \right)}^{\frac{1}{4}}}\left( {t - {t_0}} \right)} \right)}} , \\
x &= {\left( { - \frac{{{F_0}}}{{12}}} \right)^{\frac{1}{2}}} - \frac{{{{\left( { - \frac{{3{F_0}}}{4}} \right)}^{\frac{1}{2}}}}}{{{{\cosh }^2}\left( {{{\left( { - \frac{{3{F_0}}}{4}} \right)}^{\frac{1}{4}}}\left( {t - {t_0}} \right) + i\frac{\pi }{2}} \right)}} ,
\end{align}
respectively.
These motions describe a particle having exactly the energy of the local maximum, in the first case coming from the right and in the second coming from the left. The particle arrives at the position of the unstable equilibrium in infinite time.

When $E = - {\left( { - F_0 / 3} \right)^{3 / 2}}$ the double root equals ${e_0} = - {\left( { - F_0 / 12} \right)^{1 / 2}} > 0$. The imaginary period diverges, while the real one assumes the value
\begin{equation}
2{\omega _1} = \pi {\left( { - \frac{{3{F_0}}}{4}} \right)^{ - \frac{1}{4}}} .
\end{equation}
Formula \eqref{eq:two_small_roots_coincide} suggests that in this case,
\begin{equation}
\wp \left( {z; - {F_0}, + {{\left( { - \frac{{{F_0}}}{3}} \right)}^{\frac{3}{2}}}} \right) =  - {\left( { - \frac{{{F_0}}}{{12}}} \right)^{\frac{1}{2}}} + \frac{{{{\left( { - \frac{{3{F_0}}}{4}} \right)}^{\frac{1}{2}}}}}{{{{\sin }^2}\left( {{{\left( { - \frac{{3{F_0}}}{4}} \right)}^{\frac{1}{4}}}z} \right)}} ,
\end{equation}
implying that the unbounded and bounded motions degenerate to
\begin{align}
x &= {\left( { - \frac{{{F_0}}}{{12}}} \right)^{\frac{1}{2}}} + \frac{{{{\left( { - \frac{{3{F_0}}}{4}} \right)}^{\frac{1}{2}}}}}{{{{\sin }^2}\left( {{{\left( { - \frac{{3{F_0}}}{4}} \right)}^{\frac{1}{4}}}\left( {t - {t_0}} \right)} \right)}} ,\\
x &= {\left( { - \frac{{{F_0}}}{{12}}} \right)^{\frac{1}{2}}} + \mathop {\lim }\limits_{x \to \infty } \frac{{{{\left( { - \frac{{3{F_0}}}{4}} \right)}^{\frac{1}{2}}}}}{{{{\sin }^2}\left( {{{\left( { - \frac{{3{F_0}}}{4}} \right)}^{\frac{1}{4}}}\left( {t - {t_0}} \right) + ix} \right)}} = {\left( { - \frac{{{F_0}}}{{12}}} \right)^{\frac{1}{2}}} ,
\end{align}
respectively. The form of the limit of the bounded motion is also physically expected. The bounded motion ranges between $e_3$ and $e_2$. Since these two roots coincide, the bounded motion necessarily degenerates to a constant. The unbounded motion describes a point particle having exactly the energy of the local minimum, in the first case coming from the right and getting reflected by the potential barrier, while the bounded motion describes a point particle resting at the local minimum equilibrium position. 

The period of motion in this case assumes a finite value. The Taylor series of the potential at the region of the local minimum is
\begin{equation}
V\left( x \right) =  - {\left( { - \frac{{{F_0}}}{3}} \right)^{\frac{3}{2}}} + 2\sqrt { - 3{F_0}} {\left( {x + {{\left( { - \frac{{{F_0}}}{{12}}} \right)}^{\frac{1}{2}}}} \right)^2} - 4{\left( {x + {{\left( { - \frac{{{F_0}}}{{12}}} \right)}^{\frac{1}{2}}}} \right)^3} .
\end{equation}
Since the mass of the point particle has been assumed to equal to 2, the period of the small oscillations at the region of the local minimum is equal to
\begin{equation}
{T_{{\rm{small}}}} = \frac{{2\pi }}{{{{\left( { - 12{F_0}} \right)}^{\frac{1}{4}}}}} = 2{\omega _1} ,
\end{equation}
as expected.
\newline\newline
\textbf{Problem \ref{pr:real_imaginary_periods} Solution}

The cubic polynomial associated with the problem of a point particle with energy $E$ is
\begin{equation*}
Q\left( x \right) = 4{x^3} + {F_0}x + E .
\end{equation*}
Let $e_1$, $e_2$ and $e_3$ be its three roots, appropriately ordered, as described in section \ref{subsec:real_Weierstrass}. They obviously obey
\begin{equation*}
Q\left( {{e_i}} \right) = 4e_i^3 + {F_0}{e_i} + E = 0 .
\end{equation*}

The cubic polynomial associated with the problem of a point particle with energy $-E$ is
\begin{equation*}
R\left( x \right) = 4{x^3} + {F_0}x - E .
\end{equation*}
This polynomial obeys
\begin{equation}
R\left( { - {e_i}} \right) =  - 4e_i^3 - {F_0}{e_i} - E =  - Q\left( {{e_i}} \right) = 0 .
\end{equation}
In other words, the roots of the problem with inverted energy are just the opposite of the roots of the initial problem. It is obvious that the appropriate ordering of the roots of the inverted energy problem is
\begin{equation}
{e_1}\left( { - E} \right) =  - {e_3}\left( E \right),\quad {e_2}\left( { - E} \right) =  - {e_2}\left( E \right),\quad {e_3}\left( { - E} \right) =  - {e_1}\left( E \right) .
\end{equation}
In the case the three roots are real, the above is obvious, since $e_1 > e_2 > e_3 \Rightarrow - e_1 < - e_2 < - e_3$. In the case there is only one real root, this is necessarily $- e_2$, since $e_2$ is real, while $-e_1$ has a negative imaginary part, since $e_1$ has positive imaginary part.

In the case of three real roots, we use formula \eqref{eq:real_period} to find that
\begin{equation}
\begin{split}
2{\omega _1}\left( { - E} \right) &= \int_{ - {e_3}}^{ + \infty } {\frac{{dt}}{{\sqrt {\left( {t + {e_1}} \right)\left( {t + {e_2}} \right)\left( {t + {e_3}} \right)} }}} \\
 &= \int_{ - \infty }^{{e_3}} {\frac{{dt'}}{{\sqrt {\left( {{e_1} - t'} \right)\left( {{e_2} - t'} \right)\left( {{e_3} - t'} \right)} }}} \\
 &=  - 2i{\omega _2}\left( E \right) = \left| {2{\omega _2}\left( E \right)} \right| ,
\end{split}
\end{equation}
where we defined $t' = - t$. Similarly, in the case of one real root, we use formula \eqref{eq:complex_period1} to find
\begin{equation}
\begin{split}
2{\omega _1}\left( { - E} \right) + 2{\omega _2}\left( { - E} \right) &= \int_{ - {e_2}}^{ + \infty } {\frac{{dt}}{{\sqrt {\left( {t + {e_1}} \right)\left( {t + {e_2}} \right)\left( {t + {e_3}} \right)} }}} \\
 &= \int_{ - \infty }^{{e_2}} {\frac{{dt'}}{{\sqrt {\left( {t' - {e_2}} \right)\left( {{e_2} - t'} \right)\left( {t' - {e_3}} \right)} }}} \\
 &=  - 2i\left( {2{\omega _1}\left( E \right) - 2{\omega _2}\left( E \right)} \right) = \left| {2{\omega _1}\left( E \right) - 2{\omega _2}\left( E \right)} \right| .
\end{split}
\end{equation}
This completes the proof that the absolute value of the imaginary period corresponds in all cases to the ``time of flight'' or period of oscillations for a point particle with opposite energy.
\newline\newline
\textbf{Problem \ref{pr:discrete_symmetry} Solution}

The conservation of energy in the cubic potential problem reads
\begin{equation*}
{{\dot x}^2} - 4{x^3} - {F_0}x = E
\end{equation*}
or in terms of the roots
\begin{equation}
{{\dot x}^2} = 4\left( {x - {e_1}} \right)\left( {x - {e_2}} \right)\left( {x - {e_3}} \right) .
\end{equation}

We perform the change of variable $x = {e_3} + \frac{{\left( {{e_3} - {e_1}} \right)\left( {{e_3} - {e_2}} \right)}}{{y - {e_3}}}$. It is a matter of simple algebra to find that
\begin{align*}
&\dot x =  - \frac{{\left( {{e_3} - {e_1}} \right)\left( {{e_3} - {e_2}} \right)}}{{{{\left( {y - {e_3}} \right)}^2}}}\dot y , \\
&x - {e_1} = \frac{{\left( {{e_3} - {e_1}} \right)\left( {y - {e_2}} \right)}}{{y - {e_3}}},\;\, x - {e_2} = \frac{{\left( {{e_3} - {e_2}} \right)\left( {y - {e_1}} \right)}}{{y - {e_3}}},\;\, x - {e_3} = \frac{{\left( {{e_3} - {e_1}} \right)\left( {{e_3} - {e_2}} \right)}}{{y - {e_3}}} .
\end{align*}
Thus, in terms of the new variable $y$, the conservation of energy is written as
\begin{equation*}
\frac{{{{\left( {{e_3} - {e_1}} \right)}^2}{{\left( {{e_3} - {e_2}} \right)}^2}}}{{{{\left( {y - {e_3}} \right)}^4}}}{{\dot y}^2} = 4\frac{{\left( {{e_3} - {e_1}} \right)\left( {y - {e_2}} \right)}}{{y - {e_3}}}\frac{{\left( {{e_3} - {e_2}} \right)\left( {y - {e_1}} \right)}}{{y - {e_3}}}\frac{{\left( {{e_3} - {e_1}} \right)\left( {{e_3} - {e_2}} \right)}}{{y - {e_3}}}
\end{equation*}
or
\begin{equation}
{{\dot y}^2} = 4\left( {y - {e_1}} \right)\left( {y - {e_2}} \right)\left( {y - {e_3}} \right) .
\end{equation}

Therefore, the conservation of energy is invariant under this transformation. We could say that the problem has a $\mathbb{Z}_2$ symmetry, since performing the change of variable twice leads to the initial variable,
\begin{equation}
x \to {e_3} + \frac{{\left( {{e_3} - {e_1}} \right)\left( {{e_3} - {e_2}} \right)}}{{x - {e_3}}} \to {e_3} + \frac{{\left( {{e_3} - {e_1}} \right)\left( {{e_3} - {e_2}} \right)}}{{{e_3} + \frac{{\left( {{e_3} - {e_1}} \right)\left( {{e_3} - {e_2}} \right)}}{{x - {e_3}}} - {e_3}}} = x .
\end{equation}

It is trivial to show that
\begin{align*}
\mathop {\lim }\limits_{y \to  - \infty } x = {e_3},\quad \mathop {\lim }\limits_{y \to e_3^ - } x &=  - \infty ,\quad \mathop {\lim }\limits_{y \to e_3^ + } x =  + \infty ,\\
\mathop {\lim }\limits_{y \to {e_2}} x = {e_1},\quad \mathop {\lim }\limits_{y \to {e_1}} x &= {e_2},\quad \mathop {\lim }\limits_{y \to  + \infty } x = {e_3} .
\end{align*}
As a consequence, the following intervals of $x$ are mapped to intervals of $y$ as
\begin{align*}
\left( { - \infty ,{e_3}} \right] &\to \left( { - \infty ,{e_3}} \right] ,\\
\left[ {{e_3},{e_2}} \right] &\to \left[ {{e_1}, + \infty } \right) ,\\
\left[ {{e_2},{e_1}} \right] &\to \left[ {{e_2},{e_1}} \right] ,\\
\left[ {{e_1}, + \infty } \right) &\to \left[ {{e_3},{e_2}} \right] .
\end{align*}

That means that the bounded motion in one problem is mapped to the unbounded motion of the other. Since the problems are identical, the two solutions, oscillatory and scattering, are also identical,
\begin{align}
{x_{{\rm{oscillatory}}}}\left( t \right) = {y_{{\rm{oscillatory}}}}\left( t \right) &= f\left( t \right) , \\
{x_{{\rm{scattering}}}}\left( t \right) = {y_{{\rm{scattering}}}}\left( t \right) &= g\left( t \right) ,
\end{align}
which implies
\begin{equation}
f\left( t \right) = {e_3} + \frac{{\left( {{e_3} - {e_1}} \right)\left( {{e_3} - {e_2}} \right)}}{{g\left( t \right) - {e_3}}},\quad g\left( t \right) = {e_3} + \frac{{\left( {{e_3} - {e_1}} \right)\left( {{e_3} - {e_2}} \right)}}{{f\left( t \right) - {e_3}}} .
\end{equation}
Without knowing the exact form of the oscillatory and scattering motions, we know that they are connected in a specific way. Let $T$ be the period of the oscillatory motion. The above connection implies that 
\begin{align*}
f\left( 0 \right) = {e_3} , &\quad g\left( 0 \right) \to  + \infty , \\
f\left( {\frac{T}{2}} \right) = {e_2} , &\quad g\left( {\frac{T}{2}} \right) = {e_1} , \\
f\left( T \right) = {e_3} , &\quad g\left( T \right) \to  + \infty .
\end{align*}
Thus, it turns out that the period of the oscillatory motion and the ``time of flight'' of the scattering motion are equal.
\newline\newline
\textbf{Problem \ref{pr:pendulum_double_roots} Solution}

It is evident in figure \ref{fig:roots_pendulum} that there are two energies where a double root appears, namely, $E = \pm \omega^2$.

For $E = \omega^2$, the two larger roots coincide,
\begin{equation*}
{e_1} = {e_2} = {e_0} = \frac{{{\omega ^2}}}{3} .
\end{equation*}
Following the outcomes of problem \ref{pr:degenerate cases}, the real period of the Weierstrass elliptic function diverges, whereas the imaginary one assumes the value
\begin{equation*}
2{\omega _2} = i\frac{\pi }{\omega }
\end{equation*}
and the Weierstrass elliptic function is expressed in terms of hyperbolic functions as
\begin{equation*}
2\wp \left( {z;\frac{{4{\omega ^4}}}{3}, - \frac{{8{\omega ^6}}}{{27}}} \right) = \frac{{{\omega ^2}}}{3} + \frac{{{\omega ^2}}}{{{{\sinh }^2}\omega z}} .
\end{equation*}
The pendulum solution is expressed in terms of the bounded real solution in the real domain of Weierstrass equation, namely the Weierstrass elliptic function on the real axis shifted by $\omega_2$,
\begin{equation*}
2\wp \left( {t + {\omega _2};\frac{{4{\omega ^4}}}{3}, - \frac{{8{\omega ^6}}}{{27}}} \right) = \frac{{{\omega ^2}}}{3} + \frac{{{\omega ^2}}}{{{{\sinh }^2}\left( {\omega t + i\frac{\pi }{2}} \right)}} = \frac{{{\omega ^2}}}{3} - \frac{{{\omega ^2}}}{{{{\cosh }^2}\omega t}} .
\end{equation*}
Finally, equation \eqref{eq:change_variable_pendulum} implies that the degenerate solution is written as
\begin{equation}
\cos \theta  =  - 1 + \frac{2}{{{{\cosh }^2}\omega t}} .
\end{equation}

For $E = -\omega^2$, the two smaller roots coincide.
\begin{equation*}
{e_2} = {e_3} =  - {e_0} =  - \frac{{{\omega ^2}}}{3} .
\end{equation*}
Therefore, the imaginary period diverges, while the real one assumes the value
\begin{equation*}
2{\omega _1} = \frac{\pi }{{\sqrt {3{e_0}} }} = \frac{\pi }{\omega } .
\end{equation*}
As in problem \ref{pr:qubic_double_roots}, the bounded real solution in the real domain of Weierstrass equation degenerates to a constant being equal to the double root,
\begin{equation}
2\wp \left( {t + {\omega _2};\frac{{4{\omega ^4}}}{3},\frac{{8{\omega ^6}}}{{27}}} \right) =  - \frac{{{\omega ^2}}}{3} .
\end{equation}
Finally, equation \eqref{eq:change_variable_pendulum} implies 
\begin{equation}
\cos \theta  = 1 .
\end{equation}
This solution describes the pendulum lying at the stable equilibrium position. Considering this solution as the limit of the oscillatory pendulum motion when the amplitude of the oscillation goes to zero, we conclude that the limit of the period of the oscillatory motion at zero amplitude is
\begin{equation}
{T_{{\rm{oscillating}}}} = 4{\omega _1} = \frac{{2\pi }}{\omega } ,
\end{equation}
as expected.
\newline\newline
\textbf{Problem \ref{pr:hyperbolic_transmission} Solution}

In the case $E > \omega^2$, it holds that $e_1 = x_1$, $e_2 = x_2$ and $e_3 = x_3$. We start using the addition formula for Weierstrass elliptic function \eqref{pr:wp_addition}
\begin{multline*}
x\left( {- t} \right) = \ln \left[ {\frac{2}{{{\omega ^2}}}\left( {2\wp \left( {{\omega _1}/{2} - t} \right) - \frac{E}{3}} \right)} \right]\\
 = \ln \left[ {\frac{2}{{{\omega ^2}}}\left( { - 2\wp \left( {{\omega _1}} \right) - 2\wp \left( {- {\omega _1}/{2} - t} \right) + 2{{\left( {\frac{1}{2}\frac{{\wp '\left( {{\omega _1}} \right) - \wp '\left( {- {\omega _1}/{2} - t} \right)}}{{\wp \left( {{\omega _1}} \right) - \wp \left( { - {\omega _1}/{2} - t} \right)}}} \right)}^2} - 2{e_1}} \right)} \right]\\
 = \ln \left[ {\frac{2}{{{\omega ^2}}}\left( { - 4{e_1} - 2\wp \left( {\omega _1}/{2} + t \right) + \frac{1}{2}{{\left( {\frac{{\wp '\left( {\omega _1}/{2} + t \right)}}{{{e_1} - \wp \left( {\omega _1}/{2} + t \right)}}} \right)}^2}} \right)} \right] .
\end{multline*}
Then, we use the Weierstrass differential equation \eqref{eq:Weierstrass_equation} to substitute the derivative of Weierstrass function
\begin{multline*}
x\left( { - t} \right) = \ln \left[ {\frac{2}{{{\omega ^2}}}\left( { - 4{e_1} - 2\wp \left( {\omega _1}/{2} + t \right) + 2\frac{{\left( {\wp \left( {\omega _1}/{2} + t \right) - {e_2}} \right)\left( {\wp \left( {\omega _1}/{2} + t \right) - {e_3}} \right)}}{{\wp \left( {\omega _1}/{2} + t \right) - {e_1}}}} \right)} \right]\\
 = \ln \left[ {\frac{2}{{{\omega ^2}}}\left( {2\frac{{\left( { - {e_1} - {e_2} - {e_3}} \right)\wp \left( {\omega _1}/{2} + t \right) + 2e_1^2 + {e_2}{e_3}}}{{\wp \left( {\omega _1}/{2} + t \right) - {e_1}}}} \right)} \right] \\
 = \ln \left[ {\frac{2}{{{\omega ^2}}}\left( {2\frac{{2e_1^2 + {e_2}{e_3}}}{{\wp \left( {\omega _1}/{2} + t \right) - {e_1}}}} \right)} \right] .
\end{multline*}
Using the specific form of the roots, we find
\begin{multline*}
2e_1^2 + {e_2}{e_3} = 2x_1^2 + {x_2}{x_3}\\
 = 2{\left( {\frac{E}{6}} \right)^2} + \left( { - \frac{E}{{12}} + \frac{1}{4}\sqrt {{E^2} - {\omega ^4}} } \right)\left( { - \frac{E}{{12}} - \frac{1}{4}\sqrt {{E^2} - {\omega ^4}} } \right)\\
 = \frac{{{E^2}}}{{18}} + \frac{{{E^2}}}{{144}} - \frac{{{E^2} - {\omega ^4}}}{{16}} = \frac{{{\omega ^4}}}{{16}} ,
\end{multline*}
which implies that finally, we get
\begin{equation}
x\left( {- t} \right) = \ln \left[ {\frac{{{\omega ^2}}}{{4\left( {\wp \left( {\omega _1}/{2} + t \right) - {e_1}} \right)}}} \right] =  - \ln \left[ {\frac{2}{{{\omega ^2}}}\left( {2\wp \left( {\omega _1}/{2} + t \right) - \frac{E}{3}} \right)} \right] =  - x\left( t \right) .
\end{equation}

The case $E < \omega^2$ is identical with the permutation $e_1 \leftrightarrow e_2$, $\omega_1 \leftrightarrow \omega_3$.

The essential difference between the transmitting solutions and the reflecting solutions is that in the former case the time of flight equals the real half-period, while in the latter equals the whole real period. Therefore, in the case of reflecting solutions, the situation is much simpler, since,
\begin{multline}
x\left( {- t} \right) = \ln \left[ {\frac{2}{{{\omega ^2}}}\left( {2\wp \left( {{\omega _1} - t} \right) - \frac{E}{3}} \right)} \right] = \ln \left[ {\frac{2}{{{\omega ^2}}}\left( {2\wp \left( {- {\omega _1} + t} \right) - \frac{E}{3}} \right)} \right]\\
= \ln \left[ {\frac{2}{{{\omega ^2}}}\left( {2\wp \left( { {\omega _1} + t} \right) - \frac{E}{3}} \right)} \right] = x\left( {t} \right) ,
\end{multline}
as expected.
\newline\newline
\textbf{Problem \ref{pr:hyperbolic_oscillations} Solution}

The extrema of motion can be read from table \ref{tb:phi_range}. They are $\ln {\frac{4\left( {e_1 - e_2} \right)}{\omega^2}} $ and $\ln {\frac{4\left( {e_1 - e_3} \right)}{\omega^2}} $. It is a matter of simple algebra to show that
\begin{equation}
\left( {{e_1} - {e_3}} \right)\left( {{e_1} - {e_2}} \right) = e_1^2 - \left( {{e_2} + {e_3}} \right){e_1} + {e_2}{e_3} = 2e_1^2 + {e_2}{e_3} .
\end{equation}
In problem \ref{pr:hyperbolic_transmission} we found that $2e_1^2 + {e_2}{e_3} = \frac{{{\omega ^4}}}{{16}}$. Therefore, $\ln {\frac{4\left( {e_1 - e_2} \right)}{\omega^2}} = - \ln {\frac{4\left( {e_1 - e_3} \right)}{\omega^2}} $ and consequently the extrema of motion are opposite.

The periodicity property $x \left( T - t \right) = - x \left( t \right)$ can be proved easily repeating all the steps of problem \ref{pr:hyperbolic_transmission} with the substitution $t \to - t$.

The double root limit, as one can easily see in figure \ref{fig:roots} corresponds to $E = \omega^2$. In this limit the two smaller roots coincide to the value $e_2 = e_3 = - \omega^2 /12$. The oscillatory motion is described by the bounded solution. As shown in problem \ref{pr:qubic_double_roots}, in this limit, the bounded solution degenerates to a constant equal to the value of the double root. Consequently,
\begin{equation}
x\left( t \right) = \ln \left[ {- \frac{2}{{{\omega ^2}}}\left( {2\frac{{{\omega ^2}}}{{12}} - \frac{{{\omega ^2}}}{3}} \right)} \right] = \ln 1 = 0 ,
\end{equation}
which is indeed the equilibrium position. The real period degenerates to the value
\begin{equation}
T = 2\omega_1 = \frac{\pi}{\sqrt{- 3 e_2}} = \frac{2 \pi}{\omega} ,
\end{equation}
as expected.
\newpage
\subsubsection*{Problems on Quantum Mechanics Applications}
\textbf{Problem \ref{pr:lame_special} Solution}

The special solution is
\begin{equation}
\begin{split}
\tilde y\left( {x;{\omega _{1,2,3}}} \right) &= \sqrt {\wp \left( x \right) - {e_{1,3,2}}} \left( {\zeta \left( {x + {\omega _{1,2,3}}} \right) + {e_{1,3,2}}x} \right) \\
&= {y_ \pm }\left( {x;{\omega _{1,2,3}}} \right)\left( {\zeta \left( {x + {\omega _{1,2,3}}} \right) + {e_{1,3,2}}x} \right) .
\end{split}
\end{equation}
The first and second derivatives of ${y_ \pm }\left( {x;{\omega _{1,2,3}}} \right)$ are given by equations \eqref{eq:lame_first_der} and \eqref{eq:lame_second_der}. These equations for the solution modulus being equal to any of the half-periods read
\begin{align}
\frac{{d{y_ \pm }\left( {x;{\omega _{1,2,3}}} \right)}}{{dx}} &= \frac{1}{2}\frac{{\wp '\left( x \right)}}{{\wp \left( x \right) - {e_{1,3,2}}}}{y_ \pm }\left( {x;{\omega _{1,2,3}}} \right) ,\\
\frac{{{d^2}{y_ \pm }\left( {x;{\omega _{1,2,3}}} \right)}}{{d{x^2}}} &= \left( {2\wp \left( x \right) + {e_{1,3,2}}} \right){y_ \pm }\left( {x;{\omega _{1,2,3}}} \right) .
\end{align}
Therefore, we may express the first and second derivatives of  $\tilde y\left( {x;{\omega _{1,2,3}}} \right)$ as,
\begin{multline*}
\frac{{d\tilde y\left( {x;{\omega _{1,2,3}}} \right)}}{{dx}} = \frac{{d{y_ \pm }\left( {x;{\omega _{1,2,3}}} \right)}}{{dx}}\left( {\zeta \left( {x + {\omega _{1,2,3}}} \right) + {e_{1,3,2}}x} \right) \\
+ {y_ \pm }\left( {x;{\omega _{1,2,3}}} \right)\left( { - \wp \left( {x + {\omega _{1,2,3}}} \right) + {e_{1,3,2}}} \right)
\end{multline*}
and
\begin{equation}
\begin{split}
\frac{{{d^2}\tilde y\left( {x;{\omega _{1,2,3}}} \right)}}{{d{x^2}}} &= \frac{{{d^2}{y_ \pm }\left( {x;{\omega _{1,2,3}}} \right)}}{{d{x^2}}}\left( {\zeta \left( {x + {\omega _{1,2,3}}} \right) + {e_{1,3,2}}x} \right) \\
&+ 2\frac{{d{y_ \pm }\left( {x;{\omega _{1,2,3}}} \right)}}{{dx}}\left( { - \wp \left( {x + {\omega _{1,2,3}}} \right) + {e_{1,3,2}}} \right) \\
&- {y_ \pm }\left( {x;{\omega _{1,2,3}}} \right)\left( {\wp '\left( {x + {\omega _{1,2,3}}} \right)} \right)\\
 &= \left( {2\wp \left( x \right) + {e_{1,3,2}}} \right)\left( {\zeta \left( {x + {\omega _{1,2,3}}} \right) + {e_{1,3,2}}x} \right) {y_ \pm }\left( {x;{\omega _{1,2,3}}} \right)\\
 &- \frac{{\wp '\left( x \right)}}{{\wp \left( x \right) - {e_{1,3,2}}}}\left( {\wp \left( {x + {\omega _{1,2,3}}} \right) - {e_{1,3,2}}} \right) {y_ \pm }\left( {x;{\omega _{1,2,3}}} \right)\\
 &- \wp '\left( {x + {\omega _{1,2,3}}} \right) {y_ \pm }\left( {x;{\omega _{1,2,3}}} \right) .
\end{split}
\label{eq:lame_problem_special_second_der}
\end{equation}

The Weierstrass elliptic function, with an argument shifted by a half-period can be calculated with the use of the addition theorem. This has been done in problem \ref{pr:half_period_shift},
\begin{equation}
\wp \left( {x + {\omega _{1,2,3}}} \right) = {e_{1,3,2}} + \frac{{\left( {{e_{1,3,2}} - {e_{3,2,1}}} \right)\left( {{e_{1,3,2}} - {e_{2,1,3}}} \right)}}{{\wp \left( x \right) - {e_{1,3,2}}}} .
\end{equation}
It is a direct consequence that
\begin{equation}
\wp '\left( {x + {\omega _{1,2,3}}} \right) =  - \frac{{\left( {{e_{1,3,2}} - {e_{3,2,1}}} \right)\left( {{e_{1,3,2}} - {e_{2,1,3}}} \right)}}{{{{\left( {\wp \left( x \right) - {e_{1,3,2}}} \right)}^2}}}\wp '\left( x \right) .
\end{equation}
Combining the above two relations, we find
\begin{multline*}
\frac{{\wp '\left( x \right)}}{{\wp \left( x \right) - {e_{1,3,2}}}}\left( {\wp \left( {x + {\omega _{1,2,3}}} \right) - {e_{1,3,2}}} \right) + \wp '\left( {x + {\omega _{1,2,3}}} \right)\\
 = \frac{{\wp '\left( x \right)}}{{\wp \left( x \right) - {e_{1,3,2}}}}\frac{{\left( {{e_{1,3,2}} - {e_{3,2,1}}} \right)\left( {{e_{1,3,2}} - {e_{2,1,3}}} \right)}}{{\wp \left( x \right) - {e_{1,3,2}}}} \\
 - \frac{{\left( {{e_{1,3,2}} - {e_{3,2,1}}} \right)\left( {{e_{1,3,2}} - {e_{2,1,3}}} \right)}}{{{{\left( {\wp \left( x \right) - {e_{1,3,2}}} \right)}^2}}}\wp '\left( x \right) = 0
\end{multline*}

The latter implies that equation \eqref{eq:lame_problem_special_second_der} is written as
\begin{equation}
\begin{split}
\frac{{{d^2}\tilde y\left( {x;{\omega _{1,2,3}}} \right)}}{{d{x^2}}} &= \left( {2\wp \left( x \right) + {e_{1,3,2}}} \right)\left( {\zeta \left( {x + {\omega _{1,2,3}}} \right) + {e_{1,3,2}}x} \right){y_ \pm }\left( {x;{\omega _{1,2,3}}} \right) \\
&= \left( {2\wp \left( x \right) + {e_{1,3,2}}} \right)\tilde y\left( {x;{\omega _{1,2,3}}} \right) ,
\end{split}
\end{equation}
meaning that indeed the functions $\tilde y\left( {x;{\omega _{1,2,3}}} \right)$ are eigenfunctions of the $n = 1$ \Lame problem with eigenvalues
\begin{equation}
\lambda  =  - {e_{1,3,2}} .
\end{equation}
\newline\newline
\textbf{Problem \ref{pr:lame_bounded_reality} Solution}

We follow the derivation for the unbounded potential. In the following, $b$ is considered real. For $a$ in the segment $\left[ 0 , \omega_1 \right]$, $a = b$. Then,
\begin{equation*}
\begin{split}
\overline{{{\psi}_ \pm }\left( {x;b} \right)} &= \frac{{\sigma \left( {x - {\omega _2} \pm b} \right)\sigma \left( { - {\omega _2}} \right)}}{{\sigma \left( {x - {\omega _2}} \right)\sigma \left( { - {\omega _2} \pm b} \right)}}{e^{ - \zeta \left( { \pm b} \right)x}}\\
 &= \frac{{\sigma \left( {x + {\omega _2} \pm b} \right){e^{ - 2\zeta \left( {{\omega _2}} \right)\left( {x \pm b} \right)}}\sigma \left( {{\omega _2}} \right)}}{{\sigma \left( {x + {\omega _2}} \right){e^{ - 2\zeta \left( {{\omega _2}} \right)x}}\sigma \left( { + {\omega _2} \pm b} \right){e^{ - 2\zeta \left( {{\omega _2}} \right)\left( { \pm b} \right)}}}}{e^{ - \zeta \left( { \pm b} \right)x}} \\
 &= {\psi_ \pm }\left( {x;b} \right) .
\end{split}
\end{equation*}
For $a$ in the segment $\left[ 0 , \omega_2 \right]$, $a = i b$,
\begin{equation*}
\begin{split}
\overline{{{\psi}_ \pm }\left( {x;ib} \right)} &= \frac{{\sigma \left( {x - {\omega _2} \mp ib} \right)\sigma \left( { - {\omega _2}} \right)}}{{\sigma \left( {x - {\omega _2}} \right)\sigma \left( { - {\omega _2} \mp ib} \right)}}{e^{ - \zeta \left( { \mp ib} \right)x}}\\
 &= \frac{{\sigma \left( {x + {\omega _2} \mp ib} \right){e^{ - 2\zeta \left( {{\omega _2}} \right)\left( {x \mp ib} \right)}}\sigma \left( {{\omega _2}} \right)}}{{\sigma \left( {x + {\omega _2}} \right){e^{ - 2\zeta \left( {{\omega _2}} \right)x}}\sigma \left( {{\omega _2} \mp ib} \right){e^{ - 2\zeta \left( {{\omega _2}} \right)\left( { \mp ib} \right)}}}}{e^{ - \zeta \left( { \mp ib} \right)x}} \\
 &= {\psi_ \mp }\left( {x;ib} \right) .
\end{split}
\end{equation*}
For $a$ in the segment $\left[ \omega_2 , \omega_3 \right]$, $a = \omega_2 + b$,
\begin{equation*}
\begin{split}
&\overline{{{\psi}_ \pm }\left( {x;{\omega _2} + b} \right)} = \frac{{\sigma \left( {x - {\omega _2} \mp {\omega _2} \pm b} \right)\sigma \left( { - {\omega _2}} \right)}}{{\sigma \left( {x - {\omega _2}} \right)\sigma \left( { - {\omega _2} \mp {\omega _2} \pm b} \right)}}{e^{ - \zeta \left( { \mp {\omega _2} \pm b} \right)x}}\\
 &= \frac{{ - \sigma \left( {x + {\omega _2} \pm {\omega _2} \pm b} \right){e^{ - \left( {2 \pm 2} \right)\zeta \left( {{\omega _2}} \right)\left( {x \pm b} \right)}}\sigma \left( {{\omega _2}} \right)}}{{ - \sigma \left( {x + {\omega _2}} \right){e^{ - 2\zeta \left( {{\omega _2}} \right)x}}\sigma \left( {{\omega _2} \pm {\omega _2} \pm b} \right){e^{ - \left( {2 \pm 2} \right)\zeta \left( {{\omega _2}} \right)\left( { \pm b} \right)}}}}{e^{ \pm 2\zeta \left( {{\omega _2}} \right)x}}{e^{ - \zeta \left( { \pm {\omega _2} \pm b} \right)x}} \\
 &= {\psi_ \pm }\left( {x;{\omega _2} + b} \right) .
\end{split}
\end{equation*}
Finally, for $a$ in the segment $\left[ \omega_1 , \omega_3 \right]$, $a = \omega_1 + i b$ and
\begin{equation*}
\begin{split}
&\overline{{{\psi}_ \pm }\left( {x;{\omega _1} + ib} \right)} = \frac{{\sigma \left( {x - {\omega _2} \pm {\omega _1} \mp ib} \right)\sigma \left( { - {\omega _2}} \right)}}{{\sigma \left( {x - {\omega _2}} \right)\sigma \left( { - {\omega _2} \pm {\omega _1} \mp ib} \right)}}{e^{ - \zeta \left( { \pm {\omega _1} \mp ib} \right)x}}\\
 &= \frac{{\sigma \left( {x + {\omega _2} \mp {\omega _1} \mp ib} \right){e^{2\left( { \pm \zeta \left( {{\omega _1}} \right) - \zeta \left( {{\omega _2}} \right)} \right)\left( {x \mp ib} \right)}}\sigma \left( {{\omega _2}} \right)}}{{\sigma \left( {x + {\omega _2}} \right){e^{ - 2\zeta \left( {{\omega _2}} \right)x}}\sigma \left( {{\omega _2} \mp {\omega _1} \mp ib} \right){e^{2\left( { \pm \zeta \left( {{\omega _2}} \right) - \zeta \left( {{\omega _2}} \right)} \right)\left( { \mp ib} \right)}}}}{e^{ \mp 2\zeta \left( {{\omega _1}} \right)x}}{e^{ - \zeta \left( { \mp {\omega _1} \mp ib} \right)x}} \\
 &= {\psi_ \mp }\left( {x;{\omega _1} + ib} \right) ,
\end{split}
\end{equation*}
which concludes the derivation of the reality properties of the eigenfunctions of the bounded $n = 1$ \Lame problem.
\newline\newline
\textbf{Problem \ref{pr:lame_bounded_bands} Solution}

Following the derivation in the case of the unbounded potential presented in section \ref{subsec:lame}, we have
\begin{equation*}
\begin{split}
\frac{{\sigma \left( {x + 2{\omega _1} + {\omega _2} \pm a} \right)\sigma \left( {{\omega _2}} \right)}}{{\sigma \left( {x + 2{\omega _1} + {\omega _2}} \right)\sigma \left( {{\omega _2} \pm a} \right)}} &= \frac{{ - {e^{2\zeta \left( {{\omega _1}} \right)\left( {x + {\omega _1} + {\omega _2} \pm a} \right)}}\sigma \left( {x + {\omega _2} \pm a} \right)\sigma \left( {{\omega _2}} \right)}}{{ - {e^{2\zeta \left( {{\omega _1}} \right)\left( {x + {\omega _1} + {\omega _2}} \right)}}\sigma \left( {x + {\omega _2}} \right)\sigma \left( {{\omega _2} \pm a} \right)}} \\
&= {e^{ \pm 2a\zeta \left( {{\omega _1}} \right)}}\frac{{\sigma \left( {x + {\omega _2} \pm a} \right)\sigma \left( {{\omega _2}} \right)}}{{\sigma \left( {x + {\omega _2}} \right)\sigma \left( {{\omega _2} \pm a} \right)}} .
\end{split}
\end{equation*}
Thus, we can write the eigenfunctions \eqref{eq:lame_eigenstates_shifted} of the bounded problem in the form of a Bloch wave
\begin{equation}
{\psi _ \pm }\left( {x;a} \right) = {u_ \pm }\left( {x;a} \right){e^{ \pm ik\left( a \right)x}} ,
\end{equation}
where
\begin{equation}
{u_ \pm }\left( {x;a} \right) = \frac{{\sigma \left( {x + {\omega _2} \pm a} \right)\sigma \left( {{\omega _2}} \right)}}{{\sigma \left( {x + {\omega _2}} \right)\sigma \left( {{\omega _2} \pm a} \right)}}{e^{ \mp \frac{{a\zeta \left( {{\omega _1}} \right)}}{{{\omega _1}}}x}},\quad ik\left( a \right) = \frac{{a\zeta \left( {{\omega _1}} \right) - {\omega _1}\zeta \left( a \right)}}{{{\omega _1}}} ,
\end{equation}
with ${u_ \pm }\left( {x + 2 \omega_1 ; a} \right) = {u_ \pm }\left( {x ; a} \right)$. The function $k\left( a \right)$ is identical to the one in the unbounded potential. Therefore, the band structure of the bounded potential is identical to the band structure of the unbounded problem.
\newline\newline
\textbf{Problem \ref{pr:lame_normalization} Solution}

By direct computation starting from the expressions \eqref{eq:lame_eigenstates}, we find
\begin{equation*}
{y_ + }{y_ - } = \frac{{\sigma \left( {x + a} \right)\sigma \left( {x - a} \right)}}{{{\sigma ^2}\left( x \right)\sigma \left( a \right)\sigma \left( { - a} \right)}}{e^{ - \zeta \left( a \right)x}}{e^{ - \zeta \left( { - a} \right)x}} =  - \frac{{\sigma \left( {x + a} \right)\sigma \left( {x - a} \right)}}{{{\sigma ^2}\left( x \right){\sigma ^2}\left( a \right)}} .
\end{equation*}
Using the pseudo-addition formula for the Weierstrass $\sigma$ function \eqref{eq:sigma_addition}, we get
\begin{equation}
{y_ + }{y_ - } = \wp \left( x \right) - \wp \left( a \right) .
\end{equation}

Similarly, using equation \eqref{eq:lame_first_der},
\begin{equation}
\begin{split}
{y_ + }'{y_ - } - {y_ + }{y_ - }' &= \left( {\frac{1}{2}\frac{{\wp '\left( x \right) - \wp '\left( a \right)}}{{\wp \left( x \right) - \wp \left( a \right)}} - \frac{1}{2}\frac{{\wp '\left( x \right) + \wp '\left( a \right)}}{{\wp \left( x \right) - \wp \left( a \right)}}} \right){y_ + }{y_ - }\\
 &=  - \frac{{\wp '\left( a \right)}}{{\wp \left( x \right) - \wp \left( a \right)}}\left( {\wp \left( x \right) - \wp \left( a \right)} \right) =  - \wp '\left( a \right) .
\end{split}
\end{equation}

We repeat for the eigenfunctions \eqref{eq:lame_eigenstates_shifted} of the bounded $n = 1$ \Lame potential
\begin{equation}
\begin{split}
{\psi _ + }{\psi _ - } &= \frac{{\sigma \left( {x + {\omega _2} + a} \right)\sigma \left( {x + {\omega _2} - a} \right){\sigma ^2}\left( {{\omega _2}} \right)}}{{{\sigma ^2}\left( {x + {\omega _2}} \right)\sigma \left( {{\omega _2} + a} \right)\sigma \left( {{\omega _2} - a} \right)}}{e^{ - \zeta \left( a \right)x}}{e^{ - \zeta \left( { - a} \right)x}}\\
 &= \frac{{\sigma \left( {x + {\omega _2} + a} \right)\sigma \left( {x + {\omega _2} - a} \right)}}{{{\sigma ^2}\left( {x + {\omega _2}} \right){\sigma ^2}\left( a \right)}}\frac{{{\sigma ^2}\left( {{\omega _2}} \right){\sigma ^2}\left( a \right)}}{{\sigma \left( {{\omega _2} + a} \right)\sigma \left( {{\omega _2} - a} \right)}}\\
 &= \frac{{\wp \left( {x + {\omega _2}} \right) - \wp \left( a \right)}}{{{e_3} - \wp \left( a \right)}}
\end{split}
\end{equation}
and
\begin{equation}
\begin{split}
{\psi _ + }'{\psi _ - } - {\psi _ + }{\psi _ - }' &= \left( {\frac{1}{2}\frac{{\wp '\left( {x + {\omega _2}} \right) - \wp '\left( a \right)}}{{\wp \left( {x + {\omega _2}} \right) - \wp \left( a \right)}} - \frac{1}{2}\frac{{\wp '\left( {x + {\omega _2}} \right) + \wp '\left( a \right)}}{{\wp \left( {x + {\omega _2}} \right) - \wp \left( a \right)}}} \right){\psi _ + }{\psi _ - }\\
 &=  - \frac{{\wp '\left( a \right)}}{{\wp \left( {x + {\omega _2}} \right) - \wp \left( a \right)}}\frac{{\wp \left( {x + {\omega _2}} \right) - \wp \left( a \right)}}{{{e_3} - \wp \left( a \right)}} =  - \frac{{\wp '\left( a \right)}}{{{e_3} - \wp \left( a \right)}} .
\end{split}
\end{equation}

Therefore the ``normalization'' of the eigenfunctions of the bounded potential differ from the ``normalization'' of the eigenfunctions of the unbounded potential by a factor of $\sqrt{e_3 - \wp \left( a \right)}$. Of course this is just a constant and it could be included to the definition of the bounded eigenfunctions. However, such an inclusion would mess the reality properties of the eigenfunctions, as the reality of this factor depends on the value of $\wp \left( a \right)$.
\newline\newline
\textbf{Problem \ref{pr:lame_creation} Solution}

Using the formula \eqref{eq:lame_first_der} and the definition of the superpotential \eqref{eq:lame_superpotential}, we find
\begin{equation*}
\begin{split}
{A^\dag }{y_ \pm }\left( {x;a} \right) &= \left( { - \frac{d}{{dx}} + W\left( x \right)} \right)\frac{{\sigma \left( {x \pm a} \right)}}{{\sigma \left( x \right)\sigma \left( { \pm a} \right)}}{e^{ - \zeta \left( { \pm a} \right)x}}\\
 &=  - \frac{1}{2}\left( {\frac{{\wp '\left( x \right) \mp \wp '\left( a \right)}}{{\wp \left( x \right) - \wp \left( a \right)}} - \frac{{\wp '\left( x \right)}}{{\wp \left( x \right) - {e_3}}}} \right)\frac{{\sigma \left( {x \pm a} \right)}}{{\sigma \left( x \right)\sigma \left( { \pm a} \right)}}{e^{ - \zeta \left( { \pm a} \right)x}} \\
 &\equiv f_\pm \left( {x,a} \right){y_ \pm }\left( {x;a} \right) .
\end{split}
\end{equation*}
The \Lame eigenfunctions themselves are not elliptic functions. However, it is trivial that the prefactor $f_\pm \left( {x,a} \right)$ is an elliptic function both as a function of $x$ or $a$.

Using the pseudo-addition formula for the Weierstrass $\zeta$ function \eqref{eq:zeta_addition}, we find
\begin{equation*}
\begin{split}
{f_ \pm }\left( {x,a} \right) &= \zeta \left( x \right) + \zeta \left( { \pm a} \right) - \zeta \left( {x \pm a} \right) + \zeta \left( {x + {\omega _2}} \right) - \zeta \left( x \right) - \zeta \left( {{\omega _2}} \right)\\
 &= \zeta \left( { \pm a} \right) - \zeta \left( {x \pm a} \right) + \zeta \left( {x + {\omega _2}} \right) - \zeta \left( {{\omega _2}} \right) .
\end{split}
\end{equation*}

Remembering that $f_\pm \left( {x,a} \right)$ as a function of $x$ is an elliptic function, we may write it as a ratio of Weierstrass $\sigma$ functions. $f_\pm \left( {x,a} \right)$ is clearly a second order elliptic function having two first order poles at $x = - \omega_2$ and $x = \mp a$. It also obviously has a zero at $x = 0$ since,
\begin{equation*}
{f_ \pm }\left( {0,a} \right) = \zeta \left( { \pm a} \right) - \zeta \left( { \pm a} \right) + \zeta \left( {{\omega _2}} \right) - \zeta \left( {{\omega _2}} \right) = 0 .
\end{equation*}
Theorem \ref{th:average_root_congruent_to_average_pole} implies that the other zero should be congruent to $x = \mp a - \omega_2$. Indeed,
\begin{equation*}
{f_ \pm }\left( { - {\omega _2} \mp a,a} \right) = \zeta \left( { \pm a} \right) - \zeta \left( { - {\omega _2}} \right) + \zeta \left( { \mp a} \right) - \zeta \left( {{\omega _2}} \right) = 0 .
\end{equation*}

The sum of the above poles equals the sum of the zeros, and, thus, we may write $f_\pm \left( {x,a} \right)$ as,
\begin{equation*}
{f_ \pm }\left( {x,a} \right) = C\frac{{\sigma \left( x \right)\sigma \left( {x + {\omega _2} \pm a} \right)}}{{\sigma \left( {x + {\omega _2}} \right)\sigma \left( {x \pm a} \right)}} .
\end{equation*}
Requiring that ${f_ \pm }\left( {x,a} \right)$ has residue equal to one at $x = - \omega_2$ yields $C = \frac{{\sigma \left( { - {\omega _2} \pm a} \right)}}{{\sigma \left( { - {\omega _2}} \right)\sigma \left( { \pm a} \right)}}$, and, thus,
\begin{equation*}
{f_ \pm }\left( {x,a} \right) = \frac{{\sigma \left( { - {\omega _2} \pm a} \right)\sigma \left( x \right)\sigma \left( {x + {\omega _2} \pm a} \right)}}{{\sigma \left( { - {\omega _2}} \right)\sigma \left( { \pm a} \right)\sigma \left( {x + {\omega _2}} \right)\sigma \left( {x \pm a} \right)}} .
\end{equation*}

Therefore,
\begin{equation}
\begin{split}
{A^\dag }{y_ \pm }\left( {x;a} \right) &= \frac{{\sigma \left( { - {\omega _2} \pm a} \right)\sigma \left( x \right)\sigma \left( {x + {\omega _2} \pm a} \right)}}{{\sigma \left( { - {\omega _2}} \right)\sigma \left( { \pm a} \right)\sigma \left( {x + {\omega _2}} \right)\sigma \left( {x \pm a} \right)}}\frac{{\sigma \left( {x \pm a} \right)}}{{\sigma \left( x \right)\sigma \left( { \pm a} \right)}}{e^{ - \zeta \left( { \pm a} \right)x}}\\
 &= \frac{{\sigma \left( { - {\omega _2} \pm a} \right)\sigma \left( {x + {\omega _2} \pm a} \right)}}{{\sigma \left( { - {\omega _2}} \right){\sigma ^2}\left( { \pm a} \right)\sigma \left( {x + {\omega _2}} \right)}}{e^{ - \zeta \left( { \pm a} \right)x}}\\
 &=  - \frac{{\sigma \left( {{\omega _2} \pm a} \right)\sigma \left( { - {\omega _2} \pm a} \right)}}{{{\sigma ^2}\left( {{\omega _2}} \right){\sigma ^2}\left( { \pm a} \right)}}\frac{{\sigma \left( {x + {\omega _2} \pm a} \right)\sigma \left( {{\omega _2}} \right)}}{{\sigma \left( {x + {\omega _2}} \right)\sigma \left( {{\omega _2} \pm a} \right)}}{e^{ - \zeta \left( { \pm a} \right)x}}\\
 &= \left( {\wp \left( a \right) - {e_3}} \right){\psi _ \pm }\left( {x;a} \right) .
\end{split}
\end{equation}
Indeed the action of the creation operator on the eigenfunctions of the unbounded $n = 1$ \Lame problem yields the eigenfunctions of the bounded problem multiplied with a constant. This constant is the same constant appearing in the ``normalization'' properties of the eigenfunctions in problem \ref{pr:lame_normalization}.

\end{document}